\documentclass[preprint2]{aastex631}

\usepackage{amsmath}
\usepackage{multirow}
\usepackage{xspace}

\newcommand{\aumicroscopii}{\object{AU Microscopii}\xspace}
\newcommand{\aumic}{\object{AU Mic}\xspace}
\newcommand{\brier}{{\sl Brierfield}\xspace}
\newcommand{\cfht}{{\sl CFHT}\xspace}
\newcommand{\cheops}{{\sl CHEOPS}\xspace}
\newcommand{\espresso}{{\sl ESPRESSO}\xspace}
\newcommand{\irac}{{\sl IRAC}\xspace}
\newcommand{\irtf}{{\sl IRTF}\xspace}
\newcommand{\ishell}{{\sl iSHELL}\xspace}
\newcommand{\kepler}{{\sl Kepler}\xspace}
\newcommand{\kt}{{\sl K2}\xspace}
\newcommand{\lco}{{\sl LCO}\xspace}
\newcommand{\lcogt}{{\sl LCOGT}\xspace}
\newcommand{\saao}{{\sl LCO SAAO}\xspace}
\newcommand{\sso}{{\sl LCO SSO}\xspace}
\newcommand{\moravian}{{\sl Moravian 16803}\xspace}
\newcommand{\pest}{{\sl PEST}\xspace}
\newcommand{\sbig}{{\sl SBIG ST-8XME}\xspace}
\newcommand{\sinistro}{{\sl Sinistro}\xspace}
\newcommand{\spirou}{{\sl SPIRou}\xspace}
\newcommand{\spitzer}{{\sl Spitzer}\xspace}
\newcommand{\tess}{{\sl TESS}\xspace}
\newcommand{\vlt}{{\sl VLT}\xspace}
\newcommand{\aij}{{\tt AIJ}\xspace}
\newcommand{\celerite}{{\tt celerite}\xspace}
\newcommand{\celeritetwo}{{\tt celerite2}\xspace}
\newcommand{\exofast}{{\tt EXOFASTv2}\xspace}
\newcommand{\exostriker}{{\tt Exo-Striker}\xspace}
\newcommand{\mercury}{{\tt mercury6}\xspace}
\newcommand{\rebound}{{\tt rebound}\xspace}
\newcommand{\tbnm}{\tablenotemark}
\newcommand{\scze}{\scriptsize}
\newcommand{\mstar}{M$_{\star}$\xspace}
\newcommand{\rstar}{R$_{\star}$\xspace}
\newcommand{\lstar}{L$_{\star}$\xspace}
\newcommand{\msun}{M$_{\sun}$\xspace}
\newcommand{\rsun}{R$_{\sun}$\xspace}
\newcommand{\lsun}{L$_{\sun}$\xspace}
\newcommand{\mplan}{M$_{\rm p}$\xspace}
\newcommand{\rplan}{R$_{\rm p}$\xspace}
\newcommand{\mjup}{M$_{\rm J}$\xspace}
\newcommand{\rjup}{R$_{\rm J}$\xspace}
\newcommand{\mear}{M$_{\earth}$\xspace}
\newcommand{\rear}{R$_{\earth}$\xspace}
\newcommand{\teff}{T$_{\rm eff}$\xspace}
\newcommand{\porb}{P$_{\rm orb}$\xspace}
\newcommand{\tc}{T$_{\rm C}$\xspace}
\newcommand{\zs}{z$_{\rm s}$\xspace}
\newcommand{\logmstmsu}{log$_{10}\left(\frac{{\rm M}_{\rm \star}}{{\rm M}_{\rm \sun}}\right)$}
\newcommand{\logfracpd}{log$_{10}\left(\frac{\rm Period}{\rm days}\right)$}
\newcommand{\ergfrac}{$\frac{{\rm erg}}{{\rm s}~{\rm cm}^{2}}$}
\newcommand{\radroot}{$\sqrt{\left(1-\frac{{\rm R}_{\rm p}}{{\rm R}_{\star}}\right)^{2}-\mbox{b}^{2}}$}
\newcommand{\reduced}{$\chi^{2}_{\rm red}$\xspace}

\begin{document}

\title{Transit Timing Variations for \aumicroscopii b \& c}

\author[0000-0002-7424-9891]{Justin M. Wittrock}
\affiliation{Department of Physics \& Astronomy, George Mason University, 4400 University Drive MS 3F3, Fairfax, VA 22030, USA}
\email{jwittroc@gmu.edu}

\author[0000-0001-6187-5941]{Stefan Dreizler}
\affiliation{Institut f\"ur Astrophysik, Georg-August-Universit\"at, Friedrich-Hund-Platz 1, 37077 G\"ottingen, Germany}

\author[0000-0003-4701-8497]{Michael A. Reefe}
\affiliation{Department of Physics \& Astronomy, George Mason University, 4400 University Drive MS 3F3, Fairfax, VA 22030, USA}

\author[0000-0003-2528-3409]{Brett M. Morris}
\affiliation{Center for Space and Habitability, University of Bern, Gesellschaftsstrasse 6, 3012 Bern, Switzerland}

\author[0000-0002-8864-1667]{Peter P. Plavchan}
\affiliation{Department of Physics \& Astronomy, George Mason University, 4400 University Drive MS 3F3, Fairfax, VA 22030, USA}

\author[0000-0001-8014-0270]{Patrick J. Lowrance}
\affiliation{IPAC-Spitzer, California Institute of Technology, MC 314-6, 1200 E. California Blvd., Pasadena, California 91125, USA}

\author[0000-0002-9355-5165]{Brice-Olivier Demory}
\affiliation{Center for Space and Habitability, University of Bern, Gesellschaftsstrasse 6, 3012 Bern, Switzerland}

\author[0000-0003-4714-1364]{James G. Ingalls}
\affiliation{IPAC-Spitzer, California Institute of Technology, MC 314-6, 1200 E. California Blvd., Pasadena, California 91125, USA}

\author[0000-0002-0388-8004]{Emily A. Gilbert}
\affiliation{Department of Astronomy \& Astrophysics, University of Chicago, 5640 S. Ellis Ave, Chicago, IL 60637, USA}
\affiliation{University of Maryland, Baltimore County, 1000 Hilltop Circle, Baltimore, MD 21250, USA}
\affiliation{The Adler Planetarium, 1300 South Lakeshore Drive, Chicago, IL 60605, USA}
\affiliation{NASA Goddard Space Flight Center, 8800 Greenbelt Road, Greenbelt, MD 20771, USA}
\affiliation{GSFC Sellers Exoplanet Environments Collaboration}

\author[0000-0001-7139-2724]{Thomas Barclay}
\affiliation{University of Maryland, Baltimore County, 1000 Hilltop Circle, Baltimore, MD 21250, USA}
\affiliation{NASA Goddard Space Flight Center, 8800 Greenbelt Road, Greenbelt, MD 20771, USA}


\author[0000-0002-2078-6536]{Bryson L. Cale}
\affiliation{Department of Physics \& Astronomy, George Mason University, 4400 University Drive MS 3F3, Fairfax, VA 22030, USA}

\author[0000-0001-6588-9574]{Karen A. Collins}
\affiliation{Center for Astrophysics \textbar \ Harvard \& Smithsonian, 60 Garden Street, Cambridge, MA 02138, USA}

\author[0000-0003-2781-3207]{Kevin I. Collins}
\affiliation{Department of Physics \& Astronomy, George Mason University, 4400 University Drive MS 3F3, Fairfax, VA 22030, USA}

\author[0000-0002-1835-1891]{Ian J. M. Crossfield}
\affiliation{Department of Physics \& Astronomy, University of Kansas, 1251 Wescoe Hall Dr, Lawrence, KS 66045, USA}

\author[0000-0003-2313-467X]{Diana Dragomir}
\affiliation{Department of Physics \& Astronomy, University of New Mexico, Albuquerque, NM, USA}

\author[0000-0003-3773-5142]{Jason D. Eastman}
\affiliation{Center for Astrophysics \textbar \ Harvard \& Smithsonian, 60 Garden Street, Cambridge, MA 02138, USA}

\author[0000-0001-8364-2903]{Mohammed El Mufti}
\affiliation{Department of Physics \& Astronomy, George Mason University, 4400 University Drive MS 3F3, Fairfax, VA 22030, USA}

\author[0000-0002-2457-7889]{Dax Feliz}
\affiliation{Department of Physics \& Astronomy, Vanderbilt University, 6301 Stevenson Center Ln., Nashville, TN 37235, USA}

\author[0000-0002-2592-9612]{Jonathan Gagn\'e}
\affiliation{Plan\'etarium Rio Tinto Alcan, Espace pour la vie, 4801 av. Pierre-De Coubertin, Montr\'eal, QC H1V 3V4, Canada}
\affiliation{Institute for Research on Exoplanets, Universit\'e de Montr\'eal, D\'epartement de Physique, C.P. 6128 Succ. Centre-ville, Montr\'eal, QC H3C 3J7, Canada}

\author[0000-0002-5258-6846]{Eric Gaidos}
\affiliation{University of Hawai`i at M$\bar{a}$noa, 1680 East-West Road, Honolulu, HI 96822, USA}

\author[0000-0002-8518-9601]{Peter Gao}
\affiliation{Earth and Planets Laboratory, Carnegie Institution for Science, 5241 Broad Branch Rd NW, Washington, DC 20015, USA}

\author[0000-0001-9596-8820]{Claire S. Geneser}
\affiliation{Mississippi State University, 75 B. S. Hood Road, Mississippi State, MS 39762, USA}

\author[0000-0003-1263-8637]{Leslie Hebb}
\affiliation{Physics Department, Hobart and William Smith Colleges,  Geneva, NY 14456, USA}

\author{Christopher E. Henze}
\affiliation{NASA Ames Research Center, MS 244-30, Moffett Field, CA 94035, USA}

\author[0000-0003-1728-0304]{Keith D. Horne}
\affiliation{SUPA Physics and Astronomy, University of St. Andrews, Fife, KY16 9SS Scotland, UK}

\author[0000-0002-4715-9460]{Jon M. Jenkins}
\affiliation{NASA Ames Research Center, MS 244-30, Moffett Field, CA 94035, USA}

\author[0000-0002-4625-7333]{Eric L. N. Jensen}
\affiliation{Department of Physics \& Astronomy, Swarthmore College, Swarthmore PA 19081, USA}

\author[0000-0002-7084-0529]{Stephen R. Kane}
\affiliation{University of California, Riverside, 900 University Ave., Riverside, CA 92521, USA}

\author[0000-0002-4311-9250]{Laurel Kaye}
\affiliation{Sub-department of Astrophysics, Department of Physics, University of Oxford, Denys Wilkinson Building, Keble Road, Oxford, OX1 3RH, UK}

\author[0000-0002-5084-168X]{Eder Martioli}
\affiliation{Sorbonne Universit\'e, CNRS, UMR 7095, Institut d’Astrophysique de Paris, 98 bis bd Arago, 75014 Paris, France}
\affiliation{Laborat\'orio Nacional de Astrof\'isica, Rua Estados Unidos 154, Itajub\'a, MG 37504-364, Brazil}

\author[0000-0003-3896-3059]{Teresa A. Monsue}
\affiliation{NASA Goddard Space Flight Center, 8800 Greenbelt Road, Greenbelt, MD 20771, USA}

\author[0000-0003-0987-1593]{Enric Pall\'e}
\affiliation{Instituto de Astrof\'isica de Canarias, V\'ia L\'actea s/n, E-38205 La Laguna, Tenerife, Spain}
\affiliation{Departamento de Astrof\'isica, Universidad de La Laguna, E-38206 La Laguna, Tenerife, Spain}

\author[0000-0003-1309-2904]{Elisa V. Quintana}
\affiliation{NASA Goddard Space Flight Center, 8800 Greenbelt Road, Greenbelt, MD 20771, USA}

\author[0000-0002-3940-2360]{Don J. Radford}
\affiliation{Brierfield Observatory, Bowral, New South Wales, Australia}

\author[0000-0002-4650-594X]{Veronica Roccatagliata}
\affiliation{Dipartimento di Fisica “Enrico Fermi”, Universita’ di Pisa, Largo Pontecorvo 3, 56127 Pisa, Italy}
\affiliation{INFN, Sezione di Pisa, Largo Bruno Pontecorvo 3, 56127 Pisa, Italy}
\affiliation{INAF-Osservatorio Astrofisico di Arcetri, Largo E. Fermi 5, 50125 Firenze, Italy}

\author[0000-0001-5347-7062]{Joshua E. Schlieder}
\affiliation{NASA Goddard Space Flight Center, 8800 Greenbelt Road, Greenbelt, MD 20771, USA}

\author[0000-0001-8227-1020]{Richard P. Schwarz}
\affiliation{Patashnick Voorheesville Observatory, Voorheesville, NY 12186, USA}

\author[0000-0002-1836-3120]{Avi Shporer}
\affiliation{Department of Physics and Kavli Institute for Astrophysics and Space Research, Massachusetts Institute of Technology, Cambridge, MA 02139, USA}

\author[0000-0002-3481-9052]{Keivan G. Stassun}
\affiliation{Department of Physics \& Astronomy, Vanderbilt University, 6301 Stevenson Center Ln., Nashville, TN 37235, USA}

\author[0000-0003-2163-1437]{Christopher Stockdale}
\affiliation{Hazelwood Observatory, Hazelwood South, Victoria, Australia}

\author[0000-0001-5603-6895]{Thiam-Guan Tan}
\affiliation{Perth Exoplanet Survey Telescope, Perth, Western Australia, Australia}
\affiliation{Curtin Institute of Radio Astronomy, Curtin University, Bentley, Western Australia 6102}

\author[0000-0002-2903-2140]{Angelle M. Tanner}
\affiliation{Mississippi State University, 75 B. S. Hood Road, Mississippi State, MS 39762, USA}

\author[0000-0001-7246-5438]{Andrew Vanderburg}
\affiliation{Department of Physics and Kavli Institute for Astrophysics and Space Research, Massachusetts Institute of Technology, Cambridge, MA 02139, USA}

\author[0000-0002-5928-2685]{Laura D. Vega}
\affiliation{Department of Astronomy, University of Maryland, College Park, MD 20742, USA}
\affiliation{NASA Goddard Space Flight Center, 8800 Greenbelt Road, Greenbelt, MD 20771, USA}
\affiliation{Center for Research and Exploration in Space Science \& Technology, NASA/GSFC, Greenbelt, MD 20771, USA}

\author[0000-0002-7846-6981]{Songhu Wang}
\affiliation{Department of Astronomy, Indiana University, Bloomington, IN 47405, USA}

\begin{abstract}
{}
{We explore the transit timing variations (TTVs) of the young (22 Myr) nearby \aumic planetary system. For \aumic b, we introduce three \spitzer (4.5 $\mu$m) transits, five \tess transits, 11 \lco transits, one \pest transit, one \brier transit, and two transit timing measurements from Rossiter-McLaughlin observations; for \aumic c, we introduce three \tess transits.}
{We present two independent TTV analyses. First, we use \exofast to jointly model the \spitzer and ground-based transits and to obtain the midpoint transit times. We then construct an O--C diagram and model the TTVs with \exostriker. Second, we reproduce our results with an independent photodynamical analysis.}
{We recover a TTV mass for \aumic c of 10.8$^{+2.3}_{-2.2}$ \mear. We compare the TTV-derived constraints to a recent radial-velocity (RV) mass determination. We also observe excess TTVs that do not appear to be consistent with the dynamical interactions of b and c alone, and do not appear to be due to spots or flares. Thus, we present a hypothetical non-transiting ``middle-d'' candidate exoplanet that is consistent with the observed TTVs, the candidate RV signal, and would establish the AU Mic system as a compact resonant multi-planet chain in a 4:6:9 period commensurability.}
{These results demonstrate that the \aumic planetary system is dynamically interacting producing detectable TTVs, and the implied orbital dynamics may inform the formation mechanisms for this young system. We recommend future RV and TTV observations of \aumic b and c to further constrain the masses and to confirm the existence of possible additional planet(s).}
\end{abstract}

\section{Introduction}

Exoplanetary sciences have been expanding over the past few decades, with its fields increasingly diversifying thanks in large part to several successful and diligent missions, including \kepler \citep{borucki2010}, \kt \citep{howell2014}, and the {\sl Transiting Exoplanet Survey Satellite} \citep[\tess,][]{ricker2015}. \tess has detected 4\,548 transiting candidates (TOIs) as of 2021 October 7 and 159 confirmed planets as of 2021 October 4\footnote{\url{https://exoplanetarchive.ipac.caltech.edu}}. Many discoveries have challenged our theories of planet formation, such as hot Jupiters -- e.g., 51 Pegasi b \citep{mayor1995}, HD 209458 b \citep{henry2000}, TOI-628 b \citep{rodriguez2021} -- planets in highly eccentric orbit -- e.g., 16 Cygni B b \citep{cochran1997}, BD+63 1405 b \citep{dalal2021}, HD 26161 b \citep{rosenthal2021} -- and compact systems -- e.g., HD 108236 \citep{daylan2021, bonfanti2021}, TOI-178 \citep{leleu2021}, TRAPPIST-1 \citep{gillon2016, gillon2017-1}. One way to investigate how these systems form and evolve can be done by probing young stellar systems, when their characteristics and orbital dynamics are still undergoing progression. Several young exoplanet systems have recently been discovered by the \tess and \kt missions -- e.g., DS Tucanae A \citep{newton2019}, K2-25 \citep{mann2016}, K2-33 \citep{david2016}, V1298 Tauri \citep{david2019} -- and other exoplanet detection methods including direct imaging and radial velocities (RVs) -- e.g., HD 47366 \citep{sato2016}, HR 8799 \citep{marois2008, marois2010}, PDS 70 \citep{keppler2018, haffert2019}, $\beta$ Pictoris \citep{lagrange2009, lagrange2019}. Further probing of certain systems such as PDS 70 revealed a potentially moon-forming circumplanetary disk around PDS 70 c \citep{benisty2021}. This new and growing population of transiting young exoplanets has recently enabled a new frontier in the study of planet formation and evolution. Among the nearby young exoplanet systems, the nearest one is \aumicroscopii (Tables \ref{table:stellarquant} \& \ref{table:planetquant}).

\begin{deluxetable}{l|c|c|c}
    \tabletypesize{\scriptsize}
    \tablecaption{\label{table:stellarquant}Stellar properties for host star \aumic.}
    \tablehead{Property & Unit & Quantity &  Ref}
    \startdata
Spectral Type        & ...       & M1Ve                  & ... \\
m$_{\rm V}$          & ...       & 8.81 $\pm$ 0.10       & ... \\
m$_{\rm TESS}$       & ...       & 6.755 $\pm$ 0.032     & ... \\
$\alpha_{\rm J2000}$ & h:m:s     & 20:45:09.53           & 1   \\
$\delta_{\rm J2000}$ & deg:am:as & -31:20:27.24          & 1   \\
$\mu_{\rm \alpha}$   & mas/yr    & 281.424 $\pm$ 0.075   & 1   \\
$\mu_{\rm \delta}$   & mas/yr    & -359.895 $\pm$ 0.054  & 1   \\
Distance             & pc        & 9.7221 $\pm$ 0.0046   & 2   \\
Parallax             & mas       & 102.8295 $\pm$ 0.0486 & 1   \\
\mstar               & \msun     & 0.50 $\pm$ 0.03       & 3   \\
\rstar               & \rsun     & 0.75 $\pm$ 0.03       & 4   \\
\teff                & K         & 3\,700 $\pm$ 100      & 5   \\
\lstar               & \lsun     & 0.09                  & 5   \\
Age                  & Myr       & 22 $\pm$ 3            & 6   \\
P$_{\rm rot}$        & days      & 4.863 $\pm$ 0.010     & 3   \\
v $\sin{i}$          & km/s      & 8.7 $\pm$ 0.2         & 7
    \enddata
    \tablerefs{(1) \citet{gaia2018}; (2) \citet{bailer-jones2018}; (3) \citet{plavchan2020}; (4) \citet{white2019}; (5) \citet{plavchan2009}; (6) \citet{mamajek2014}; (7) \citet{lannier2017}}
\end{deluxetable}

\begin{deluxetable*}{l|c|c|c|c|c}
    \tablecaption{\label{table:planetquant}Planetary properties for \aumic system.}
    \tablehead{Property & Description & Unit & \aumic b & \aumic c & Ref}
    \startdata
\porb                   & Orbital Period                    & days       & 8.4630004$^{+0.0000058}_{-0.0000060}$ & 18.858982$^{+0.000053}_{-0.000050}$ & 1                  \\
a                       & Semi-Major Axis                   & au         & 0.0645 $\pm$ 0.0013                   & 0.1101 $\pm$ 0.0022                 & 2                  \\
e                       & Eccentricity                      & ...        & 0.12$^{+0.16}_{-0.08}$                & 0.13$^{+0.16}_{-0.09}$              & 1                  \\
i                       & Inclination                       & deg        & 89.5 $\pm$ 0.3                        & 89.0$^{+0.5}_{-0.4}$                & 2                  \\
$\omega$                & Argument of Periastron            & deg        & -0.3$^{+2.4}_{-2.3}$                  & -0.3$^{+2.5}_{-2.2}$                & 1                  \\
\multirow{2}{*}{\mplan} & \multirow{2}{*}{Planetary Mass}   & \mjup      & 0.054 $\pm$ 0.015                     & 0.007 $<$ M$_{c}$ $<$ 0.079         & \multirow{2}{*}{2} \\
                        &                                   & \mear      & 17 $\pm$ 5                            & 2 $<$ M$_{c}$ $<$ 25                &                    \\
\multirow{2}{*}{\rplan} & \multirow{2}{*}{Planetary Radius} & \rjup      & 0.374$^{+0.021}_{-0.020}$             & 0.249$^{+0.028}_{-0.027}$           & \multirow{2}{*}{1} \\
                        &                                   & \rear      & 4.19$^{+0.24}_{-0.22}$                & 2.79$^{+0.31}_{-0.30}$              &                    \\
$\rho_{\rm p}$          & Planetary Density                 & g/cm$^{3}$ & 1.4 $\pm$ 0.4                         & 0.4 $<$ $\rho_{c}$ $<$ 4.1          & 2                  \\
K                       & RV Semi-Amplitude                 & m/s        & 8.5$^{+2.3}_{-2.2}$                   & 0.8 $<$ K$_{c}$ $<$ 9.5             & 2                  \\
\tc $-$ 2\,458\,000     & Time of Conjunction               & BJD        & 330.39051 $\pm$ 0.00015               & 342.2223 $\pm$ 0.0005               & 2                  \\
t$_{\rm duration}$      & Transit Duration                  & hours      & 3.50 $\pm$ 0.08                       & 4.5 $\pm$ 0.8                       & 2                  \\
\rplan/\rstar           & ...                               & ...        & 0.0512 $\pm$ 0.0020                   & 0.0340$^{+0.0034}_{-0.0033}$        & 1                  \\
a/\rstar                & ...                               & ...        & 19.1 $\pm$ 0.3                        & 29 $\pm$ 3                          & 2                  \\
b                       & Impact Parameter                  & ...        & 0.16$^{+0.13}_{-0.11}$                & 0.30$^{+0.21}_{-0.20}$              & 1
    \enddata
    \tablerefs{(1) \citet{gilbert2021}; (2) \citet{martioli2021}}
\end{deluxetable*}

\aumicroscopii (TOI-2221, TIC 441420236, HD 197481, GJ 803) is a young \citep[$22 \pm 3$ Myr,][]{mamajek2014}, nearby \citep[9.7 pc,][]{bailer-jones2018} BY Draconis variable star with spectral type M1Ve and relative brightness m$_{\rm V}$=8.81 mag. It is known to be fairly active, with numerous flares having been observed and studied at various wavelengths \citep{butler1981, kundu1987, cully1993, tsikoudi2000, gilbert2021}. \citet{kalas2004} observed the presence of a large dust disk having a radius between 50 and 210 au from the young star, having first been detected as a mid-infrared flux excess with {\sl IRAS} \citep{fajardo-acosta2000, zuckerman2001, song2002, liu2004, plavchan2005}. Later, \citet{plavchan2020} discovered a Neptune-sized transiting planet \aumic b interior to a spatially-resolved debris disc and with orbital period of 8.46 days. Recently, \citet{gilbert2021} and \citet{martioli2021} confirmed the existence of another planet \aumic c with orbital period of 18.86 days, which put the planets near a 4:9 orbital commensurability. The aforementioned traits of \aumic and its planets make this system a unique, viable laboratory for studying the stellar activity of a young M dwarf, the planetary formation, the evolution of exoplanet radii as a function of age, orbital architectures of young giant planet systems, atmospheric characteristics of young exoplanets, and the interplay between planets and disks.

One method that serves as a useful tool for probing the exoplanetary systems is Transit Timing Variations (TTVs). Compared to other detection methods, TTVs can detect terrestrial-mass planets with greater ease \citep{holman2005}. The planets that are in orbital resonance with each other can amplify the TTV signals \citep{agol2005}, so TTVs can be used to search for and measure the masses of other planets within a given stellar system \citep[e.g., including noteworthy systems presented in ][]{mazeh2013, becker2015, gillon2017-1, grimm2018}. Many systems have been characterized with TTVs, such as HIP 41378 \citep{bryant2021}, K2-146 \citep{lam2020}, TOI-216 \citep{dawson2021}, TOI-1266 \citep{demory2020}, TrES-3 \citep{mannaday2020}, and many Kepler systems \citep{lithwick2012, mazeh2013, hadden2014}. \citet{martioli2021} searched for the TTVs of \aumic transits from \tess light curves but did not identify any significant TTVs. \citet{szabo2021} did a TTV joint model with \tess and \cheops data; they found \aumic b's $\sim$3.9-minute variation across 33 days and attributed \aumic c as the potential source of this perturbation. \citet{gilbert2021} performed an independent analysis of \aumic transits from \tess light curves and were able to detect the TTVs on the order of $\sim$80 seconds.

For this paper, we examine the TTVs of \aumic planets by incorporating additional ground and space observations to our analysis. We present the TTVs of \aumic b and c to recover constraints on the mass for \aumic c and which indicates the presence of TTV excess that cannot be accounted by both planets b and c alone. In $\S$\ref{sec:dataobs}, we list the light curve data we include for TTV analysis and elaborate on some of the processes that were involved in data reduction. $\S$\ref{sec:exofastmod} covers the two critical developments: joint-modeling both the ground-based photometric and \spitzer light curves and extracting the midpoint transit times from these sets using the \exofast package \citep{eastman2019}, and constructing the O--C diagram using the extracted midpoint times from the observations. Then, as explained in $\S$\ref{sec:ttvs}, we model the extracted TTVs using the \exostriker package \citep{trifonov2019}. Next, we attempt to reproduce our results with an independent and direct photodynamical analysis as described in $\S$\ref{sec:photodyn}. Lastly, we discuss the results in $\S$\ref{sec:result} and close this paper in $\S$\ref{sec:conclude}.

\section{Data from Observations}\label{sec:dataobs}

We obtained 23 \aumic b transits and 3 \aumic c transits from three years worth of observations with multiple telescopes and have included them in the analysis (Table \ref{table:datasets}). In addition to space-based observations, with original transit observations from \tess and follow-ups from \spitzer, we have utilized several ground-based facilities in conducting follow-ups of \aumic, including \brier, \saao, \sso, \& \pest for photometric observations and \cfht equipped with \spirou, \irtf equipped with \ishell, \& \vlt equipped with \espresso for Rossiter-McLaughlin (R-M) observations (Tables \ref{table:facilities} \& \ref{table:instruments}). The \tess transits and one of the \spitzer transits have been previously presented in \citet{plavchan2020, gilbert2021, martioli2021}, and the R-M observation in \citet{martioli2020, palle2020}. The following subsections detail each telescope and the methodology employed upon its respective data sets.

\begin{deluxetable*}{c|l|c|c|c|c|c|r|c}
    \tablecaption{\label{table:datasets}List of \aumic observation data incorporated for TTV analysis. All ground-based photometric observations listed here were organized via TESS Follow-up Observing Program (TFOP) Working Group (WG)\tbnm{a}.}
    \tablehead{\multirow{2}{*}{Planet} & \multirow{2}{*}{Telescope} & \multirow{2}{*}{Date (UT)} & \multirow{2}{*}{Filter} & Exposure & No. of & Obs. Dur. & Transit & Ref \\
        & & & & Time (sec) & Images & (min) & Coverage & }
    \startdata
b                  & \brier 0.36 m                        & 2020-08-13 & I              & 16    & 398    & 379   & full    & ...                  \\
\hline
b                  & \cfht (\spirou)                      & 2019-06-17 & 955-2\,515 nm  & 122.6 & 116    & 302.8 & egress  & 1                    \\
\hline
b                  & \irtf (\ishell)                      & 2019-06-17 & 2.18-2.47 nm   & 120   & 47     & 105.2 & egress  & 1                    \\
\hline
\multirow{6}{*}{b} & \multirow{6}{*}{\saao 1.0 m}         & 2020-05-20 & Pan-STARRS Y   & 35    & 99     & 262   & egress  & \multirow{6}{*}{...} \\
                   &                                      & 2020-05-20 & Pan-STARRS \zs & 15    & 333    & 266   & egress  &                      \\
                   &                                      & 2020-06-06 & Pan-STARRS \zs & 15    & 266    & 218   & egress  &                      \\
                   &                                      & 2020-06-23 & Pan-STARRS \zs & 15    & 223    & 183   & egress  &                      \\
                   &                                      & 2020-09-07 & Pan-STARRS \zs & 15    & 211    & 172   & ingress &                      \\
                   &                                      & 2020-10-11 & Pan-STARRS \zs & 15    & 311    & 266   & ingress &                      \\
\hline
\multirow{5}{*}{b} & \multirow{5}{*}{\sso 1.0 m}          & 2020-04-25 & Pan-STARRS Y   & 35    & 40     & 104   & egress  & \multirow{5}{*}{...} \\
                   &                                      & 2020-04-25 & Pan-STARRS \zs & 15    & 212    & 172   & egress  &                      \\
                   &                                      & 2020-08-13 & Pan-STARRS \zs & 15    & 379    & 312   & full    &                      \\
                   &                                      & 2020-09-16 & Pan-STARRS \zs & 15    & 408    & 340   & full    &                      \\
                   &                                      & 2020-10-03 & Pan-STARRS \zs & 15    & 248    & 219   & egress  &                      \\
\hline
b                  & \pest 0.30 m                         & 2020-07-10 & V              & 15    & 1\,143 & 556   & full    & ...                  \\
\hline
\multirow{3}{*}{b} & \multirow{3}{*}{\spitzer (\irac)}    & 2019-02-10 & 4.5 $\mu$m     & 0.08  & 3\,020 & 475.7 & full    & \multirow{3}{*}{...} \\
                   &                                      & 2019-02-27 & 4.5 $\mu$m     & 0.08  & 3\,377 & 475.7 & egress  &                      \\
                   &                                      & 2019-09-09 & 4.5 $\mu$m     & 0.08  & 6\,002 & 990.9 & full    &                      \\
\hline
\multirow{5}{*}{b} & \multirow{5}{*}{\tess\tbnm{\scze b}} & 2018-07-26 & TESS           & 120   & 329    & 718.0 & full    & \multirow{5}{*}{2}   \\
                   &                                      & 2018-08-12 & TESS           & 120   & 296    & 708.0 & full    &                      \\
                   &                                      & 2020-07-10 & TESS           & 20    & 2\,132 & 719.7 & full    &                      \\
                   &                                      & 2020-07-19 & TESS           & 20    & 2\,137 & 719.7 & full    &                      \\
                   &                                      & 2020-07-27 & TESS           & 20    & 2\,120 & 719.7 & full    &                      \\
\hline
\multirow{3}{*}{c} & \multirow{3}{*}{\tess\tbnm{\scze b}} & 2018-08-11 & TESS           & 120   & 342    & 718.0 & full    & \multirow{3}{*}{2}   \\
                   &                                      & 2020-07-09 & TESS           & 20    & 2\,138 & 719.7 & full    &                      \\
                   &                                      & 2020-07-28 & TESS           & 20    & 2\,133 & 719.7 & full    &                      \\
\hline
b                  & \vlt (\espresso)                     & 2019-08-07 & 378.2-788.7 nm & 200   & 88     & 359   & full    & 3
    \enddata
    \tablenotetext{a}{\url{https://tess.mit.edu/followup}}
    \tablenotetext{b}{$\sim$12-hour snippets of the $\sim$27-day duration \tess Cycle 1 and 3 light curves were extracted for our analysis, centered approximately on each transit.}
    \tablerefs{(1) \citet{martioli2020}; (2) \citet{gilbert2021}; (3) \citet{palle2020}}
\end{deluxetable*}

\begin{deluxetable*}{l|l|l|r|c|c|c|c}
    \tablecaption{\label{table:facilities}List of facilities utilized for photometric and Rossiter-McLaughlin follow-up observations of \aumic.}
    \tablehead{\multirow{2}{*}{Telescope} & \multirow{2}{*}{Instrument} & \multirow{2}{*}{Location} & Aperture & Pixel Scale & Resolution & FOV & \multirow{2}{*}{Ref} \\
        & & & (m) & (arcsec) & (pixels) & (arcmin) &}
    \startdata
\brier   & \moravian & Bowral, New South Wales       & 0.36   & 0.732 & 4\,096$\times$4\,096 & 50$\times$50     & 1 \\
\cfht    & \spirou   & Maunakea, Hawai`i             & 3.58   & ...   & ...                  & ...              & 2 \\
\irtf    & \ishell   & Maunakea, Hawai`i             & 3.2    & ...   & ...                  & ...              & 3 \\
\saao    & \sinistro & Sutherland, South Africa      & 1.0    & 0.389 & 4\,096$\times$4\,096 & 26.5$\times$26.5 & 4 \\
\sso     & \sinistro & Mount Woorut, New South Wales & 1.0    & 0.389 & 4\,096$\times$4\,096 & 26.5$\times$26.5 & 4 \\
\pest    & \sbig     & Perth, Western Australia      & 0.3048 & 1.23  & 1\,530$\times$1\,020 & 31$\times$21     & 5 \\
\spitzer & \irac     & ...                           & 0.85   & 1.22  & 256$\times$256       & 5.2$\times$5.2   & 6 \\
\vlt     & \espresso & Cerro Paranal, Chile          & 8.2    & ...   & ...                  & ...              & 7
    \enddata
    \tablerefs{(1) \url{https://www.brierfieldobservatory.com}; (2) \url{https://www.cfht.hawaii.edu}; (3) \url{http://irtfweb.ifa.hawaii.edu}; (4) \url{https://lco.global/observatory}; (5) \url{http://pestobservatory.com}; (6) \url{https://www.spitzer.caltech.edu}; (7) \url{https://www.eso.org/public/teles-instr/paranal-observatory/vlt}}
\end{deluxetable*}

\begin{deluxetable*}{l|l|l|c|r|l|r|c}
    \tablecaption{\label{table:instruments}Specifications of instruments used for ground-based Rossiter-McLaughlin follow-up observations of \aumic.}
    \tablehead{\multirow{2}{*}{Instrument} & \multirow{2}{*}{Telescope} & \multirow{2}{*}{Observing Mode} & $\lambda$ Range & Resolving & Aperture & Average & \multirow{2}{*}{Ref} \\
        & & & (nm) & Power & (arcsec) & SNR &}
    \startdata
\espresso & \vlt  & HR (1-UT)                      & 378.2-788.7 & 140\,000 & 1.0   & 93.9 & 1    \\
\ishell   & \irtf & K$_{\rm gas}$                  & 2.18-2.47   & 75\,000  & 0.125 & 65   & 2, 4 \\
\spirou   & \cfht & Stokes $V$ Spectropolarimetric & 955-2\,515  & 70\,000  & 1.29  & 242  & 3, 4
    \enddata
    \tablerefs{(1) \citet{donati2020}, \citet{palle2020}; (2) \citet{rayner2016}; (3) \citet{pepe2021}; (4) \citet{martioli2020}}
\end{deluxetable*}

\subsection{\tess Photometry}

\tess\footnote{\url{https://tess.mit.edu} \\ \url{https://heasarc.gsfc.nasa.gov/docs/tess}} is a space-based telescope designed to scan nearby bright F5-M5 stars for transiting exoplanets \citep{ricker2015}. Since its launch on 2018 April 18 and the start of its primary mission on 2018 July 25, \tess has been probing the sky for $\sim$3 years as of this writing and has made numerous groundbreaking contributions to planetary detection -- e.g., DS Tuc A \citep{newton2019}, TOI-700 \citep{gilbert2020}, TOI-1338 \citep{kostov2020}. Its two-year primary mission divided the sky into the Southern and the Northern Ecliptic Hemispheres, with each being divided further into 13 sectors. \tess began its search in the Southern Ecliptic Hemisphere and probed each sector for 28 days. Within each 28-day span, a subset of primary exoplanet transit search target stars in a given sector were monitored at 2-minute cadence, and Full Frame Images (FFIs) were collected at 30-minute cadence. The data collected by \tess are then processed by the Science Processing Operations Center (SPOC), which functions to generate the calibrated images, perform aperture photometry, remove systematic artifacts, and searches the light curves for transiting planet signatures \citep{jenkins2016}. \tess successfully completed its two-year primary mission and is now in its extended mission by repeating its observation in each of 26 sectors, with some notable differences: \tess is probing or will probe new targets along with the old targets, the 2-minute cadence for 20\,000 targets per sector is boosted with 20-second cadence for 1\,000 targets per sector, and FFIs are retrieved at a shorter 10-minute cadence.

\begin{figure*}
    \centering
    \includegraphics[width=\textwidth]{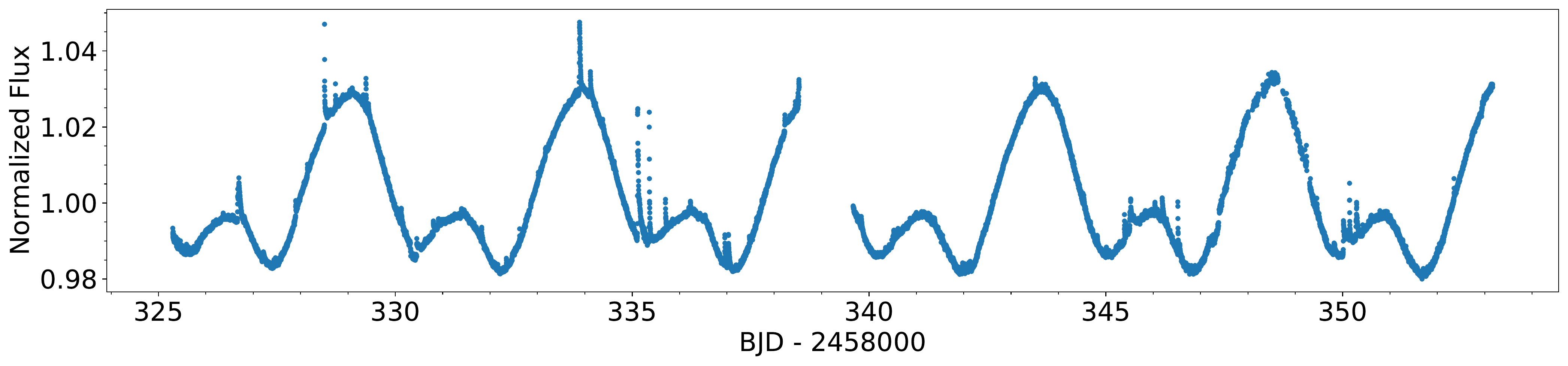} \\
    \includegraphics[width=\textwidth]{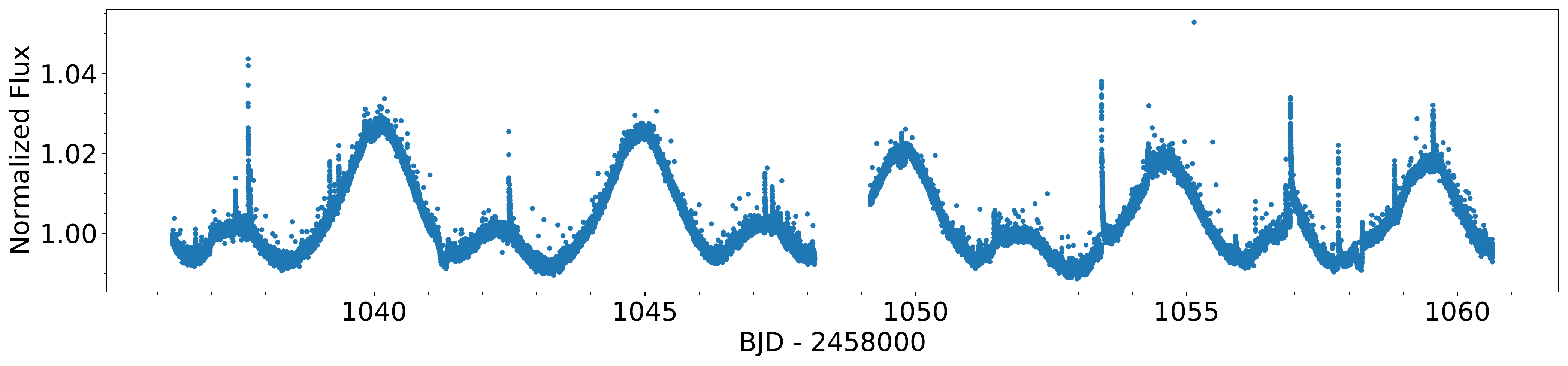}
    \caption{\tess photometry of \aumic. The blue points are \tess measurements. The top plot is from Cycle 1 (Sector 1, 2018 July 25 19:00:27 UT $-$ 2018 August 22 16:14:51 UT), 2-minute cadence. The bottom plot is from Cycle 3 (Sector 27, 2020 July 05 18:31:16 UT $-$ 2020 July 30 03:21:15 UT), 20-second cadence.}
    \label{fig:tess}
\end{figure*}

\tess observed \aumic (Figure \ref{fig:tess}) at 2-minute cadence during Cycle 1 (Sector 1, 2018 July 25 19:00:27 UT to 2018 August 22 16:14:51 UT)\footnote{The following Guest Investigator (GI) proposals were awarded for \aumic's Cycle 1 observations: G011176/PI Czekala, G011185/PI Davenport, G011264/PI Davenport, G011180/PI Dressing, G011239/PI Kowalski, G011175/PI Mann, \& G011266/PI Schlieder.} using Camera 1 CCD 4. This set has a 1.13-day gap due to data downlink. During its observation between 2018 August 16 16:00 UT and 2018 August 18 16:00 UT, the Fine Pointing mode calibration was not configured optimally, culminating in poorer data quality due to excessive spacecraft pointing jitter. \citet{gilbert2021} employed the data quality flags to filter out the problematic part of the data set, resulting in some additional gaps in data. \tess observed \aumic again during Cycle 3 (Sector 27, 2020 July 05 18:31:16 UT to 2020 July 30 03:21:15 UT)\footnote{The following GI proposals were awarded for \aumic's Cycle 3 observations: G03272/PI Burt, G03227/PI Davenport, G03063/PI Llama, G03228/PI Million, G03205/PI Monsue, G03141/PI Newton, G03202/PI Paudel, G03263/PI Plavchan, G03226/PI Silverstein, \& G03273/PI Vega.}, this time at both 20-second and 2-minute cadences, with the latter constructed by co-added six 20-second exposures. The data downlink during this period led to a 1.02-day gap in the data.

We use the \aumic\ \tess Cycle 1 and 3 transit light curves from \citet{gilbert2021} for our primary TTV analysis ($\S$\ref{sec:ttvs}), and herein we summarize their analysis. In our independent photodynamical analysis ($\S$\ref{sec:photodyn}), we reanalyze the \tess light curves directly. The \aumic\ \tess Cycle 1 and 3 light curves were retrieved from the Mikulski Archive for Space Telescopes (MAST)\footnote{\url{https://archive.stsci.edu}} archive using the {\tt lightkurve} package \citep{lightkurve} while setting its bitmask filter to ``default''. The Presearch Data Conditioning (PDC SAP) light curves were chosen since they addressed crowding and instrumental systematics \citep{smith2012, stumpe2012, stumpe2014}. After filtering the NaNs out of the data sets, 25.07 days of Cycle 1, 23.29 days of 2-minute Cycle 3, and 22.57 days of 20-second Cycle 3 are left. Next, the Savitzky-Golay filter is applied to the light curves to smooth out \aumic's spot modulation. Since the flares are abundant in \tess light curves, especially during most of \aumic b's and c's transits, the {\tt bayesflare} \citep{pitkin2014} and {\tt xoflares} packages were used to extract and model the flares instead of trimming the flares out as in \citet{plavchan2020}. Then, the \celeritetwo package \citep{foreman-mackey2017, foreman-mackey2018} was applied to model the stellar variability of \aumic, and the {\tt exoplanet} package \citep{foreman-mackey2021} was utilized to model the transits of \aumic b and c.

\subsection{\spitzer (\irac) Photometry}\label{sec:spitzer}

The {\sl Spitzer Space Telescope}\footnote{\url{https://www.spitzer.caltech.edu}} was constructed as part of NASA's Great Observatories Program's final mission to probe various astrophysical objects at infrared wavelengths \citep{werner2004}. Launched in 2003 August 25, \spitzer carried out its primary mission along with NASA's Astronomical Search for Origins Program for 5.75 years until the liquid helium coolant was depleted on 2009 May 15; afterward, it continued under several extended missions -- starting with the \spitzer Warm mission, then the \spitzer Beyond mission, and finally the \spitzer Final Voyage mission -- for the next 10.5 years starting from 2009 July 27 until its decommission on 2020 January 30. \spitzer fulfilled an indispensable role in characterizing exoplanets \citep{deming2020} -- e.g. including some benchmark systems observed, HD 189733 b \citep{grillmair2007, todorov2014}, HD 209458 b \citep{zellem2014}, HD 219134 b \& c \citep{gillon2017-2}, TRAPPIST-1 b through h \citep{morris2018, zhang2018, ducrot2018, ducrot2020}, WASP-26 b \citep{mahtani2013}.

To collect more data on the planetary object detected by \tess, \spitzer Director’s Discretionary Time (DDT) observations were proposed and awarded (PID 14214 \& 14241) in the final year of operations for observations of \aumic with the {\sl Infrared Array Camera} \citep[\irac,][]{fazio2004} due to possible calculated transits on 2019 February 10 \& 27 and September 10, which are presented in this paper. \spitzer observed \aumic with \irac on five occasions during the \spitzer Beyond and Final Voyage missions (2019 February 10 10:58:58 to 18:54:38 UT, 2019 February 27 09:46:23 to 17:42:02 UT, 2019 September 9 13:50:57 to 21:46:45 UT, 2019 September 9 23:26:02 UT to 2019 September 10 06:21:54 UT, \& 2019 September 14 03:40:07 to 12:33:36 UT). The first two observations were originally considered to be eclipses from the initially assumed orbital period for \aumic b from the \tess mission Cycle 1 observations. However, these observations detected additional transits of \aumic b, establishing a true period to be half as long as originally thought \citep{plavchan2020}. The third observation is of a transit search for an originally estimated but incorrect period for \aumic c, and the fourth observation is of a third transit of \aumic b; these two observations have been combined into one light curve set for this analysis. The final observation is a secondary eclipse search of \aumic b which will be described and analyzed in a separate paper and is not included in this work.

All of the observations were taken using the 32$\times$32 pixel sub-array mode with an exposure time of 0.08 seconds to avoid saturation on the star (measurement cadence is 0.1 seconds). After placing the star on the ``sweet-spot'' pixel, using the pointing calibration and reference sensor (PCRS) peak-up mode \citep{ingalls2012}, we exposed with continuous staring (no dithers). The observations were all taken at 4.5 $\mu$m, as this channel has a lower systematics due to the intra-pixel sensitivity. The coordinates were adjusted for the high parallax and proper motion of \aumic for the proposed observation dates. Each observation set consisted of a 30-minute pre-stare dither pattern, an 8-hour stare, and a 10-minute post-staring dither pattern. All data were calibrated by the \spitzer pipeline S19.2 and can be accessed using the \spitzer Heritage Archive (SHA)\footnote{\url{http://sha.ipac.caltech.edu}}.

For each of the three transit observations discussed here, the following data reduction steps were performed on each \aumic stellar image measured on 0.1-second intervals. We used the IDL routine {\tt box\_centroider}\footnote{\url{https://irsa.ipac.caltech.edu/data/SPITZER/docs/irac/calibrationfiles/pixelphase/box_centroider.pro}} supplied by the \spitzer Science Center to measure the location of \aumic on the pixel. We then performed aperture photometry on each image using the IDL Astronomy Users Library routine {\tt aper}\footnote{\url{https://idlastro.gsfc.nasa.gov/ftp/pro/idlphot/aper.pro}} with a fixed aperture of 2.25 pixels and subtracting a sky annulus of 3-7 pixels about the centroid.

All \irac photometry at 4.5 $\mu$m contain instrumental systematics caused by the coupling between spacecraft pointing fluctuations and drifts with intra-pixel sensitivity variations. For the three transit observations analyzed here, we take three approaches for detrending the instrument systematics and compare the results. First, as in \citet{plavchan2020}, we detrended this effect using an independent pixel mapping dataset measured for non-variable star BD+67 1044 \citep{ingalls2018}. Because this calibration star doesn't intrinsically vary, we take its photometric variations to reflect the pixel sensitivity map. We estimated the relative pixel sensitivity at the $(x,y)$ centroid locations of each \aumic observation using the K-Nearest Neighbors with Kernel Regression technique described by \citet{ingalls2018} and divided all \aumic measurements by the sensitivities. This approach was published for the first \spitzer transit in \citet{plavchan2020}. However, we noticed that additional high-frequency (shorter than the transit duration) photometric variability remained in the light curve that looked like astrophysical ``hot spot'' crossings. But we subsequently identified a strong correlation of these light curve features with the \spitzer PSF FWHM. Therefore, second, we detrend the light curve using both the trend time series for the pixel centroid motion and the PSF FWHM; this is the systematic-detrended time series that we adopt in this work for further analysis. Third, we also tested the noise-pixel technique detailed by \citet{lewis2013} and achieve qualitatively similar results to our second approach.

We modeled the \irac intra-pixel sensitivity \citep{ingalls2016} using a modified implementation of the {\tt BLISS} ({\tt BiLinearly-Interpolated Sub-pixel Sensitivity}) mapping algorithm \citep{stevenson2012}. We used a modified version of the {\tt BLISS} mapping (BM) approach to mitigate the correlated noise associated with intra-pixel sensitivity. In our photometric baseline model, we complement the BM correction with a linear function of the Point Response Function (PRF) Full Width at Half-Maximum (FWHM). In addition to the BM, our baseline model includes the PRF's FWHM along the $x$ and $y$ axes, which significantly reduces the level of correlated noise as shown in previous studies (e.g., \citealt{lanotte2014}, \citealt{demory2016-1}, \citealt{demory2016-2}, \citealt{gillon2017-1}, \citealt{mendonca2018}). Our baseline model does not include time-dependent parameters. Our implementation of this baseline model is included in a Markov Chain Monte Carlo (MCMC) framework already presented in the literature \citep{gillon2012}. We run two chains of 200\,000 steps each and check for convergence and efficient mixing using the Gelman-Rubin statistic \citep{gelman1992}; all of the chains have converged with their GR statistic $<$ 1.01.

We next construct a 2nd-order polynomial model fit to account for the rotational modulation of stellar activity present in the three \spitzer light curves. \aumic is active with its rotational modulation of stellar activity, and on the timescale of a transit duration can be described by a 2nd-order polynomial \citep{addison2021}; longer timescales would necessitate a Gaussian Process or similar analysis as undertaken in \citet{plavchan2020} and \citet{gilbert2021} for the \tess transits. These polynomial coefficients are marginalized over in our TTV analysis to account for the timing uncertainties introduced from the rotational modulation of stellar activity. We cross-check our approach to that using a Gaussian Process model in \citet{plavchan2020} and derive consistent TTVs and corresponding uncertainties.

The second \spitzer transit also has an unusual ``jump'' feature during the middle of the transit that was thought to be caused by either a flare or a transit egress of another planet; we do not identify any systematic indicator that this jump coincides with. We explored the timing of \aumic c's transits but found that none line up with the \spitzer's transit of \aumic b. So instead we constructed and fit a custom flare model for this feature, consisting of a linear rapid rise followed by an exponential decay. The amplitude of the flare model is marginalized over in our analysis of the TTVs to account for the impact it has on the transit times; however, the flare rise, peak, and decay times are fixed in our analysis. The adopted flare rise and decay times are informed by and consistent with the characterization of the flares in the \tess light curve analyzed in \citet{gilbert2021}. Here, the 2nd-order polynomial coefficients are degenerate with the flare times and are marginalized over to account for the impact the flare has in our derived transit time and uncertainty. Since the flare did not occur during ingress or egress, it has minimal impact on our derived transit midpoint time and corresponding uncertainty.

The third \spitzer light curve was additionally detrended with an ad hoc Gaussian model given the presence of a low-level Gaussian-like trend coincident with the transit midpoint time (note, not a Gaussian Process, but a Gaussian change in brightness with time). Again, we do not identify a systematic indicator correlated with this brightness variation in the light curve and associate it with an astrophysical origin for \aumic. The amplitude of the Gaussian model is marginalized over in our analysis to assess its impact on the TTVs, but the width and peak time were fixed; in this case, marginalizing over the 2nd-order polynomial coefficients in our model again compensates for and is degenerate with any error in the fixed Gaussian centroid time and width. The astrophysical origins of this brightness change and its coincidence with the transit midpoint time, as well as other remaining residuals present in the \aumic\ \spitzer light curves as seen in Figure \ref{fig:exofastv2models}, are beyond the scope of this work and are the subject of a future publication.

For all but one detrending time series, the additive coefficients were set to zero and the multiplicative coefficients were set to one. The exception is the flare detrending from the second \spitzer set, with the additive coefficient set to one and the multiplicative coefficient set to zero. Afterward, we did a joint model of the \spitzer and ground-based photometric transits using the \exofast package \citep{eastman2019}. This process to extract the midpoint transit times from the \spitzer light curves is explained in more details in $\S$\ref{sec:exofastmod}.

\subsection{Rossiter-McLaughlin Spectroscopy}

The Rossiter-McLaughlin (R-M) technique is an advantageous tool in detecting and characterizing transiting exoplanets, including determining their spin-orbit alignments \citep{ohta2005, winn2007, triaud2018}. The R-M effect is observed when the planet crosses the host star during the RV observation, blocking a portion of the star's rotational signal and generating a characteristic feature on the time series RV profile \citep{holt1893, rossiter1924, mclaughlin1924, ohta2005, winn2007}. Many exoplanets have been characterized with R-M's -- e.g. including some benchmark systems, CoRoT-3 b \& HD 189733 b \citep{triaud2009}, KELT-20 b \citep{rainer2021}, K2-232 b \citep{wang2021}, TOI-1431 b \citep{stangret2021}, WASP-17 b \citep{anderson2010}.

R-M observation also offer an additional way in which to derive transit mid-point times independent of photometric observations, a method that has not previously been commonly used for TTV analysis because it is resource-intensive with its use of high-resolution spectrometers on large aperture telescopes. Today, however, \tess mission candidates are relatively brighter and nearby compared to \kepler systems, and more amenable to R-M observations. We include transit midpoint times derived from two R-M observations of \aumic b's transits that were obtained at relatively important epochs shortly after the \spitzer observations and between the two-year gap in the \tess observations. The first was collected using the \spirou and \ishell instruments, and the second the \espresso instrument. We retrieved the two transit midpoint times from \citet[][private communication]{martioli2020-pc} and \citet[][private communication]{palle2020-pc}, respectively. The following sections summarize the work done by \cite{martioli2020} and \cite{palle2020} on processing the respective \spirou + \ishell and \espresso data.

\subsubsection{\cfht (\spirou) \& \irtf (\ishell) Spectroscopy}

The {\sl SpectroPolarim\`etre InfraRouge} (\spirou)\footnote{\url{https://www.cfht.hawaii.edu/Instruments/SPIRou}}, mounted on the 3.6 m {\sl Canada-France-Hawai`i Telescope} (\cfht) located atop Maunakea, Hawai`i, is a high-resolution near-infrared (NIR) spectrometer that is capable of imaging in YJHK bands (0.95-2.5 $\mu$m) with a resolving power of $\sim$70\,000 and is equipped with a fiber-fed cryogenic high-resolution \'echelle spectrograph that can perform high-precision velocimetry and spectropolarimetry, which allows it to simultaneously observe magnetic features and stellar activities of the host stars \citep{donati2020}. The \ishell\footnote{\url{http://irtfweb.ifa.hawaii.edu/~ishell}}, installed on the 3.2 m {\sl NASA Infrared Telescope Facility} (\irtf) also located atop Maunakea, Hawai`i, is a high-resolution 1.1-5.3 $\mu$m spectrometer with a resolving power of 75\,000 and was designed to replace {\sl CSHELL} as an instrument with enhanced spectroscopic capabilities \citep{rayner2016}. These qualities make \spirou and \ishell useful tools to carry out follow-up observations on transiting exoplanets of young, active M dwarfs \citep{morin2010, afram2019}, such as \aumic.

As part of the \spirou Legacy Survey's Work Package 2 (WP2) \citep{donati2020}, \citet{martioli2020} observed \aumic on 2019 June 17 10:10:56 to 15:13:45 UT with \spirou set in Stokes V spectropolarimetric mode and captured an egress of \aumic b. 116 spectra were collected that night, each taken at 122.6 seconds exposure, with the average SNR of 242. \citet{martioli2020} also used \ishell set at K$_{\mbox{\scze gas}}$ mode for simultaneous but shorter observation of \aumic (2019 June 17 11:08:19 to 12:53:32 UT); 47 120-second spectra were collected with that instrument, with SNR of $\sim$60-70. The typical seeing condition was 0.96 $\pm$ 0.13'', the initial and final airmass were 2.9 and 1.8, respectively, with its minimum being 1.59, and the Moon was 99\% illuminated and 40.3$^{\circ}$ from the target.

\citet{martioli2020} implemented the reduction pipeline APERO (A PipelinE to Reduce Observations, Cook et al. in prep.) to reduce and process the \spirou data and to calculate the cross-correlation functions (CCFs). Next, the “M2\_weighted\_RV\_-5.mas” line mask was applied to the spectra, and the lines were masked if their telluric absorption is deeper than 40\%. The line mask then underwent further refinement by removing lines not present in \aumic's Stokes-I spectrum using the technique from \citet{moutou2020}. The CCFs were calculated from each spectral order and summed to achieve greater precision, then the RVs were measured from the CCF by using the velocity shift's least-square fit. The \ishell RVs were extracted using the {\tt pychell} pipeline, and yielded consistent RVs and precision with the \spirou RVs \citep{cale2019}.

Next, \citet{martioli2020} constructed the R-M model using the {\tt emcee} Markov chain Monte Carlo (MCMC) package \citep{foreman-mackey2013} and the stellar activity model using the approach from \citet{donati1997}. The stellar activity model was then subtracted from the measured RVs, and the R-M model was then applied as a correction to the subtracted RVs. Finally, \citet[][private communication]{martioli2020-pc} modeled \aumic b's midpoint time from \spirou + \ishell's best-fit subtracted and corrected RV model using \tc = $\mathcal{N}$(2458330.39153, 0.00070) as a prior.

\subsubsection{\vlt (\espresso) Spectroscopy}

{\sl Echelle Spectrograph for Rocky Exoplanet and Stable Spectroscopic Observations} (\espresso)\footnote{\url{https://www.eso.org/sci/facilities/paranal/instruments/espresso.html}} is a high-precision RV spectrometer situated in the Combined-Coud\'e Laboratory (CCL) at the focus of the {\sl Very Large Telescope} (\vlt) atop Cerro Paranal, Chile \citep{pepe2021}. Its spectrograph probes the sky at 378.2-788.7 nm range, and it can use either four 8.2 m telescopes (4-UT) with lower resolution of $\sim$70\,000 or only one of them (1-UT) with higher resolutions of $\sim$140\,000 in the High-Resolution (HR) mode or $>$190\,000 in the Ultra-High-Resolution (UHR) mode.

\citet{palle2020} observed \aumic on 2019 August 7 3:24 to 9:23 UT with \espresso set at the standard HR (1-UT) mode and captured a full transit of \aumic b. 88 spectra were collected that night, each taken at 200 seconds exposure. The SNR averaged around 93.9, the initial and final airmass were 1.03 and 2.37, respectively, with its minimum being 1.007, and the sky was clear.

\citet{palle2020} applied several separate approaches in reducing and modeling the \espresso data; however, we only highlight one of those approaches that provided us the midpoint time for this paper. The {\tt SERVAL} package \citep{zechmeister2018} was implemented to extract and calibrate the spectra and generate the RV profile of \aumic b. Next, the R-M effect was modeled using the combination of the \celerite \citep{foreman-mackey2017} package's Gaussian process (GP) and {\tt PyAstronomy} package \citep{czesla2019}, and the stellar activity was modeled with GP described by a Mat\'ern 3/2 kernel implemented by \celerite. These models are then applied as corrections to the \espresso RV profile. \citet[][private communication]{palle2020-pc} extracted the midpoint time from the {\tt SERVAL} GP + {\tt PyAstronomy} best-fit RV profile using \tc = $\mathcal{N}$(2458702.77277, 0.00189) as a prior.

\subsection{Ground-Based Photometry}

The TESS Follow-up Observing Program (TFOP) Working Group (WG)\footnote{\url{https://tess.mit.edu/followup}} coordinated numerous ground-based follow-ups for various TOIs, including \aumic. As a result, 13 \aumic follow-up photometric transit observations were made using different observatories: one \brier 0.36 m, six \saao 1.0 m, five \sso 1.0 m, and one \pest 0.30 m. The light curves from these observations are available online through ExoFOP-TESS\footnote{\url{https://exofop.ipac.caltech.edu/tess}} \citep{akeson2013}. The follow-up observation schedules were conducted with the online version of the {\tt TAPIR} package \citep{jensen2013}. We utilized the {\tt AstroImageJ} \citep[\aij,][]{collins2017} to process the ground-based light curves (except \pest, which was processed through their own pipeline) and then create a subset table containing only BJD\_TDB, normalized detrended flux, flux uncertainty, and detrending columns from the ground-based light curves to prepare them for \exofast modeling and extraction of midpoint times ($\S$\ref{sec:exofastmod}). The following sub-subsections describe the role each telescope played in collecting and processing the light curves.

\subsubsection{\lcogt (\sinistro) Photometry}

We made use of two 1.0 m \lco Ritchey-Chretien Cassegrain telescopes, both equipped with \sinistro, that are part of the {\sl Las Cumbres Observatory Global Telescope} network \citep[\lcogt,][]{brown2013}\footnote{\url{https://lco.global/observatory}}. Two of \sinistro's filters used for \aumic observations were Pan-STARRS Y and Pan-STARRS \zs, with their central wavelength peaks at 1004.0 and 870.0 nm, respectively. The third filter B was used for simultaneous observation with Y; however, the data collected with B are omitted from this paper due to non-detection of \aumic b's transits and more pronounced stellar activity in the bluer band.

\sso, located on Mount Woorut near Coonabarabran, New South Wales, Australia, observed \aumic on four separate nights (2020 April 25 16:03:03 to 17:49:38 UT at 35 seconds exposure with Y and 2020 April 25 16:00:47 to 18:52:43 UT, 2020 August 13 12:41:53 to 17:50:19 UT, 2020 September 16 09:12:31 to 14:41:47 UT, \& 2020 October 3 09:17:36 to 12:36:33 UT at 15 seconds exposure with \zs). 40 images from the first night were collected with Y, and 212, 379, 408, \& 248 images from the respective first, second, third, \& fourth nights were collected with \zs. An egress was captured on the first and fourth nights while a full transit was captured on the second and third nights.

\saao, located in Sutherland, South Africa, observed \aumic on five separate nights (2020 May 20 22:53:29 UT to 2020 May 21 03:15:29 UT at 35 seconds exposure with Y and 2020 May 20 22:51:11 UT to 2020 May 21 03:17:38 UT, 2020 June 6 21:44:20 UT to 2020 June 7 01:22:36 UT, 2020 June 23 20:37:29 to 23:36:35 UT, 2020 September 7 21:55:46 UT to 2020 September 8 00:43:57 UT, \& 2020 October 11 18:08:39 to 22:29:54 UT at 15 seconds exposure with \zs). 99 images from the first night were collected with Y, and 333, 266, 223, 211, \& 311 images from the respective first, second, third, fourth, \& fifth nights were collected with \zs. The second night's photometric quality was impacted by a combination of clouds and a full Moon. An egress was captured on the first through third nights while an ingress was captured on the fourth and fifth nights.

All light curves from \lcogt were reduced and detrended with \aij. For each \lcogt night, the following detrending parameters were applied: AIRMASS for UT2020-04-25 (Y), UT2020-05-20 (Y \& \zs), UT2020-06-06, \& UT2020-06-23; Width\_T1 for UT2020-04-25 (\zs), UT2020-09-16, UT2020-10-03, \& UT2020-10-11; AIRMASS + Width\_T1 for UT2020-08-13; and Width\_T1 + Sky/Pixel\_T1 for UT2020-09-07. We also used \aij to generate a subset table for each light curve.

\subsubsection{\pest (\sbig) Photometry}

{\sl Perth Exoplanet Survey Telescope} (\pest)\footnote{\url{http://pestobservatory.com}}, based in Perth, Western Australia, is a 12'' (0.3048 m) Meade LX200 Schmidt-Cassegrain Telescope that was equipped with SBIG ST-8XME camera at the time of the \aumic observation. \pest observed \aumic on 2020 July 10 13:26:53 to 22:42:33 UT with V and captured a full transit. 1143 images were collected, each at 15 seconds exposure. The \pest light curve was reduced and processed through the {\tt PEST} pipeline\footnote{\url{http://pestobservatory.com/the-pest-pipeline}}. We then use \aij to create a subset table that included {\tt PEST}-generated detrending parameters comp\_flux + x\_coord + y\_coord + dist\_center + fwhm + airmass + sky.

\subsubsection{\brier (\moravian) Photometry}

The {\sl Brierfield Observatory}\footnote{\url{https://www.brierfieldobservatory.com}}, located in Bowral, New South Wales, Australia, houses the 14'' (0.36 m) Planewave Corrected Dall-Kirkham Astrograph telescope mounted with the instrument {\sl Moravian G4-16000 KAF-16803}. \brier observed \aumic on 2020 August 13 11:35:21 to 17:54:35 UT with I and captured a full transit. 398 images were collected, each at 16 seconds exposure. The \brier light curve was reduced and detrended with \aij; the detrending parameters were Meridian\_Flip + X(FITS)\_T1 + Y(FITS)\_T1 + tot\_C\_cnts. Afterward, a subset table was generated from this light curve with \aij.

\section{\tess, \spitzer, \& Ground-Based Photometric Joint Modeling}\label{sec:exofastmod}

We use the \exofast package \citep{eastman2019} to model the transits and characterize our light curves. \exofast estimates the posterior probabilities through the Markov chain Monte Carlo (MCMC) to both determine the statistical significance of our ground-based \& \spitzer detections and the confidence in the time of conjunction measurements to assess for the presence of detectable TTVs. Five \tess transits, three \spitzer transits and 13 ground-based photometric transits of \aumic b and three \tess transits of \aumic c are included in the model. The following detrending parameters are treated as additive: flare (\spitzer), sky (\spitzer \& \pest), \& Sky/Pixel\_T1 (\saao); the remaining detrending parameters are treated as multiplicative.

\begin{deluxetable}{l|c|cc}
    \tablecaption{\label{table:exofastv2priors}Stellar, planetary, and transit priors for \exofast modeling with \aumic's eight \tess, three \spitzer, \& 13 ground transits.}
    \tablehead{\multirow{2}{*}{Prior} & \multirow{2}{*}{Unit} & \multicolumn{2}{c}{Input} \\
        & & \aumic b & \aumic c}
    \startdata
\logmstmsu    & ...      & \multicolumn{2}{c}{$\mathcal{N}$(-0.301, 0.026)}              \\
\rstar        & \rsun    & \multicolumn{2}{c}{$\mathcal{N}$(0.75, 0.03)}                 \\
\teff         & K        & \multicolumn{2}{c}{$\mathcal{N}$(3700, 100)}                  \\
Age           & Gyr      & \multicolumn{2}{c}{$\mathcal{N}$(0.022, 0.003)}               \\
Parallax      & mas      & \multicolumn{2}{c}{$\mathcal{N}$(102.8295, 0.0486)}           \\
\hline
\tc           & BJD\_TDB & $\mathcal{N}$(2458330.39051)  & $\mathcal{N}$(2458342.2223)   \\
\logfracpd    & ...      & $\mathcal{N}$(0.92752436)     & $\mathcal{N}$(1.2755182)      \\
\rplan/\rstar & ...      & $\mathcal{N}$(0.0512, 0.0020) & $\mathcal{N}$(0.0340, 0.0034) \\
e             & ...      & $\mathcal{N}$(0.12, 0.16)     & $\mathcal{N}$(0.13, 0.16)     \\
\hline
TTV Offset    & days     & \multicolumn{2}{c}{$\mathcal{U}$(-0.02, 0.02)}                \\
Depth Offset  & ...      & \multicolumn{2}{c}{$\mathcal{U}$(-0.01, 0.01)}
    \enddata
    \tablecomments{$\mathcal{N}$ denotes the Gaussian priors, and $\mathcal{U}$ denotes the uniform priors. The Gaussian priors were taken from Tables \ref{table:stellarquant} \& \ref{table:planetquant}}, while the logarithmic functions were calculated. TTV and depth offsets are arbitrary and applied as constraints to all transits.
\end{deluxetable}

\begin{deluxetable}{l|c|c}
    \tablecaption{\label{table:exofastv2sed}Apparent magnitude priors for \exofast's spectral energy distribution (SED) fitting of \aumic. This is intended to place constraints on \aumic's MIST evolutionary models \citep{choi2016, dotter2016} and is applied only to the first of the two \exofast runs}.
    \tablehead{Band & Apparent Magnitude & Ref}
    \startdata
Gaia            & 7.84 $\pm$ 0.02  & 1   \\
Gaia$_{\rm BP}$ & 8.94 $\pm$ 0.02  & 1   \\
Gaia$_{\rm RP}$ & 6.81 $\pm$ 0.02  & 1   \\
J$_{\rm 2M}$    & 5.44 $\pm$ 0.02  & 2   \\
H$_{\rm 2M}$    & 4.83 $\pm$ 0.02  & 2   \\
K$_{\rm 2M}$    & 4.53 $\pm$ 0.02  & 3   \\
B               & 10.06 $\pm$ 0.02 & ... \\
V               & 8.89 $\pm$ 0.18  & ... \\
g$_{\rm SDSS}$  & 9.58 $\pm$ 0.05  & 4   \\
r$_{\rm SDSS}$  & 8.64 $\pm$ 0.09  & 4   \\
i$_{\rm SDSS}$  & 7.36 $\pm$ 0.14  & 4
    \enddata
    \tablerefs{(1) \citet{gaia2018}; (2) \citet{cutri2003}; (3) \citet{stauffer2010}; (4) \citet{zacharias2012}}
\end{deluxetable}

The Gaussian priors from Table \ref{table:exofastv2priors} were taken from Tables \ref{table:stellarquant} and \ref{table:planetquant}, while the logarithmic functions were calculated; the logarithmic version of stellar mass and orbital period were used because they are the fitted priors in \exofast. The TTV and depth offset priors were implemented to place constraints on the variation of transit timing and depth of all light curves; any transit depth variability is not investigated further herein. Since both Pan-STARRS Y and Pan-STARRS \zs are not available among the filters in \exofast, y and z' (Sloan z) were used as respective approximate substitutes.

\begin{deluxetable*}{l|c|c|c|c}
    \tablecaption{\label{table:detrend}Detrending parameters incorporated into \exofast modeling of \aumic b transits. The flare (\spitzer), sky (\spitzer \& \pest), \& Sky/Pixel\_T1 (\saao) were implemented as additives; the remaining detrending parameters were implemented as multiplicative. See $\S$\ref{sec:dataobs} for details on detrending parameters used for each observation. Since both Pan-STARRS Y and Pan-STARRS \zs are not available among the filters in \exofast, y and z' (Sloan z) were used as respective approximate substitutes.}
    \tablehead{Telescope & Date (UT) & Filter & Detrending Parameter(s) & Note}
    \startdata
\spitzer & 2019-02-10 & 4.5 $\mu$m & x, y, noise/pixel, FWHM\_x, FWHM\_y, sky, linear, quadratic                 & 1 \\
\spitzer & 2019-02-27 & 4.5 $\mu$m & x, y, noise/pixel, FWHM\_x, FWHM\_y, sky, linear, quadratic, flare          & 1 \\
\spitzer & 2019-09-09 & 4.5 $\mu$m & x, y, noise/pixel, FWHM\_x, FWHM\_y, sky, linear, quadratic, Gaussian       & 1 \\
\sso     & 2020-04-25 & z'         & Width\_T1                                                                   & 2 \\
\sso     & 2020-04-25 & y          & AIRMASS                                                                     & 2 \\
\saao    & 2020-05-20 & z'         & AIRMASS                                                                     & 2 \\
\saao    & 2020-05-20 & y          & AIRMASS                                                                     & 2 \\
\saao    & 2020-06-06 & z'         & AIRMASS                                                                     & 2 \\
\saao    & 2020-06-23 & z'         & AIRMASS                                                                     & 2 \\
\pest    & 2020-07-10 & V          & comp\_flux, x\_coord, y\_coord, dist\_center, fwhm, airmass, sky            & 3 \\
\brier   & 2020-08-13 & I          & Meridian\_Flip, X(FITS)\_T1, Y(FITS)\_T1, tot\_C\_cnts                      & 2 \\
\sso     & 2020-08-13 & z'         & AIRMASS, Width\_T1                                                          & 2 \\
\saao    & 2020-09-07 & z'         & Width\_T1, Sky/Pixel\_T1                                                    & 2 \\
\sso     & 2020-09-16 & z'         & Width\_T1                                                                   & 2 \\
\sso     & 2020-10-03 & z'         & Width\_T1                                                                   & 2 \\
\saao    & 2020-10-11 & z'         & Width\_T1                                                                   & 2
    \enddata
    \tablenotetext{1}{See $\S$\ref{sec:spitzer} for details on the detrending parameters applied to \spitzer data.}
    \tablenotetext{2}{Detrending parameters generated from \aij \citep{collins2017}.}
    \tablenotetext{3}{Detrending parameters generated from {\tt PEST} pipeline (\url{http://pestobservatory.com/the-pest-pipeline}).}
\end{deluxetable*}

\begin{figure*}
    \centering
    \begin{tabular}{ccc}
    \includegraphics[width=46mm]{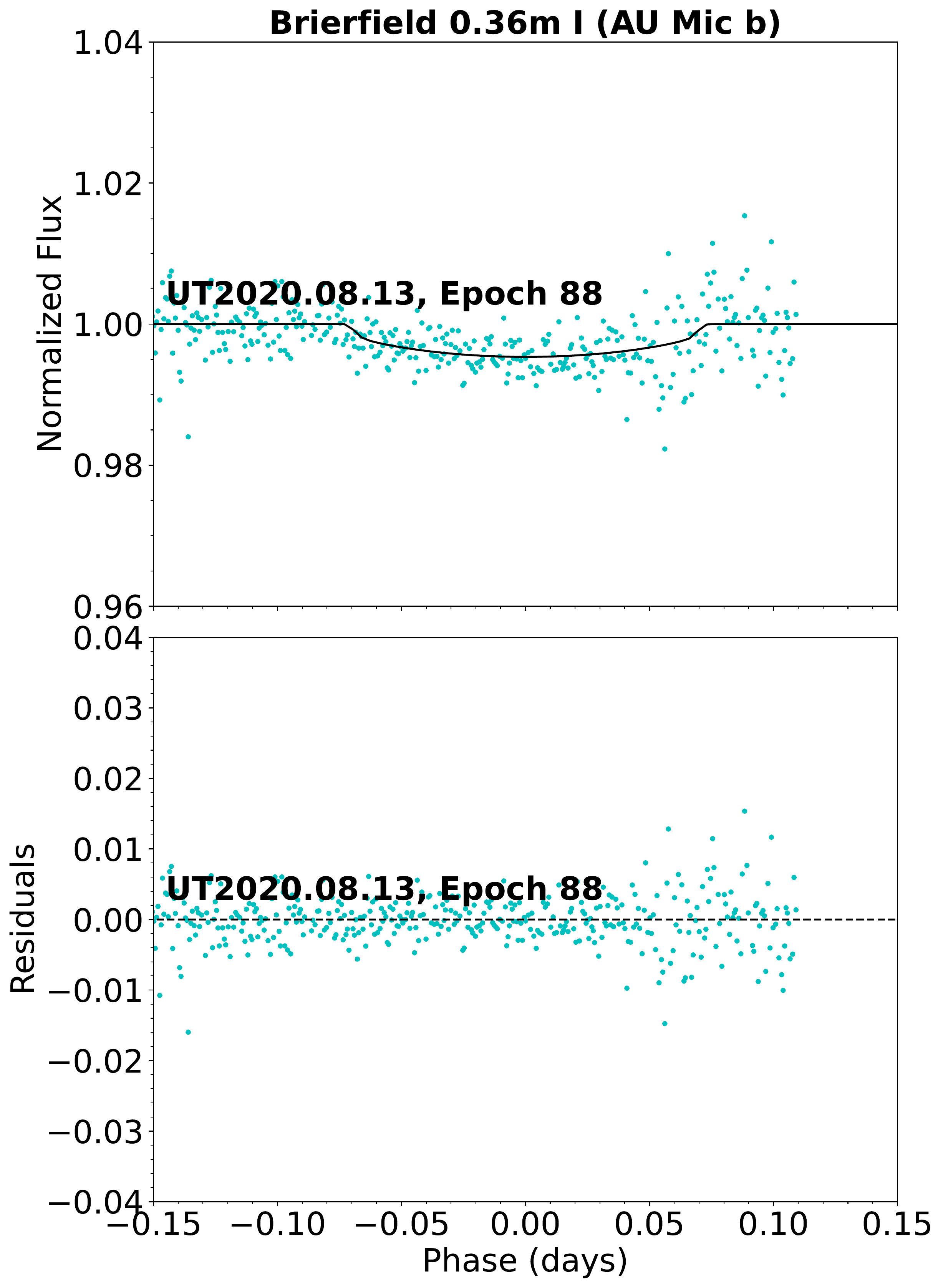} &
    \includegraphics[width=46mm]{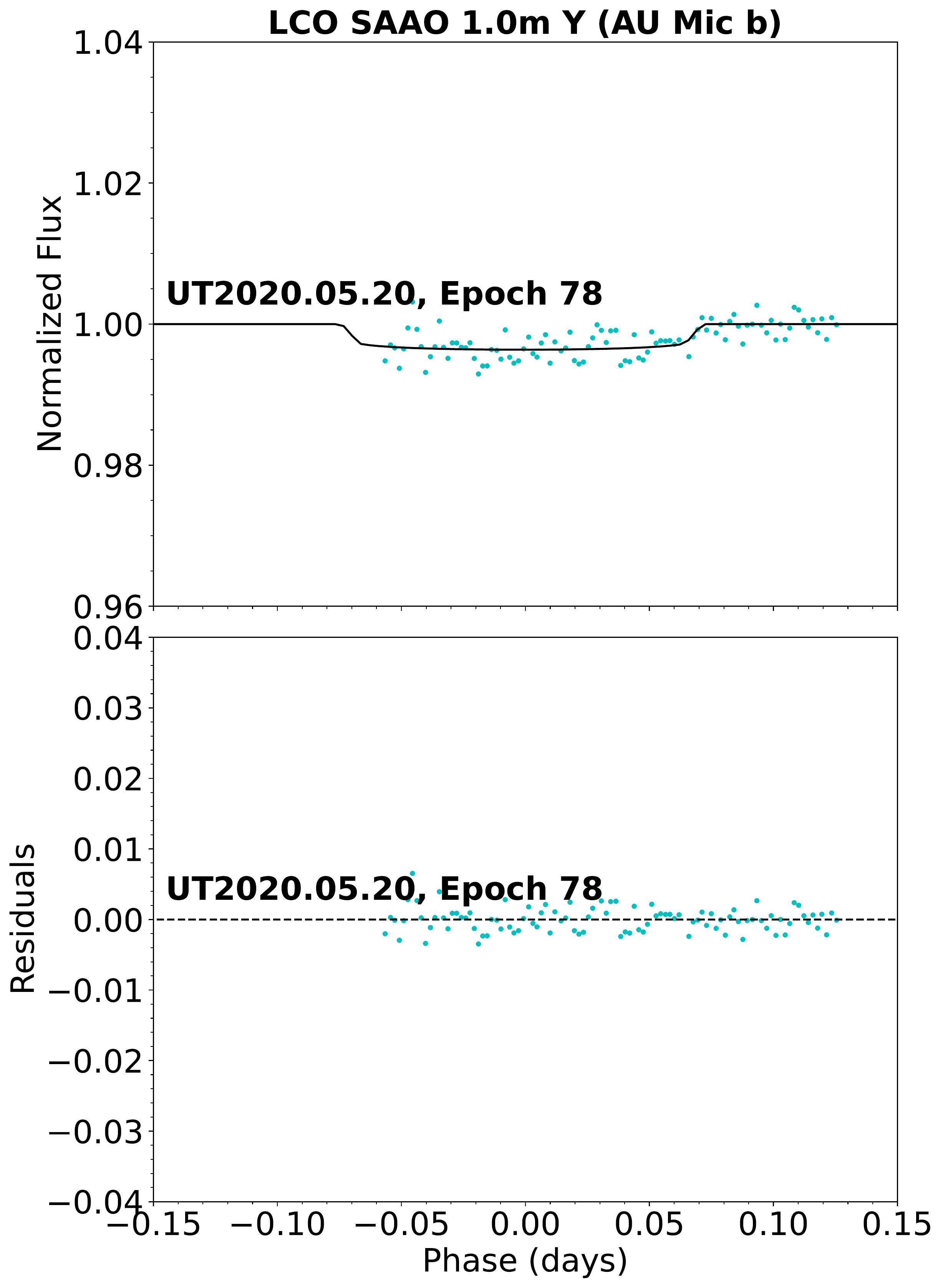} &
    \includegraphics[width=46mm]{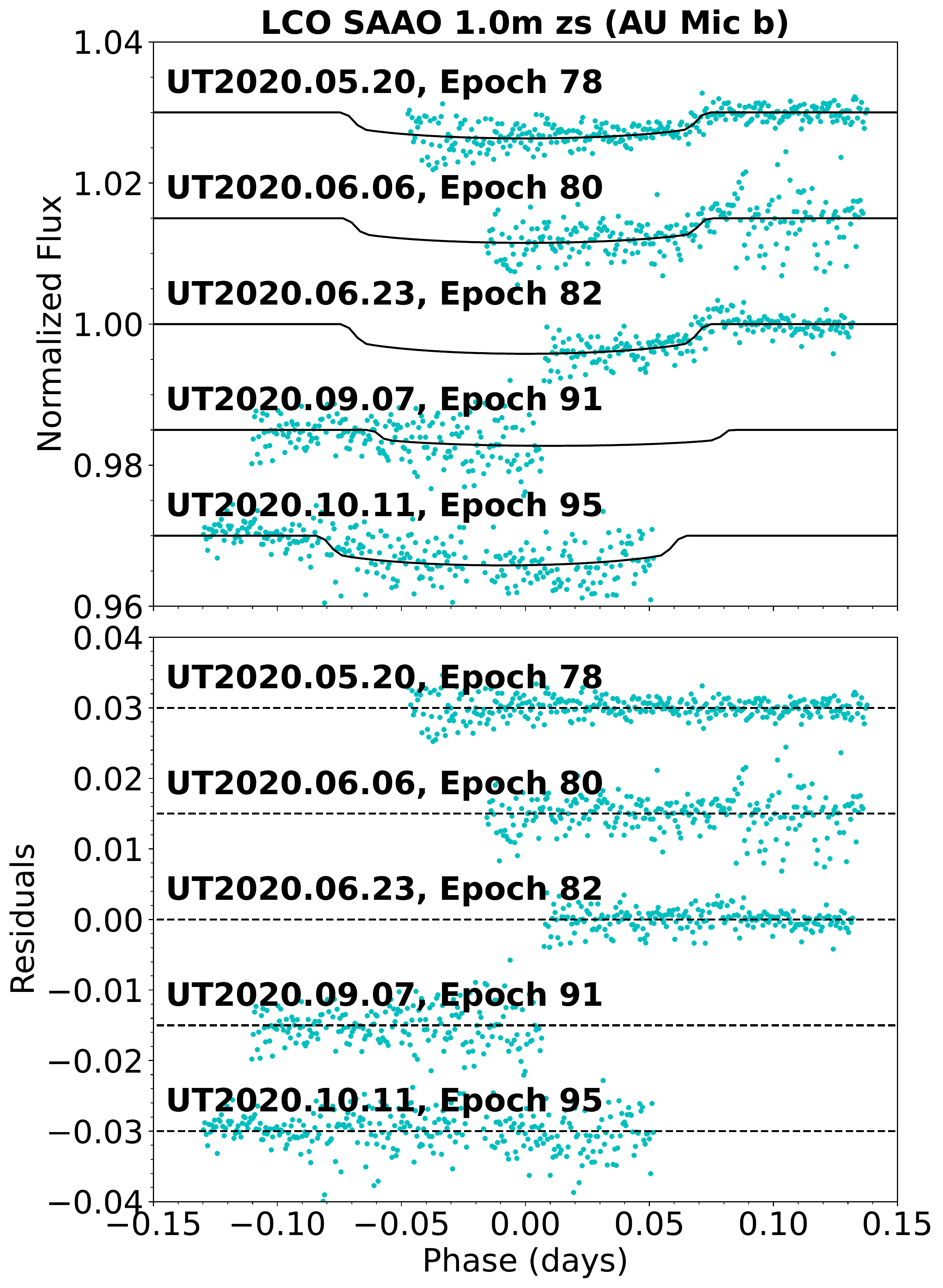} \\
    \includegraphics[width=46mm]{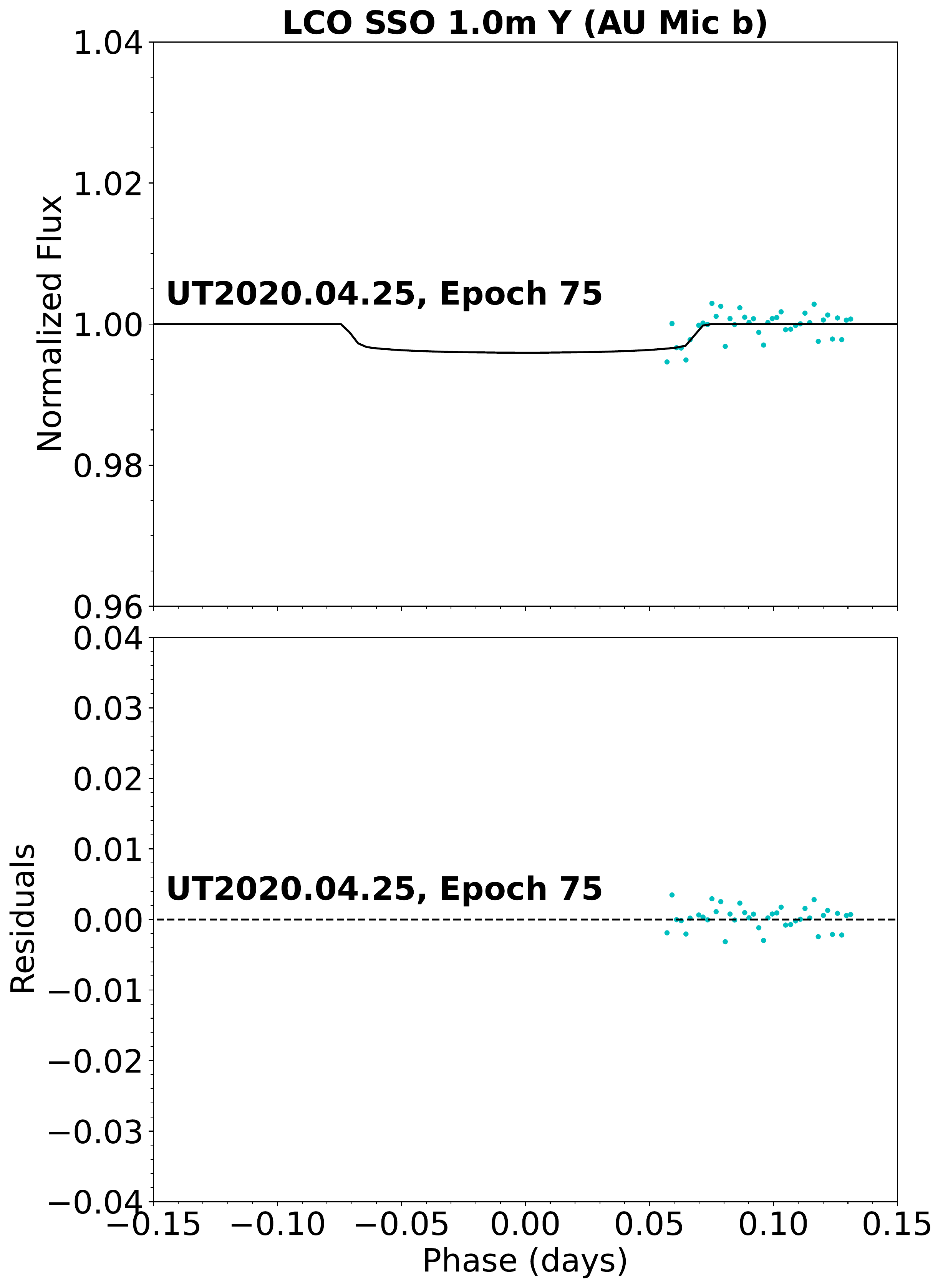} &
    \includegraphics[width=46mm]{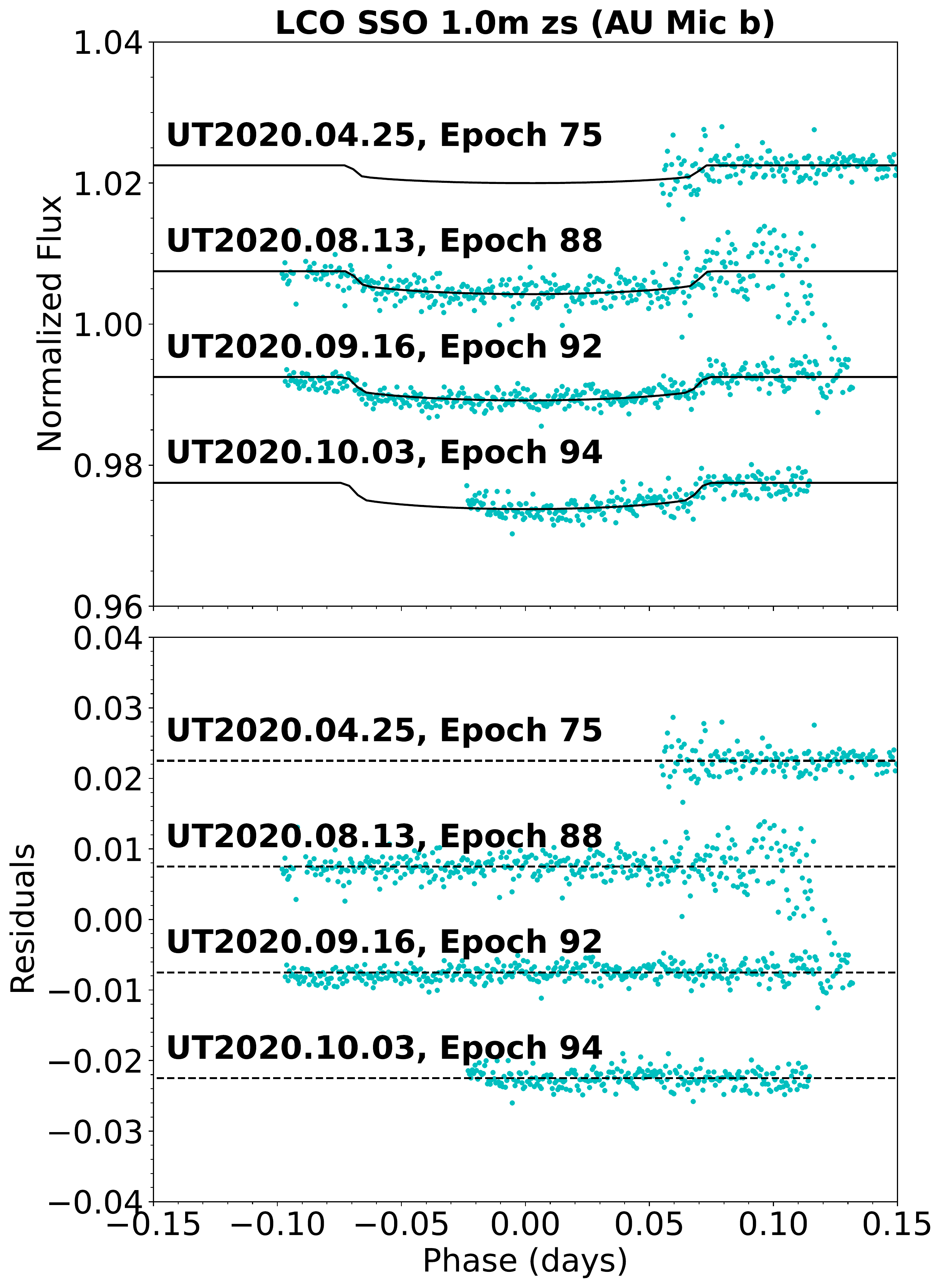} &
    \includegraphics[width=46mm]{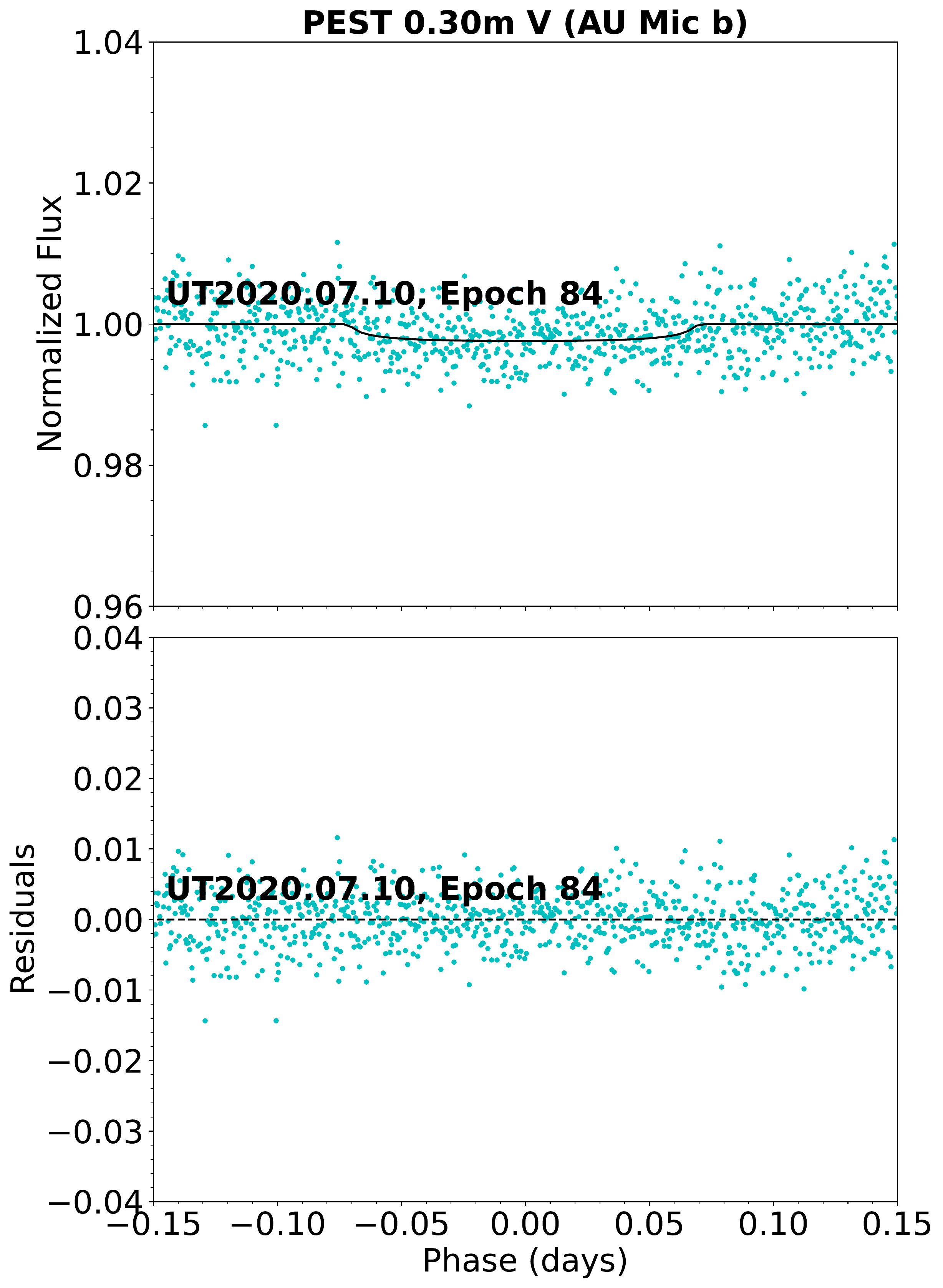} \\
    \includegraphics[width=46mm]{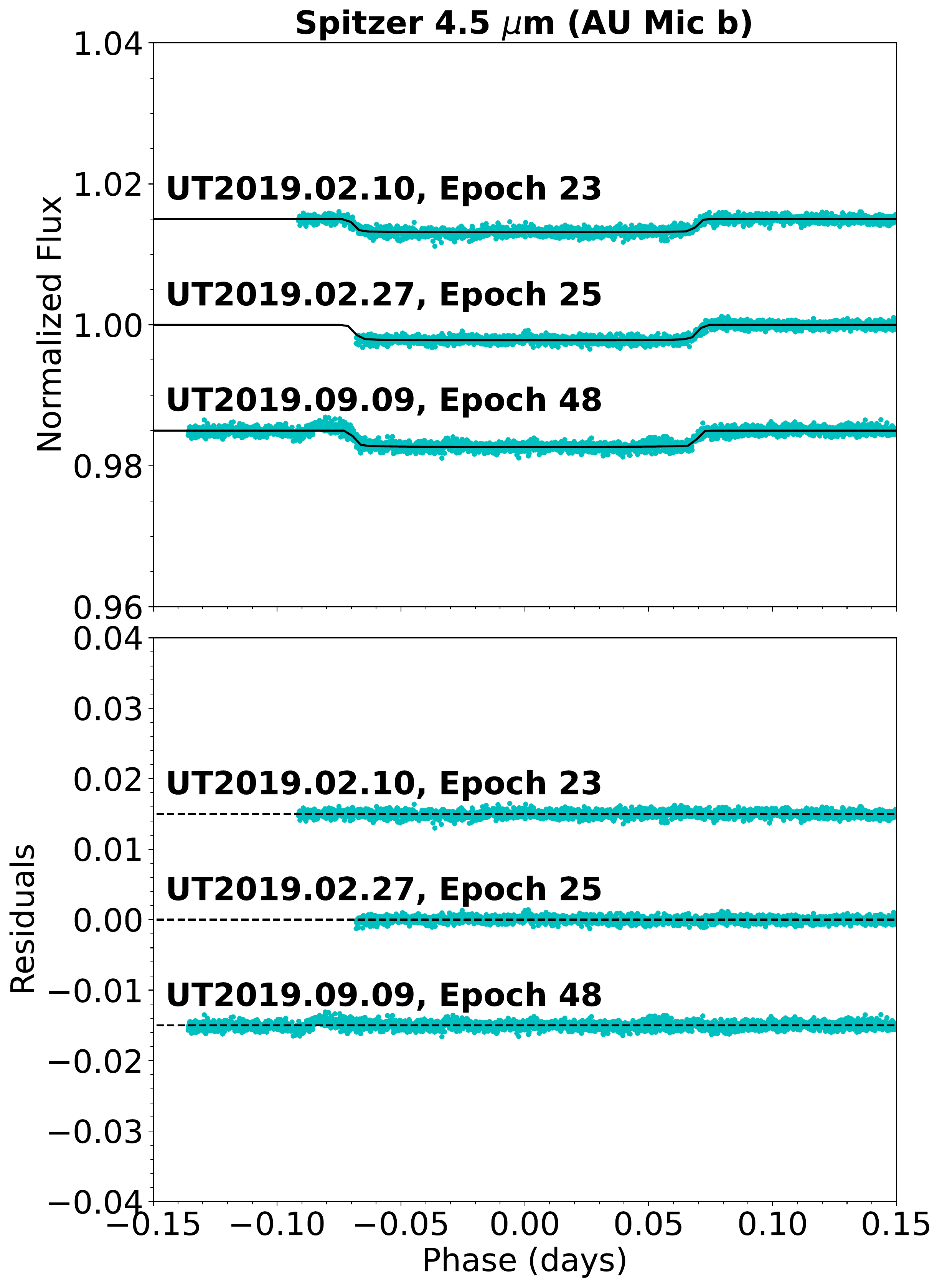} &
    \includegraphics[width=46mm]{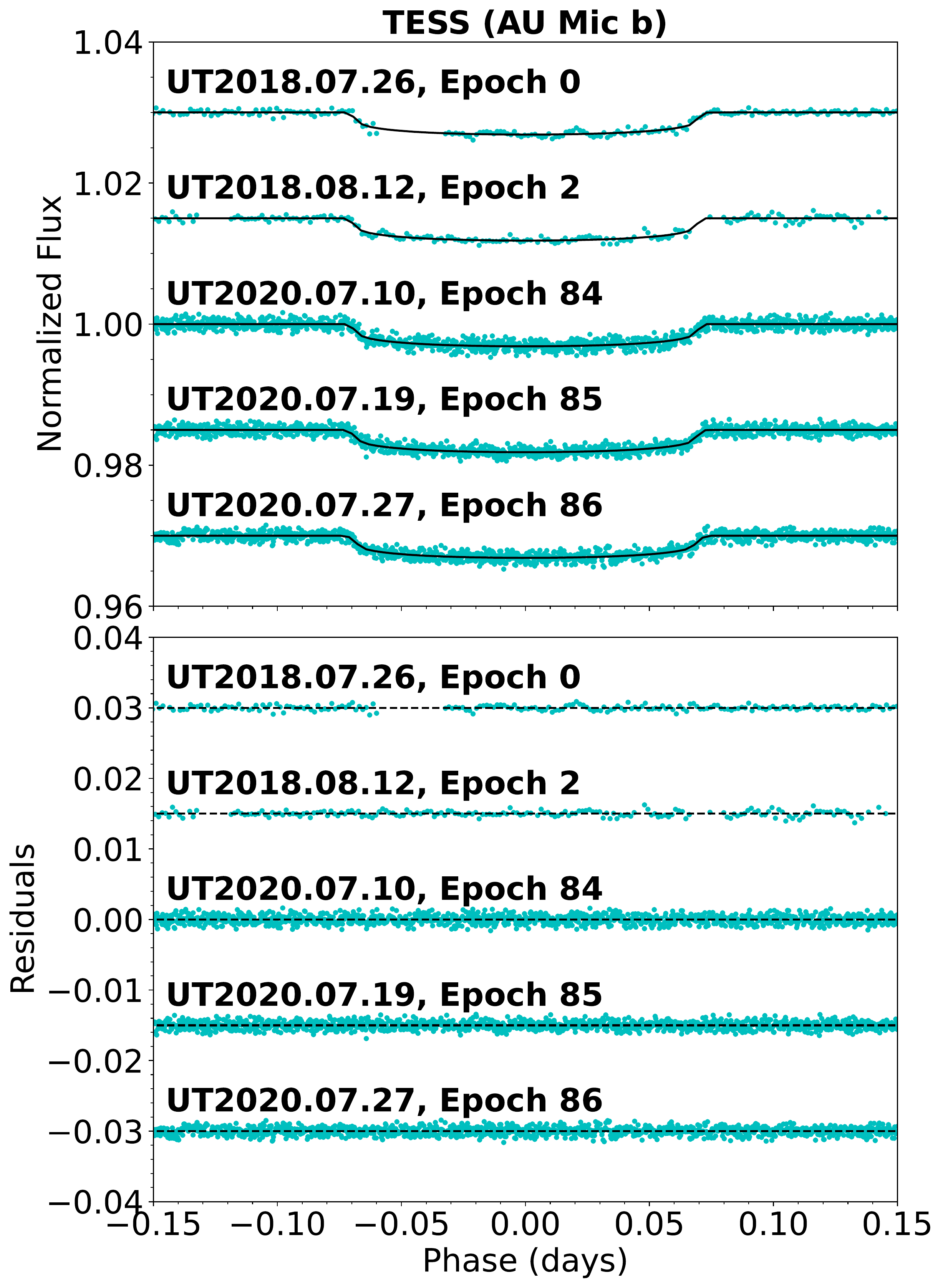} &
    \includegraphics[width=46mm]{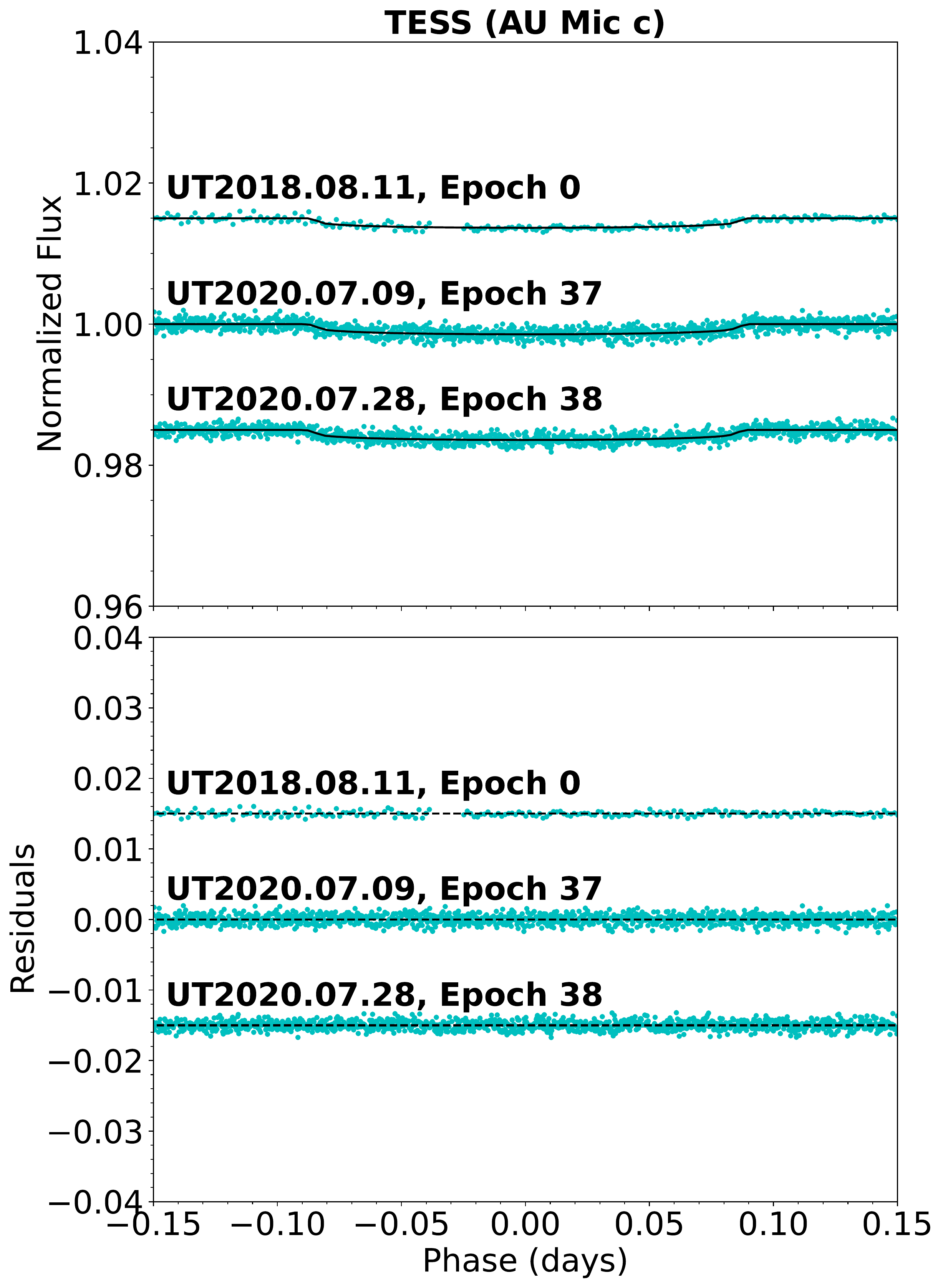}
    \end{tabular}
    \caption{Two-panel plots of comparison between ground-based + \spitzer + \tess transits (cyan) and \exofast's best-fit model (black) for AU Mic b \& c. The modelings are in the upper panels, and the residuals are in the lower panels. Herein epoch refers to the number of transits since the first transit of either b or c. The \brier transit is obtained at epoch 88; \saao at 78, 80, 82, 91, \& 95; \sso at 75, 88, 92, \& 94; \pest at 84; \spitzer at 23, 25, \& 48; and \tess b at 0, 2, 84, 85, \& 86, relative to the first \tess b transit. The \tess c transit is obtained at epoch 0, 37, \& 38 relative to the first \tess c transit.}
    \label{fig:exofastv2models}
\end{figure*}

Given that \aumic is a low-mass red dwarf, we configured \exofast to use MIST for evolutionary models \citep{choi2016, dotter2016} and ignore the Claret \& Bloemen limb darkening tables \citep{claret2011}. Additionally, we incorporate the spectral energy distribution (SED) to place constraint on MIST evolutionary models; the bands and their corresponding magnitude priors are presented in Table \ref{table:exofastv2sed}. We also assume the orbit of both \aumic b \& c to be non-circular. For the \exofast modeling, each of the 16 observations are detrended as indicated in Table \ref{table:detrend}; $\S$\ref{sec:dataobs} describes additional details on detrending parameters used for each data set. We split up the \exofast modeling into two sequential MCMC runs. For the first run, we integrate up to 15\,000 steps while setting {\tt NTHIN} = 12; the first run was also configured to integrate the priors from Tables \ref{table:exofastv2priors} \& \ref{table:exofastv2sed} and to invoke the {\tt rejectflatmodel} option for all light curves with {\tt NTEMPS} = 8 to aid in faster convergence. After the first run, \exofast generates the new {\tt prior.2} file, which we then implement while repeating the process to achieve better convergence. For the second run, we integrate up to 20\,000 steps while setting {\tt NTHIN} = 5, the {\tt rejectflatmodel} option was turned off, and the MIST SED file was omitted. After these runs were completed, \exofast generated the posteriors, including the transit models (Figure \ref{fig:exofastv2models} and Tables \ref{table:exfout}, \ref{table:exfobs1}, \ref{table:exfobs2}, \& \ref{table:exfobs3}) and midpoint times (see Table \ref{table:ttvpriors}). Of particular note are our eccentricity posteriors of 0.079$^{+0.160}_{-0.058}$ for \aumic b and 0.114$^{+0.120}_{-0.074}$ for \aumic c, which exclude moderate to high eccentricities. Additional analyses of the transits individually indicates that this posterior is most constrained by the \spitzer transits presented herein.

\clearpage
\startlongtable
\begin{deluxetable*}{l|c|c|cc}
    \tablecaption{\label{table:exfout}\exofast-generated median values and 68\% confidence interval for AU Mic system.}
    \tablehead{\multirow{2}{*}{Posterior} & \multirow{2}{*}{Description} & \multirow{2}{*}{Unit} & \multicolumn{2}{c}{Quantity} \\
        & & & \aumic b & \aumic c}
    \startdata
\mstar                     & Stellar Mass                                       & \msun             & \multicolumn{2}{c}{0.558$^{+0.068}_{0.090}$}                             \\
\rstar                     & Stellar Radius                                     & \rsun             & \multicolumn{2}{c}{0.749$^{+0.030}_{-0.037}$}                            \\
\lstar                     & Stellar Luminosity                                 & \lsun             & \multicolumn{2}{c}{0.094$^{+0.014}_{-0.015}$}                            \\
$\rho_{\star}$             & Stellar Density                                    & g/cm$^{3}$        & \multicolumn{2}{c}{1.89$^{+0.24}_{-0.21}$}                               \\
$\log{g}$                  & Surface Gravity                                    & ...               & \multicolumn{2}{c}{4.443$^{+0.044}_{-0.048}$}                            \\
\teff                      & Effective Temperature                              & K                 & \multicolumn{2}{c}{3701$^{+100}_{-98}$}                                  \\
$[{\rm Fe/H}]$             & Metallicity                                        & ...               & \multicolumn{2}{c}{0.17$^{+0.27}_{-0.37}$}                               \\
$[{\rm Fe/H}]_{0}$         & Initial Metallicity\tbnm{\scze a}                  & ...               & \multicolumn{2}{c}{0.12$^{+0.24}_{-0.36}$}                               \\
Age                        & ...                                                & Gyr               & \multicolumn{2}{c}{0.0216$^{+0.0034}_{-0.0031}$}                         \\
EEP                        & Equal Evolutionary Phase\tbnm{\scze b}             & ...               & \multicolumn{2}{c}{164.9$^{+5.4}_{-5.8}$}                                \\
\hline
\porb                      & Orbital Period                                     & days              & 8.462993$^{+0.000048}_{-0.000039}$ & 18.859005$^{+0.000084}_{-0.000072}$ \\
\mplan                     & Planetary Mass\tbnm{\scze c}                       & \mjup             & 0.058$^{+0.021}_{-0.016}$          & 0.0291$^{+0.0110}_{0.0080}$         \\
\rplan                     & Planetary Radius                                   & \rjup             & 0.373$^{+0.017}_{-0.021}$          & 0.250$^{+0.013}_{-0.015}$           \\
\tc                        & Time of Conjunction\tbnm{\scze d}                  & BJD\_TDB          & 2458330.3916$^{+0.0031}_{-0.0030}$ & 2458342.2238$^{+0.0026}_{-0.0023}$  \\
T$_{\rm T}$                & Time of Minimum Projected Separation\tbnm{\scze e} & BJD\_TDB          & 2458330.3916$^{+0.0031}_{-0.0030}$ & 2458342.2238$^{+0.0026}_{-0.0023}$  \\
T$_{0}$                    & Optimal Conjunction Time\tbnm{\scze f}             & BJD\_TDB          & 2458457.3365$^{+0.0031}_{-0.0029}$ & 2458455.3777$^{+0.0027}_{-0.0022}$  \\
a                          & Semi-Major Axis                                    & au                & 0.0669$^{+0.0026}_{-0.0038}$       & 0.1141$^{+0.0044}_{-0.0065}$        \\
e                          & Eccentricity                                       & ...               & 0.079$^{+0.160}_{-0.058}$          & 0.114$^{+0.120}_{-0.074}$           \\
i                          & Inclination                                        & deg               & 89.72$^{+0.22}_{-0.29}$            & 89.39$^{+0.47}_{-0.28}$             \\
$\omega$                   & Argument of Periastron                             & deg               & -90 $\pm$ 100                      & -90 $\pm$ 100                       \\
T$_{\rm eq}$               & Equilibrium Temperature\tbnm{\scze g}              & K                 & 595$^{+18}_{-21}$                  & 456$^{+14}_{-16}$                   \\
$\tau_{\rm circ}$          & Tidal Circularization Timescale                    & Gyr               & 152$^{+81}_{-55}$                  & 16400$^{+9000}_{-7500}$             \\
K                          & RV Semi-Amplitude\tbnm{\scze c}                    & m/s               & 8.6$^{+3.2}_{-2.4}$                & 3.33$^{+1.30}_{-0.92}$              \\
\rplan/\rstar              & ...                                                & ...               & 0.05137$^{+0.00099}_{-0.00120}$    & 0.03429$^{+0.00100}_{-0.00099}$     \\
a/\rstar                   & ...                                                & ...               & 19.28$^{+0.79}_{-0.73}$            & 32.9 $\pm$ 1.3                      \\
$\delta$                   & Transit Depth (\rplan/\rstar)$^2$                  & ...               & 0.00264$^{+0.00010}_{-0.00012}$    & 0.001176$^{+0.000070}_{-0.000067}$  \\
$\delta_{\rm I}$           & Transit Depth in I                                 & ...               & 0.00366$^{+0.00130}_{-0.00072}$    & 0.00152$^{+0.00048}_{-0.00027}$     \\
$\delta_{\rm z'}$          & Transit Depth in z'                                & ...               & 0.00409$^{+0.00053}_{-0.00044}$    & 0.00169$^{+0.00020}_{-0.00018}$     \\
$\delta_{\rm 4.5\mu m}$    & Transit Depth in $4.5\mu m$                        & ...               & 0.00265$^{+0.00010}_{-0.00012}$    & 0.001182$^{+0.000069}_{-0.000068}$  \\
$\delta_{\rm TESS}$        & Transit Depth in TESS                              & ...               & 0.00339$^{+0.00019}_{-0.00022}$    & 0.001462$^{+0.000077}_{-0.000093}$  \\
$\delta_{\rm V}$           & Transit Depth in V                                 & ...               & 0.00341$^{+0.0014}_{-0.00058}$     & 0.00146$^{+0.00051}_{-0.00024}$     \\
$\delta_{\rm y}$           & Transit Depth in y                                 & ...               & 0.00357$^{+0.00160}_{-0.00068}$    & 0.00149$^{+0.00057}_{-0.00025}$     \\
$\tau$                     & Ingress/Egress Transit Duration                    & days              & 0.00721$^{+0.00022}_{-0.00025}$    & 0.00664$^{+0.00180}_{-0.00084}$     \\
T$_{14}$                   & Total Transit Duration                             & days              & 0.14585$^{+0.00031}_{-0.00036}$    & 0.1773$^{+0.0017}_{-0.0015}$        \\
T$_{\rm FWHM}$             & FWHM Transit Duration                              & days              & 0.13861$^{+0.00020}_{-0.00023}$    & 0.1703$^{+0.0011}_{-0.0013}$        \\
b                          & Transit Impact Parameter                           & ...               & 0.098$^{+0.092}_{-0.065}$          & 0.35$^{+0.19}_{-0.26}$              \\
b$_{\rm S}$                & Eclipse Impact Parameter                           & ...               & 0.093$^{+0.091}_{-0.063}$          & 0.346$^{+0.095}_{-0.260}$           \\
$\tau_{\rm S}$             & Ingress/Egress Eclipse Duration                    & days              & 0.00715$^{+0.00059}_{-0.00070}$    & 0.00647$^{+0.00061}_{-0.00065}$     \\
T$_{\rm S,14}$             & Total Eclipse Duration                             & days              & 0.145$^{+0.011}_{-0.014}$          & 0.173$^{+0.020}_{-0.027}$           \\
T$_{\rm S,FWHM}$           & FWHM Eclipse Duration                              & days              & 0.137$^{+0.010}_{-0.014}$          & 0.166$^{+0.020}_{-0.026}$           \\
$\delta_{\rm S,2.5 \mu m}$ & Blackbody Eclipse Depth at 2.5 $\mu$m              & ppm               & 0.63$^{+0.18}_{-0.16}$             & 0.0147$^{+0.0060}_{-0.0051}$        \\
$\delta_{\rm S,5.0 \mu m}$ & Blackbody Eclipse Depth at 5.0 $\mu$m              & ppm               & 24.9$^{+3.3}_{-3.4}$               & 2.55$^{+0.48}_{-0.49}$              \\
$\delta_{\rm S,7.5 \mu m}$ & Blackbody Eclipse Depth at 7.5 $\mu$m              & ppm               & 74.3$^{+6.7}_{-7.4}$               & 12.2$^{+1.5}_{-1.7}$                \\
$\rho_{\rm p}$             & Planetary Density\tbnm{\scze c}                    & g/cm$^{3}$        & 1.42$^{+0.60}_{-0.35}$             & 2.28$^{+0.94}_{-0.62}$              \\
$\log{g_{\rm p}}$          & Surface Gravity\tbnm{\scze c}                      & ...               & 3.02$^{+0.14}_{-0.12}$             & 3.07$^{+0.14}_{-0.11}$              \\
$\Theta$                   & Safronov Number                                    & ...               & 0.0373$^{+0.0140}_{-0.0097}$       & 0.048$^{+0.019}_{-0.013}$           \\
$\langle F \rangle$        & Incident Flux                                      & 10$^{9}$ \ergfrac & 0.0282$^{+0.0035}_{-0.0039}$       & 0.0096 $\pm$ 0.0013                 \\
T$_{\rm P}$                & Time of Periastron                                 & BJD\_TDB          & 2458323.3$^{+2.7}_{-3.0}$          & 2458332.8$^{+6.1}_{-6.2}$           \\
T$_{\rm S}$                & Time of Eclipse                                    & BJD\_TDB          & 2458334.62$^{+0.54}_{-0.50}$       & 2458332.8 $\pm$ 1.3                 \\
T$_{\rm A}$                & Time of Ascending Node                             & BJD\_TDB          & 2458328.27$^{+0.30}_{-0.48}$       & 2458337.41$^{+0.72}_{-1.20}$        \\
T$_{\rm D}$                & Time of Descending Node                            & BJD\_TDB          & 2458332.51$^{+0.50}_{-0.28}$       & 2458347.03$^{+1.20}_{-0.72}$        \\
V$_{\rm c}$/V$_{\rm e}$    & {\it h}                                            & ...               & 1.000$^{+0.045}_{-0.038}$          & 1.009$^{+0.140}_{-0.060}$           \\
\radroot                   & Transit Chord                                      & ...               & 1.0466$^{+0.0042}_{-0.0130}$       & 0.974$^{+0.050}_{-0.092}$           \\
sign                       & {\it i}                                            & ...               & 1.40$^{+0.40}_{-0.33}$             & 1.48$^{+0.33}_{-0.34}$              \\
e$\cos{\omega}$            & ...                                                & ...               & -0.001$^{+0.100}_{-0.093}$         & $0.00^{+0.10}_{-0.11}$              \\
e$\sin{\omega}$            & ...                                                & ...               & -0.004$^{+0.037}_{-0.054}$         & -0.015$^{+0.060}_{-0.14}$           \\
\mplan$\sin{i}$            & Minimum Mass\tbnm{\scze c}                         & \mjup             & 0.058$^{+0.021}_{-0.016}$          & 0.0291$^{+0.0110}_{-0.0080}$        \\
\mplan/\mstar              & Mass Ratio\tbnm{\scze c}                           & ...               & 0.000099$^{+0.000039}_{-0.000028}$ & 0.000050$^{+0.000021}_{-0.000014}$  \\
d/\rstar                   & Separation at Mid Transit                          & ...               & 19.2$^{+1.6}_{-1.5}$               & 33.1$^{+4.2}_{-2.7}$                \\
P$_{\rm T}$                & A Priori Non-grazing Transit Probability           & ...               & 0.0495$^{+0.0043}_{-0.0038}$       & 0.0291$^{+0.0026}_{-0.0033}$        \\
P$_{\rm T,G}$              & A Priori Transit Probability                       & ...               & 0.0548$^{+0.0047}_{-0.0041}$       & 0.0312$^{+0.0028}_{-0.0035}$        \\
P$_{\rm S}$                & A Priori Non-grazing Eclipse Probability           & ...               & 0.04958$^{+0.00210}_{-0.00067}$    & 0.0299$^{+0.0038}_{-0.0020}$        \\
P$_{\rm S,G}$              & A Priori Eclipse Probability                       & ...               & 0.05492$^{+0.00240}_{-0.00073}$    & 0.0320$^{+0.0042}_{-0.0021}$        \\
\hline
u$_{\rm 1,I}$              & Linear Limb-darkening Coefficient in I             & ...               & \multicolumn{2}{c}{0.57$^{+0.39}_{-0.36}$}                               \\
u$_{\rm 1,z'}$             & Linear Limb-darkening Coefficient in z'            & ...               & \multicolumn{2}{c}{0.72$^{+0.13}_{-0.14}$}                               \\
u$_{\rm 1,4.5\mu m}$       & Linear Limb-darkening Coefficient in 4.5 $\mu$m    & ...               & \multicolumn{2}{c}{0.0086$^{+0.0130}_{-0.0065}$}                         \\
u$_{\rm 1,TESS}$           & Linear Limb-darkening Coefficient in TESS          & ...               & \multicolumn{2}{c}{0.453$^{+0.059}_{-0.064}$}                            \\
u$_{\rm 1,V}$              & Linear Limb-darkening Coefficient in V             & ...               & \multicolumn{2}{c}{0.47$^{+0.45}_{-0.33}$}                               \\
u$_{\rm 1,y}$              & Linear Limb-darkening Coefficient in y             & ...               & \multicolumn{2}{c}{0.54$^{+0.45}_{-0.36}$}                               \\
u$_{\rm 2,I}$              & Quadratic Limb-darkening Coefficient in I          & ...               & \multicolumn{2}{c}{0.01$^{+0.42}_{-0.33}$}                               \\
u$_{\rm 2,z'}$             & Quadratic Limb-darkening Coefficient in z'         & ...               & \multicolumn{2}{c}{-0.28$^{+0.17}_{-0.10}$}                              \\
u$_{\rm 2,4.5\mu m}$       & Quadratic Limb-darkening Coefficient in 4.5 $\mu$m & ...               & \multicolumn{2}{c}{0.140$^{+0.037}_{-0.038}$}                            \\
u$_{\rm 2,TESS}$           & Quadratic Limb-darkening Coefficient in TESS       & ...               & \multicolumn{2}{c}{0.207$^{+0.100}_{-0.097}$}                            \\
u$_{\rm 2,V}$              & Quadratic Limb-darkening Coefficient in V          & ...               & \multicolumn{2}{c}{0.09$^{+0.45}_{-0.37}$}                               \\
u$_{\rm 2,y}$              & Quadratic Limb-darkening Coefficient in y          & ...               & \multicolumn{2}{c}{0.02$^{+0.40}_{-0.34}$}
    \enddata
    \tablecomments{See Table 3 in \citet{eastman2019} for a detailed description of all parameters. Since both Pan-STARRS Y and Pan-STARRS \zs are not available among the filters in \exofast, y and z' (Sloan z) were used as respective approximate substitutes. Additionally, the Claret \& Bloemen limb darkening tables \citep{claret2011} default option has been disabled since \aumic is a low-mass red dwarf.}
    \tablenotetext{a}{The metallicity of the star at birth.}
    \tablenotetext{b}{Corresponds to static points in a star's evolutionary history. See $\S$2 in \citet{dotter2016}.}
    \tablenotetext{c}{Uses measured radius and estimated mass from \citet{chen2017}.}
    \tablenotetext{d}{Time of Conjunction is commonly reported as the ``transit time".}
    \tablenotetext{e}{Time of Minimum Projected Separation is a more correct ``transit time".}
    \tablenotetext{f}{Optimal Time of Conjunction minimizes the covariance between \tc and Period.}
    \tablenotetext{g}{Assumes no albedo and perfect redistribution.}
    \tablenotetext{h}{The velocity at \tc of an assumed circular orbit divided by the velocity of the modeled eccentric orbit.}
    \tablenotetext{i}{The sign of the solution to the quadratic mapping from V$_{\rm c}$/V$_{\rm e}$ to e.}
\end{deluxetable*}

\movetabledown=30mm
\begin{rotatetable*}
\begin{deluxetable*}{c|l|c|c|c|c|c|c}
    \tablecaption{\label{table:exfobs1}\exofast-generated median values and 68\% confidence interval for follow-up observations of AU Mic transits (Part I).}
    \tablehead{Planet & Telescope & Date (UT) & Filter & $\sigma^{2}$ (Added Variance) & TTV\tbnm{a} (days) & T$\delta$ V\tbnm{b} & F$_{0}$\tbnm{c}}
    \startdata
b & \tess    & 2018-07-26 & TESS       & 0.0000000019$^{+0.0000000100}_{-0.0000000088}$ & -0.0024$^{+0.0028}_{-0.0031}$ & -0.0006$^{+0.0013}_{-0.0012}$    & 1.000002$^{+0.000024}_{-0.000023}$ \\
c & \tess    & 2018-08-11 & TESS       & 0.0000000030 $\pm$ 0.0000000097                & -0.0005$^{+0.0020}_{-0.0026}$ & -0.00006$^{+0.00080}_{-0.00074}$ & 0.999994$^{+0.000026}_{-0.000027}$ \\
b & \tess    & 2018-08-12 & TESS       & 0.000000005$^{+0.000000016}_{-0.000000013}$    & -0.0006$^{+0.0029}_{-0.0030}$ & -0.0006$^{+0.0014}_{-0.0012}$    & 0.999999 $\pm$ 0.000030            \\
b & \spitzer & 2019-02-10 & 4.5 $\mu$m & 0.0000000600$^{+0.0000000050}_{-0.0000000047}$ & 0.0039 $\pm$ 0.0030           & -0.00853$^{+0.00130}_{-0.00100}$ & 1.000075 $\pm$ 0.000022            \\
b & \spitzer & 2019-02-27 & 4.5 $\mu$m & 0.0000000374$^{+0.0000000040}_{-0.0000000036}$ & 0.0049$^{+0.0029}_{-0.0031}$  & -0.0055$^{+0.0015}_{-0.0012}$    & 1.000220$^{+0.000019}_{-0.000017}$ \\
b & \spitzer & 2019-09-09 & 4.5 $\mu$m & 0.0000000933$^{+0.0000000042}_{-0.0000000037}$ & 0.0020$^{+0.0032}_{-0.0035}$  & -0.0039$^{+0.0014}_{-0.0011}$    & 0.999893$^{+0.000013}_{-0.000014}$ \\
b & \sso     & 2020-04-25 & z'         & 0.00000234$^{+0.00000029}_{-0.00000028}$       & -0.0028$^{+0.0040}_{-0.0048}$ & -0.0054$^{+0.0047}_{-0.0032}$    & 1.00000$^{+0.00013}_{-0.00014}$    \\
b & \sso     & 2020-04-25 & y          & 0.00000104$^{+0.00000100}_{-0.00000067}$       & -0.0037$^{+0.0048}_{-0.0059}$ & 0.0037$^{+0.0046}_{-0.0076}$     & 0.99964$^{+0.00031}_{-0.00038}$    \\
b & \saao    & 2020-05-20 & z'         & 0.00000188 $\pm$ 0.00000018                    & -0.0019$^{+0.0044}_{-0.0049}$ & 0.0005$^{+0.0025}_{-0.0027}$     & 1.00034 $\pm$ 0.00018              \\
b & \saao    & 2020-05-20 & y          & 0.00000140$^{+0.00000060}_{-0.00000047}$       & 0.0084$^{+0.0057}_{-0.0066}$  & 0.0000$^{+0.0047}_{-0.0052}$     & 0.99993$^{+0.00044}_{-0.00039}$    \\
b & \saao    & 2020-06-06 & z'         & 0.00000677$^{+0.00000081}_{-0.00000066}$       & -0.0053$^{+0.0050}_{-0.0056}$ & -0.0018$^{+0.0053}_{-0.0050}$    & 1.00005$^{+0.00041}_{-0.00034}$    \\
b & \saao    & 2020-06-23 & z'         & 0.00000185$^{+0.00000025}_{-0.00000021}$       & 0.0003$^{+0.0040}_{-0.0049}$  & 0.0079$^{+0.0016}_{-0.0031}$     & 0.99987$^{+0.00018}_{-0.00021}$    \\
c & \tess    & 2020-07-09 & TESS       & 0.000000001 $\pm$ 0.000000015                  & -0.0024$^{+0.0031}_{-0.0047}$ & 0.00023$^{+0.00100}_{-0.00063}$  & 1.000001$^{+0.000020}_{-0.000021}$ \\
b & \tess    & 2020-07-10 & TESS       & 0.000000002$^{+0.000000012}_{-0.000000011}$    & -0.0018$^{+0.0038}_{-0.0047}$ & -0.0006$^{+0.0013}_{-0.0011}$    & 0.999999$^{+0.000015}_{-0.000018}$ \\
b & \pest    & 2020-07-10 & V          & 0.00000600$^{+0.00000074}_{-0.00000076}$       & -0.0038 $\pm$ 0.0061          & -0.0081$^{+0.0029}_{-0.0014}$    & 0.99817$^{+0.00018}_{-0.00015}$    \\
b & \tess    & 2020-07-19 & TESS       & 0.0000000015$^{+0.0000000100}_{-0.0000000099}$ & -0.0009$^{+0.0036}_{-0.0050}$ & -0.0006$^{+0.0013}_{-0.0011}$    & 1.000000 $\pm$ 0.000015            \\
b & \tess    & 2020-07-27 & TESS       & 0.0000000011 $\pm$ 0.0000000097                & 0.0002$^{+0.0037}_{-0.0050}$  & -0.0007$^{+0.0013}_{-0.0011}$    & 0.999999 $\pm$ 0.000015            \\
c & \tess    & 2020-07-28 & TESS       & 0.000000001$^{+0.000000013}_{-0.000000012}$    & 0.0001$^{+0.0032}_{-0.0039}$  & 0.00022$^{+0.00097}_{-0.00058}$  & 1.000001$^{+0.000017}_{-0.000019}$ \\
b & \brier   & 2020-08-13 & I          & 0.0000117$^{+0.0000011}_{-0.0000010}$          & 0.0085$^{+0.0047}_{-0.0055}$  & 0.0070$^{+0.0022}_{-0.0036}$     & 0.99935$^{+0.00025}_{-0.00028}$    \\
b & \sso     & 2020-08-13 & z'         & 0.00000406$^{+0.00000039}_{-0.00000037}$       & -0.0018$^{+0.0044}_{-0.0062}$ & -0.0013$^{+0.0034}_{-0.0036}$    & 1.00009$^{+0.00030}_{-0.00026}$    \\
b & \saao    & 2020-09-07 & z'         & 0.00000688$^{+0.00000083}_{-0.00000070}$       & 0.0064$^{+0.0092}_{-0.0068}$  & -0.0057$^{+0.0042}_{-0.0030}$    & 0.99988$^{+0.00045}_{-0.00035}$    \\
b & \sso     & 2020-09-16 & z'         & 0.00000131$^{+0.00000013}_{-0.00000011}$       & -0.0020$^{+0.0041}_{-0.0054}$ & 0.0000$^{+0.0021}_{-0.0023}$     & 0.99996 $\pm$ 0.00013              \\
b & \sso     & 2020-10-03 & z'         & 0.00000125$^{+0.00000017}_{-0.00000014}$       & 0.0011$^{+0.0041}_{-0.0056}$  & 0.0020$^{+0.0022}_{-0.0024}$     & 1.00216$^{+0.00017}_{-0.00016}$    \\
b & \saao    & 2020-10-11 & z'         & 0.00000651$^{+0.00000069}_{-0.00000058}$       & 0.0100$^{+0.0070}_{-0.0074}$  & 0.0039$^{+0.0030}_{-0.0033}$     & 1.00099$^{+0.00048}_{-0.00049}$
    \enddata
    \tablecomments{See Table 3 in \citet{eastman2019} for a detailed description of all parameters. Since both Pan-STARRS Y and Pan-STARRS \zs are not available among the filters in \exofast, y and z' (Sloan z) were used as respective approximate substitutes.}
    \tablenotetext{a}{Transit Timing Variation}
    \tablenotetext{b}{Transit Depth Variation}
    \tablenotetext{c}{Baseline Flux}
\end{deluxetable*}
\end{rotatetable*}

\movetabledown=30mm
\begin{rotatetable*}
\begin{deluxetable*}{c|l|c|c|c|c|c|c|c}
    \tablecaption{\label{table:exfobs2}\exofast-generated median values and 68\% confidence interval for follow-up observations of AU Mic b (Part II).}
    \tablehead{Planet & Telescope & Date (UT) & Filter & C$_{0}$\tbnm{a} & C$_{1}$\tbnm{a} & M$_{0}$\tbnm{b} & M$_{1}$\tbnm{b} & M$_{2}$\tbnm{b}}
    \startdata
b & \tess    & 2018-07-26 & TESS       & ...                                 & ...                                & ...                              & ...                             & ...                              \\
c & \tess    & 2018-08-11 & TESS       & ...                                 & ...                                & ...                              & ...                             & ...                              \\
b & \tess    & 2018-08-12 & TESS       & ...                                 & ...                                & ...                              & ...                             & ...                              \\
b & \spitzer & 2019-02-10 & 4.5 $\mu$m & -0.000165$^{+0.000043}_{-0.000041}$ & ...                                & -0.00110$^{+0.00034}_{-0.00033}$ & -0.0112 $\pm$ 0.0010            & 0.0108 $\pm$ 0.0010              \\
b & \spitzer & 2019-02-27 & 4.5 $\mu$m & -0.000247$^{+0.000044}_{-0.000042}$ & 0.000048$^{+0.000041}_{-0.000032}$ & 0.00070$^{+0.00033}_{-0.00037}$  & 0.00037$^{+0.00037}_{-0.00041}$ & -0.00303$^{+0.00081}_{-0.00076}$ \\
b & \spitzer & 2019-09-09 & 4.5 $\mu$m & -0.000576$^{+0.000093}_{-0.000110}$ & ...                                & -0.00030 $\pm$ 0.00016           & 0.00194$^{+0.00020}_{-0.00021}$ & -0.00022$^{+0.00061}_{-0.00063}$ \\
b & \sso     & 2020-04-25 & z'         & ...                                 & ...                                & 0.00000$^{+0.00024}_{-0.00026}$  & ...                             & ...                              \\
b & \sso     & 2020-04-25 & y          & ...                                 & ...                                & -0.00111$^{+0.00074}_{-0.00079}$ & ...                             & ...                              \\
b & \saao    & 2020-05-20 & z'         & ...                                 & ...                                & 0.00102 $\pm$ 0.00035            & ...                             & ...                              \\
b & \saao    & 2020-05-20 & y          & ...                                 & ...                                & -0.00007$^{+0.00071}_{-0.00070}$ & ...                             & ...                              \\
b & \saao    & 2020-06-06 & z'         & ...                                 & ...                                & 0.00025$^{+0.00081}_{-0.00074}$  & ...                             & ...                              \\
b & \saao    & 2020-06-23 & z'         & ...                                 & ...                                & -0.00021$^{+0.00038}_{-0.00049}$ & ...                             & ...                              \\
c & \tess    & 2020-07-09 & TESS       & ...                                 & ...                                & ...                              & ...                             & ...                              \\
b & \tess    & 2020-07-10 & TESS       & ...                                 & ...                                & ...                              & ...                             & ...                              \\
b & \pest    & 2020-07-10 & V          & -0.0037$^{+0.0050}_{-0.0049}$       & ...                                & 0.1038 $\pm$ 0.0040              & -0.1074 $\pm$ 0.0043            & -0.0002$^{+0.0017}_{-0.0016}$    \\
b & \tess    & 2020-07-19 & TESS       & ...                                 & ...                                & ...                              & ...                             & ...                              \\
b & \tess    & 2020-07-27 & TESS       & ...                                 & ...                                & ...                              & ...                             & ...                              \\
c & \tess    & 2020-07-28 & TESS       & ...                                 & ...                                & ...                              & ...                             & ...                              \\
b & \brier   & 2020-08-13 & I          & ...                                 & ...                                & -0.0038 $\pm$ 0.0052             & -0.0026$^{+0.0045}_{-0.0047}$   & -0.0009$^{+0.0022}_{-0.0023}$    \\
b & \sso     & 2020-08-13 & z'         & ...                                 & ...                                & -0.00002$^{+0.00043}_{-0.00052}$ & 0.00028$^{+0.00029}_{-0.00033}$ & ...                              \\
b & \saao    & 2020-09-07 & z'         & -$0.00020^{+0.00063}_{-0.00066}$    & ...                                & -0.00035 $\pm$ 0.00096           & ...                             & ...                              \\
b & \sso     & 2020-09-16 & z'         & ...                                 & ...                                & 0.00014$^{+0.00033}_{-0.00029}$  & ...                             & ...                              \\
b & \sso     & 2020-10-03 & z'         & ...                                 & ...                                & -0.00179 $\pm$ 0.00028           & ...                             & ...                              \\
b & \saao    & 2020-10-11 & z'         & ...                                 & ...                                & -0.0014 $\pm$ 0.0016             & ...                             & ...
    \enddata
    \tablecomments{See Table 3 in \citet{eastman2019} for a detailed description of all parameters. Since both Pan-STARRS Y and Pan-STARRS \zs are not available among the filters in \exofast, y and z' (Sloan z) were used as respective approximate substitutes. Also see $\S$\ref{sec:dataobs} of this paper for details on detrending parameters used for each observation; the flare (\spitzer), sky (\spitzer \& \pest), \& Sky/Pixel\_T1 (\saao) were set as additive while the remaining detrending parameters were set as multiplicative.}
    \tablenotetext{a}{Additive Detrending Coefficient}
    \tablenotetext{b}{Multiplicative Detrending Coefficient}
\end{deluxetable*}
\end{rotatetable*}

\movetabledown=30mm
\begin{rotatetable*}
\begin{deluxetable*}{c|l|c|c|c|c|c|c|c}
    \tablecaption{\label{table:exfobs3}\exofast-generated median values and 68\% confidence interval for follow-up observations of AU Mic b (Part III).}
    \tablehead{Planet & Telescope & Date (UT) & Filter & M$_{3}$\tbnm{a} & M$_{4}$\tbnm{a} & M$_{5}$\tbnm{a} & M$_{6}$\tbnm{a} & M$_{7}$\tbnm{a}}
    \startdata
b & \tess    & 2018-07-26 & TESS       & ...                              & ...                              & ...                                 & ...                                 & ...                                 \\
c & \tess    & 2018-08-11 & TESS       & ...                              & ...                              & ...                                 & ...                                 & ...                                 \\
b & \tess    & 2018-08-12 & TESS       & ...                              & ...                              & ...                                 & ...                                 & ...                                 \\
b & \spitzer & 2019-02-10 & 4.5 $\mu$m & -0.00623$^{+0.00075}_{-0.00074}$ & 0.00496 $\pm$ 0.00048            & -0.000407$^{+0.000041}_{-0.000042}$ & -0.000270$^{+0.000057}_{-0.000053}$ & ...                                 \\
b & \spitzer & 2019-02-27 & 4.5 $\mu$m & 0.00326$^{+0.00073}_{-0.00080}$  & 0.00118$^{+0.00065}_{-0.00060}$  & 0.000873$^{+0.000095}_{-0.000110}$  & -0.00088$^{+0.00015}_{-0.00013}$    & ...                                 \\
b & \spitzer & 2019-09-09 & 4.5 $\mu$m & 0.00049 $\pm$ 0.00056            & -0.00174$^{+0.00046}_{-0.00045}$ & -0.000003$^{+0.000044}_{-0.000046}$ & -0.000403$^{+0.000061}_{-0.000057}$ & -0.000519$^{+0.000059}_{-0.000058}$ \\
b & \sso     & 2020-04-25 & z'         & ...                              & ...                              & ...                                 & ...                                 & ...                                 \\
b & \sso     & 2020-04-25 & y          & ...                              & ...                              & ...                                 & ...                                 & ...                                 \\
b & \saao    & 2020-05-20 & z'         & ...                              & ...                              & ...                                 & ...                                 & ...                                 \\
b & \saao    & 2020-05-20 & y          & ...                              & ...                              & ...                                 & ...                                 & ...                                 \\
b & \saao    & 2020-06-06 & z'         & ...                              & ...                              & ...                                 & ...                                 & ...                                 \\
b & \saao    & 2020-06-23 & z'         & ...                              & ...                              & ...                                 & ...                                 & ...                                 \\
c & \tess    & 2020-07-09 & TESS       & ...                              & ...                              & ...                                 & ...                                 & ...                                 \\
b & \tess    & 2020-07-10 & TESS       & ...                              & ...                              & ...                                 & ...                                 & ...                                 \\
b & \pest    & 2020-07-10 & V          & 0.0004$^{+0.0012}_{-0.0014}$     & -0.0023$^{+0.0028}_{-0.0026}$    & 0.0013 $\pm$ 0.0011                 & -0.00284$^{+0.00088}_{-0.00087}$    & ...                                 \\
b & \tess    & 2020-07-19 & TESS       & ...                              & ...                              & ...                                 & ...                                 & ...                                 \\
b & \tess    & 2020-07-27 & TESS       & ...                              & ...                              & ...                                 & ...                                 & ...                                 \\
c & \tess    & 2020-07-28 & TESS       & ...                              & ...                              & ...                                 & ...                                 & ...                                 \\
b & \brier   & 2020-08-13 & I          & -0.0009$^{+0.0015}_{-0.0014}$    & ...                              & ...                                 & ...                                 & ...                                 \\
b & \sso     & 2020-08-13 & z'         & ...                              & ...                              & ...                                 & ...                                 & ...                                 \\
b & \saao    & 2020-09-07 & z'         & ...                              & ...                              & ...                                 & ...                                 & ...                                 \\
b & \sso     & 2020-09-16 & z'         & ...                              & ...                              & ...                                 & ...                                 & ...                                 \\
b & \sso     & 2020-10-03 & z'         & ...                              & ...                              & ...                                 & ...                                 & ...                                 \\
b & \saao    & 2020-10-11 & z'         & ...                              & ...                              & ...                                 & ...                                 & ...
    \enddata
    \tablecomments{See Table 3 in \citet{eastman2019} for a detailed description of all parameters. Since both Pan-STARRS Y and Pan-STARRS \zs are not available among the filters in \exofast, y and z' (Sloan z) were used as respective approximate substitutes. Also see $\S$\ref{sec:dataobs} of this paper for details on detrending parameters used for each observation; the flare (\spitzer), sky (\spitzer \& \pest), \& Sky/Pixel\_T1 (\saao) were set as additive while the remaining detrending parameters were set as multiplicative.}
    \tablenotetext{a}{Multiplicative Detrending Coefficient}
\end{deluxetable*}
\end{rotatetable*}

\section{TTV Analysis}\label{sec:ttvs}

In this section, we present our O--C diagram and TTV dynamical modeling with \exostriker, considering both a two-planet model and an example three-planet model to account for the observed TTVs.

\subsection{O--C Diagram}\label{sec:ocdiagram}

\begin{figure}
    \centering
    \includegraphics[width=\linewidth]{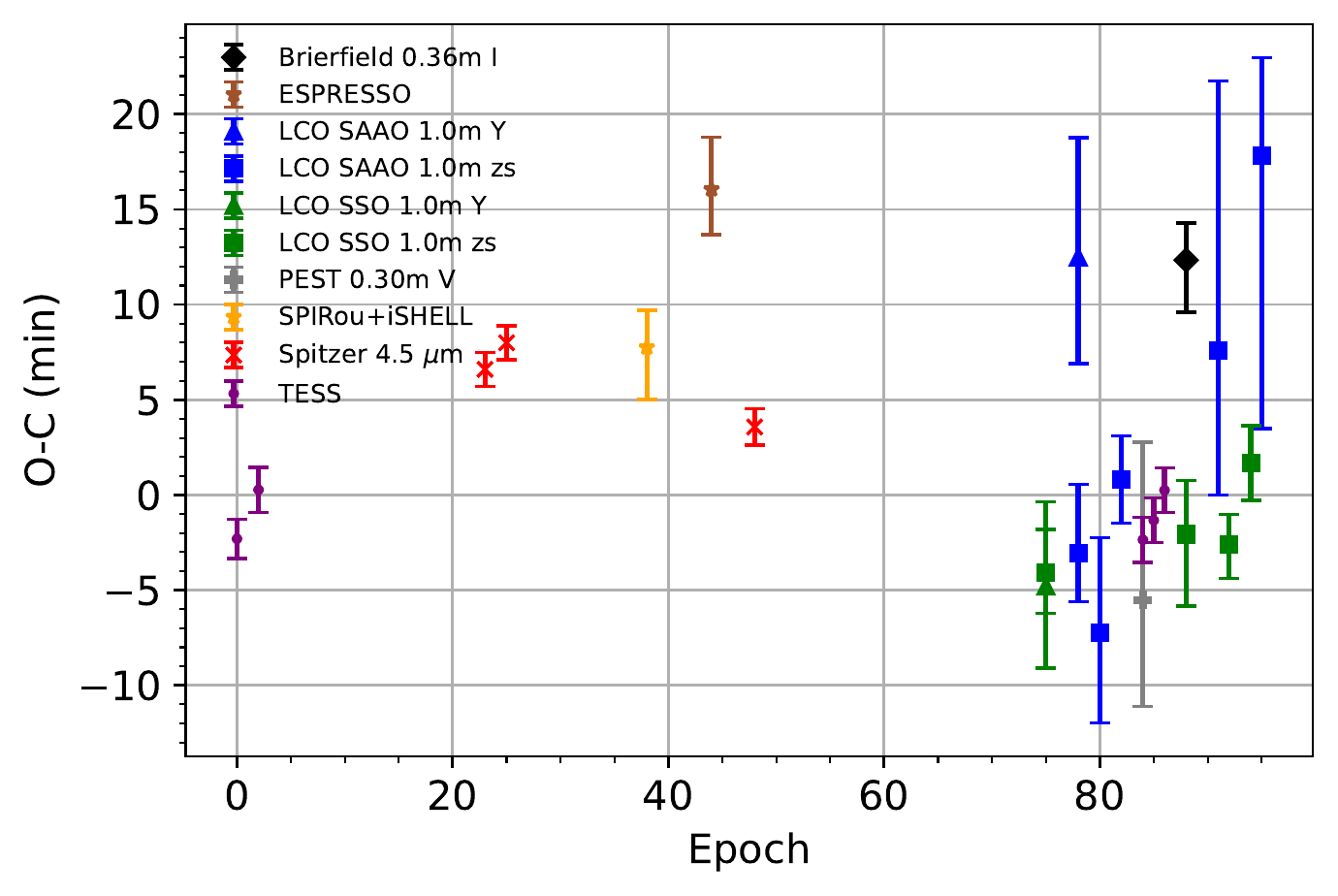}
    \caption{O--C Diagram of \aumic b, using the \exofast-generated measured midpoint times and the calculated expected midpoint times for all 23 \aumic b transit data sets from Table \ref{table:datasets}. \aumic b's period and \tc from \citet{gilbert2021} were used for the calculation of expected midpoint times.}
    \label{fig:ocdiagram}
\end{figure}

We calculate the expected midpoint times using \aumic b's period and \tc from \citet{gilbert2021}; the period and \tc from \citet{martioli2021} yield similar results. Using the measured and expected midpoint times, we construct the O--C diagram in Figure \ref{fig:ocdiagram}. We make use of the combined transit midpoint times (Table \ref{table:ttvpriors}), now extracted from all light curves and R-M observations through the processes described in the previous $\S$\ref{sec:dataobs} for data from \citet{gilbert2021} and the R-M observations, and from our own analysis in $\S$\ref{sec:exofastmod}. With a $\chi^{2}\gg1$, it is readily apparent by eye that the precise \spitzer transit times are significantly deviant from those expected from a linear ephemeris from the \tess transit times alone. Note, the \spitzer transit times are in BJD and corrected for the relative light-time travel delay between \spitzer and the Solar System barycenter. Additionally, the ground-based R-M transit midpoints are similarly late and consistent with the \spitzer transit times. The ground-based photometric transits show large scatter and correspondingly larger timing uncertainties relative to the space-based transit midpoint times.

\begin{deluxetable*}{c|l|c|c|c|c}
    \tabletypesize{\scze}
    \tablecaption{\label{table:ttvpriors}\aumic planets' midpoint time priors for \exostriker models.}
    \tablehead{Planet & Telescope & Transit N & Date (UT) & Band & T$_{0}$ (BJD)}
    \startdata
\multirow{23}{*}{b} & \tess             & 1  & 2018-07-26 & TESS           & 2458330.38920 $\pm$ 0.00041 \\
                    & \tess             & 3  & 2018-08-12 & TESS           & 2458347.31699 $\pm$ 0.00060 \\
                    & \spitzer          & 24 & 2019-02-10 & 4.5 $\mu$m     & 2458525.04439 $\pm$ 0.00017 \\
                    & \spitzer          & 26 & 2019-02-27 & 4.5 $\mu$m     & 2458541.97136 $\pm$ 0.00016 \\
                    & \spirou + \ishell & 39 & 2019-06-17 & (a)            & 2458651.99020 $\pm$ 0.00180 \\
                    & \espresso         & 45 & 2019-08-07 & 378.2-788.7 nm & 2458702.77397 $\pm$ 0.00178 \\
                    & \spitzer          & 49 & 2019-09-09 & 4.5 $\mu$m     & 2458736.61730 $\pm$ 0.00015 \\
                    & \sso              & 76 & 2020-04-25 & Pan-STARRS Y   & 2458965.11250 $\pm$ 0.00300 \\
                    & \sso              & 76 & 2020-04-25 & Pan-STARRS \zs & 2458965.11300 $\pm$ 0.00140 \\
                    & \saao             & 79 & 2020-05-20 & Pan-STARRS \zs & 2458990.50270 $\pm$ 0.00240 \\
                    & \saao             & 79 & 2020-05-20 & Pan-STARRS Y   & 2458990.51350 $\pm$ 0.00430 \\
                    & \saao             & 81 & 2020-06-06 & Pan-STARRS \zs & 2459007.42580 $\pm$ 0.00340 \\
                    & \saao             & 83 & 2020-06-23 & Pan-STARRS \zs & 2459024.35740 $\pm$ 0.00140 \\
                    & \pest             & 85 & 2020-07-10 & V              & 2459041.27900 $\pm$ 0.00570 \\
                    & \tess             & 85 & 2020-07-10 & TESS           & 2459041.28120 $\pm$ 0.00030 \\
                    & \tess             & 86 & 2020-07-19 & TESS           & 2459049.74491 $\pm$ 0.00028 \\
                    & \tess             & 87 & 2020-07-27 & TESS           & 2459058.20901 $\pm$ 0.00026 \\
                    & \sso              & 89 & 2020-08-13 & Pan-STARRS \zs & 2459075.13340 $\pm$ 0.00250 \\
                    & \brier            & 89 & 2020-08-13 & I              & 2459075.14340 $\pm$ 0.00240 \\
                    & \saao             & 92 & 2020-09-07 & Pan-STARRS \zs & 2459100.52910 $\pm$ 0.00980 \\
                    & \sso              & 93 & 2020-09-16 & Pan-STARRS \zs & 2459108.98502 $\pm$ 0.00092 \\
                    & \sso              & 95 & 2020-10-03 & Pan-STARRS \zs & 2459125.91400 $\pm$ 0.00110 \\
                    & \saao             & 96 & 2020-10-11 & Pan-STARRS \zs & 2459134.38820 $\pm$ 0.00990 \\
\hline
\multirow{3}{*}{c}  & \tess             & 1  & 2018-08-11 & TESS           & 2458342.22330 $\pm$ 0.00110 \\
                    & \tess             & 38 & 2020-07-09 & TESS           & 2459040.00432 $\pm$ 0.00082 \\
                    & \tess             & 39 & 2020-07-28 & TESS           & 2459058.86603 $\pm$ 0.00072
    \enddata
    \tablecomments{(a) 955-2\,515 nm (\spirou) and 2.18-2.47 nm (\ishell)}
\end{deluxetable*}

\subsection{TTV \exostriker Dynamical Modeling}

Motivated by the apparent TTV variability deviating from a linear ephemeris in $\S$\ref{sec:ocdiagram}, we utilize the \exostriker package \citep{trifonov2019} to model the variation in transit timings of \aumic b and c. Like \exofast, \exostriker utilizes a Markov chain Monte Carlo (MCMC) to assess the statistical significance of the measured TTVs and the confidence in the dynamical model posteriors that can be inferred from the TTVs. Additionally, we incorporate the stellar priors from Table \ref{table:exostriker_stellar_priors} and the planet priors from Table \ref{table:exostriker_planet_priors}.

\begin{deluxetable}{l|c|c}
    \tablecaption{\label{table:exostriker_stellar_priors}\aumic's stellar priors for \exostriker best-fit and MCMC modeling.}
    \tablehead{Prior & Unit & AU Mic}
    \startdata
Mass       & \msun & $\mathcal{N}$(0.50, 0.03)     \\
Radius     & \rsun & $\mathcal{N}$(0.75, 0.03)     \\
Luminosity & \lsun & $\mathcal{N}$(0.0900, 0.0001) \\
\teff      & K     & $\mathcal{N}$(3700, 100)      \\
v sin i    & km/s  & $\mathcal{N}$(8.7, 0.2)
    \enddata
    \tablecomments{These priors are taken from \citet{plavchan2020}.}
\end{deluxetable}

\begin{deluxetable*}{l|c|c|c|c}
    \tablecaption{\label{table:exostriker_planet_priors}\aumic's planetary priors for \exostriker best-fit and MCMC modeling.}
    \tablehead{Prior & Unit & \aumic b & \aumic c & \aumic d\tbnm{\scriptsize a}}
    \startdata
K        & m/s & $\mathcal{N}$(8.5, 2.3)         & $\mathcal{U}$(0.8, 9.5)         & $\mathcal{U}$(0.0, 10000.0)  \\
\porb    & day & $\mathcal{N}$(8.463, 0.001)     & $\mathcal{N}$(18.859, 0.001)    & $\mathcal{N}$(12.742, 0.020) \\
e        & ... & $\mathcal{U}$(0.00000, 0.58038) & $\mathcal{U}$(0.00000, 0.37308) & $\mathcal{U}$(0.000, 0.999)  \\
$\omega$ & deg & $\mathcal{U}$(0.0, 360.0)       & $\mathcal{U}$(0.0, 360.0)       & $\mathcal{U}$(0.0, 360.0)    \\
M$_{0}$  & deg & $\mathcal{U}$(0.0, 360.0)       & $\mathcal{U}$(0.0, 360.0)       & $\mathcal{U}$(0.0, 360.0)    \\
i        & deg & $\mathcal{N}$(89.5, 0.3)        & $\mathcal{N}$(89.0, 0.5)        & $\mathcal{U}$(0.0, 180.0)    \\
$\Omega$ & deg & $\mathcal{U}$(0.0, 360.0)       & $\mathcal{U}$(0.0, 360.0)       & $\mathcal{U}$(0.0, 360.0)
    \enddata
    \tablecomments{M$_{\rm 0}$ $\equiv$ mean anomaly, and $\Omega$ $\equiv$ longitude of ascending node. \\
    (a) For 3-planet model only.}
\end{deluxetable*}

We configured \exostriker to use the {\tt Simplex} and {\tt Dynamical} algorithms during both the model fitting and MCMC runs; the justification for using {\tt Dynamical} instead of {\tt Keplerian} is that the {\tt Keplerian} algorithm does not work as well in a compact system with orbital resonances \citep{fabrycky2010} like \aumic. The dynamical model time steps has been set to the lowest possible 0.01 day given the short orbital periods of both planets. To find the best-fit TTV model, we employ \exostriker's built-in {\tt scipy} minimizer algorithms; we use truncated Newton algorithm (TNC)\footnote{\url{https://docs.scipy.org/doc/scipy/reference/optimize.minimize-tnc.html}} as a primary minimizer and Nelder-Mead algorithm\footnote{\url{https://docs.scipy.org/doc/scipy/reference/optimize.minimize-neldermead.html}} as a secondary minimizer, with both set at one consecutive integration and 5\,000 integration steps; the rest of the configurations for those minimizers were left to default settings. After we find a best-fit model, we compute an MCMC with 50\,000 burn-ins, 200\,000 integration steps, and 4 walkers; the other settings for MCMC were left to defaults, including adopting 68.300\% confidence intervals for estimating one-$\sigma$ posterior uncertainties. These best-fit and MCMC configurations apply to both the 2-planet and 3-planet dynamical models presented herein.

We first explore the 2-planet scenario (\aumic b and c). Then, we explore a representative 3-planet scenario. The following sub-subsections detail the process in exploring these cases.

\subsubsection{2-Planet Dynamical Modeling}\label{sec:2planetmod}

We explored a best-fit scenario for a 2-planet model with \exostriker. The eccentricity posteriors from the \exofast analysis (Table \ref{table:exfout}) provided us $\sim$4.056$\sigma$ upper limits on both planets' eccentricity. Analysis of the transit light curves themselves exclude such high eccentricities as in \citet{gilbert2021} and \citet{plavchan2020}, but we are only modeling the transit midpoint times herein.  We also explored a best-fit scenario for a ``mass-less'' no-TTV 2-planet model with \exostriker, as a control on testing for the presence and statistical significance of TTVs.

\subsubsection{3-Planet Dynamical Modeling}\label{sec:3planetmod}

\citet{cale2021} explores the analysis of RVs of \aumic, and searches for additional candidate non-transiting planet signals.  In particular, \citet{cale2021} identifies a candidate RV signal in-between the orbits of \aumic b and c with an orbital period of 12.742 days, which to date is unconfirmed, which we call a hypothetical ``d'' planet. This ``middle-d'' non-transiting planet scenario, if confirmed, would establish the \aumic planetary system in a 4:6:9 orbital resonant chain.  This would not be the first known non-transiting planet to exist between two transiting planets; \citet{christiansen2017}, \citet{buchhave2016}, \citet{sun2019}, and \citet{osborn2021} identified similar exoplanet configurations for the HD 3167, Kepler-20, Kepler-411, and TOI-431 planetary systems respectively. Similarly to the 2-planet model, we imposed 4$\sigma$ upper limits on planets b's and c's eccentricity for this paper.

\subsection{TTV Results}

Here we present the results of both the no-TTV, 2-planet and 3-planet TTV modeling using the \exostriker package. We also calculate the mass of \aumic c using the generated posteriors from \exostriker.

\subsubsection{Results on \exostriker 2-Planet Modeling}

\begin{figure*}
    \centering
    \begin{tabular}{cc}
    \includegraphics[width=86mm]{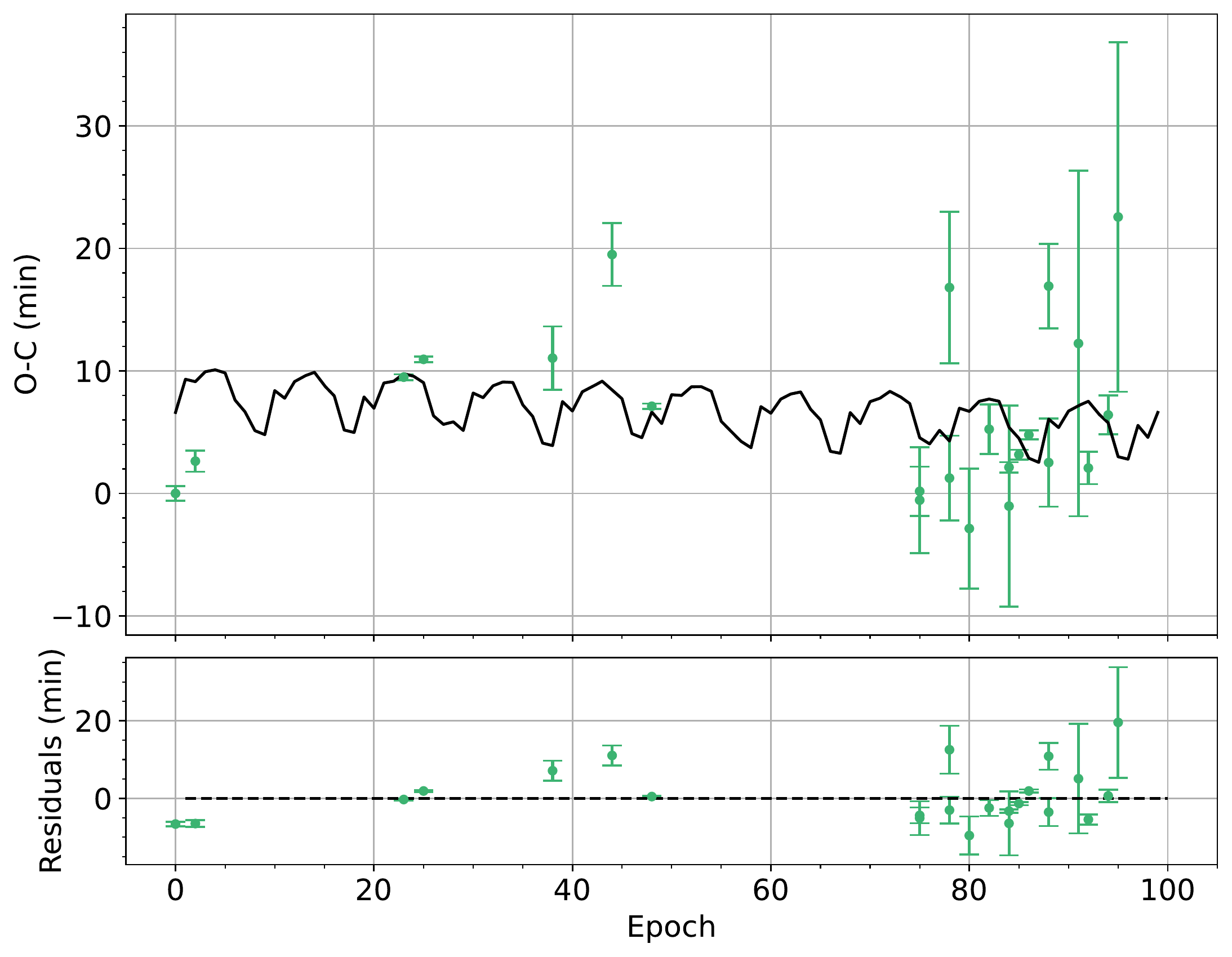} &
    \includegraphics[width=86mm]{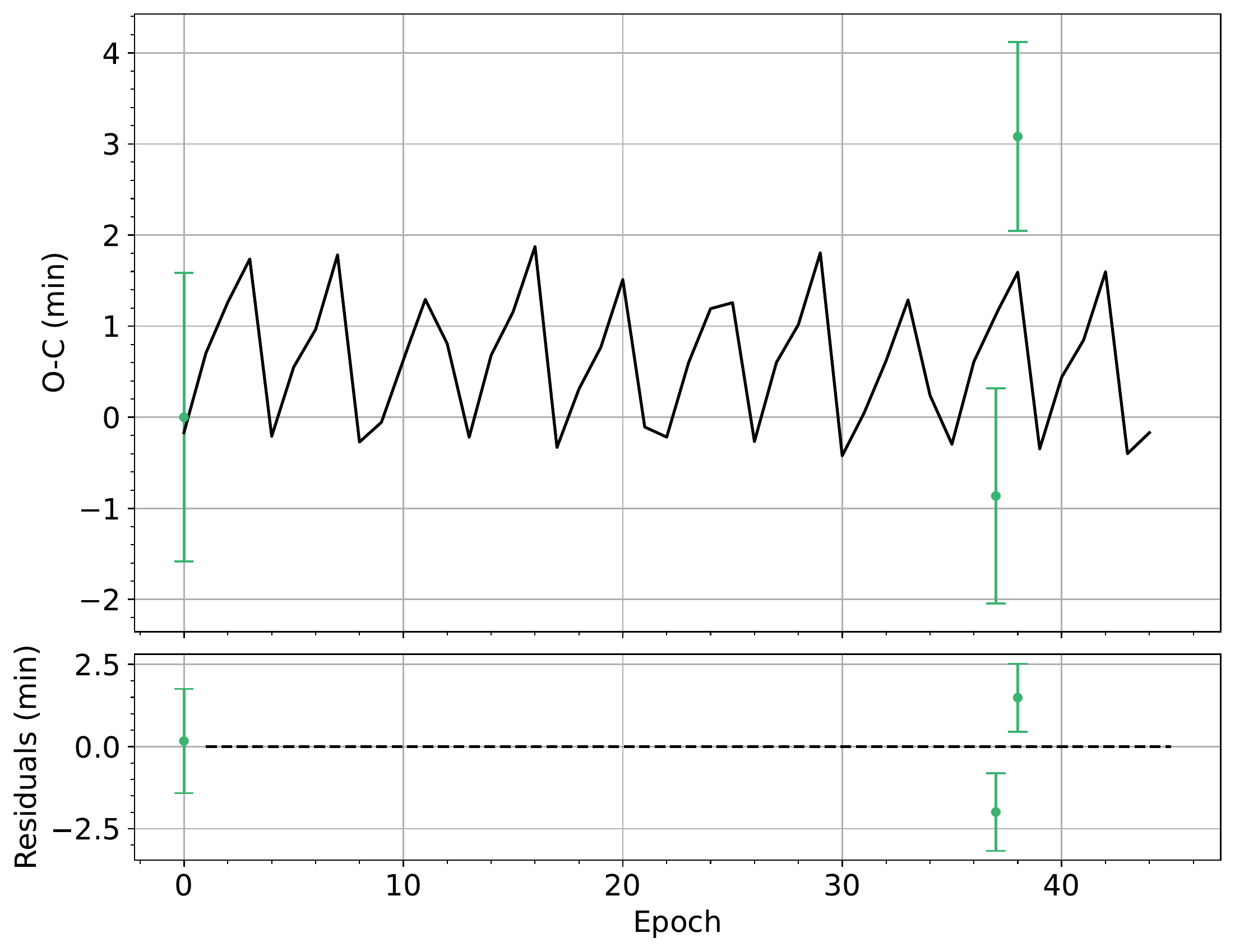}
    \end{tabular}
    \caption{2-planet O--C diagram of \aumic b (left) and \aumic c (right), with comparison between TTVs (green) and \exostriker-generated best-fit models (black).}
    \label{fig:2p_all_model}
\end{figure*}

\begin{deluxetable*}{l|c|cc|cc}
    \tablecaption{\label{table:2p_all_parameters}\exostriker-generated parameters for \aumic 2-planet best-fit and MCMC modeling.}
    \tablehead{\multirow{2}{*}{Parameter} & \multirow{2}{*}{Unit} & \multicolumn{2}{c|}{Best-fit} & \multicolumn{2}{c}{MCMC} \\
        & & \aumic b & \aumic c & \aumic b & \aumic c}
    \startdata
K                 & m/s & 4.03906  & 9.49903             & 4.10830 $\pm$ 1.47951     & 9.31429 $\pm$ 0.27661     \\
\porb             & day & 8.46255  & 18.86109            & 8.46257 $\pm$ 0.00003     & 18.86112 $\pm$ 0.00076    \\
e                 & ... & 0.00000  & 0.08329             & 0.00089 $\pm$ 0.00065     & 0.08220 $\pm$ 0.00473     \\
$\omega$          & deg & 89.79621 & 216.71379           & 61.89388 $\pm$ 36.98216   & 215.56678 $\pm$ 1.90214   \\
M$_{0}$           & deg & 0.00000  & 0.00032             & 27.96061 $\pm$ 20.80132   & 1.12536 $\pm$ 0.80464     \\
i                 & deg & 89.50262 & 88.99146            & 89.51412 $\pm$ 0.31257    & 88.95986 $\pm$ 0.51613    \\
$\Omega$          & deg & 0.00000  & 0.00006             & 185.70005 $\pm$ 116.59810 & 184.90569 $\pm$ 116.58777 \\
\hline
$-\ln\mathcal{L}$ & ... & \multicolumn{2}{c|}{-61.93923} & \multicolumn{2}{c}{-74.96317}                         \\
$\chi^{2}$        & ... & \multicolumn{2}{c|}{428}       & \multicolumn{2}{c}{455}                               \\
\reduced          & ... & \multicolumn{2}{c|}{36}        & \multicolumn{2}{c}{38}                                \\
$\Delta AICc$     & ... & \multicolumn{2}{c|}{-57.69664} & \multicolumn{2}{c}{-83.74452}
    \enddata
\end{deluxetable*}

\begin{figure*}
    \centering
    \begin{tabular}{cc}
    \includegraphics[width=86mm]{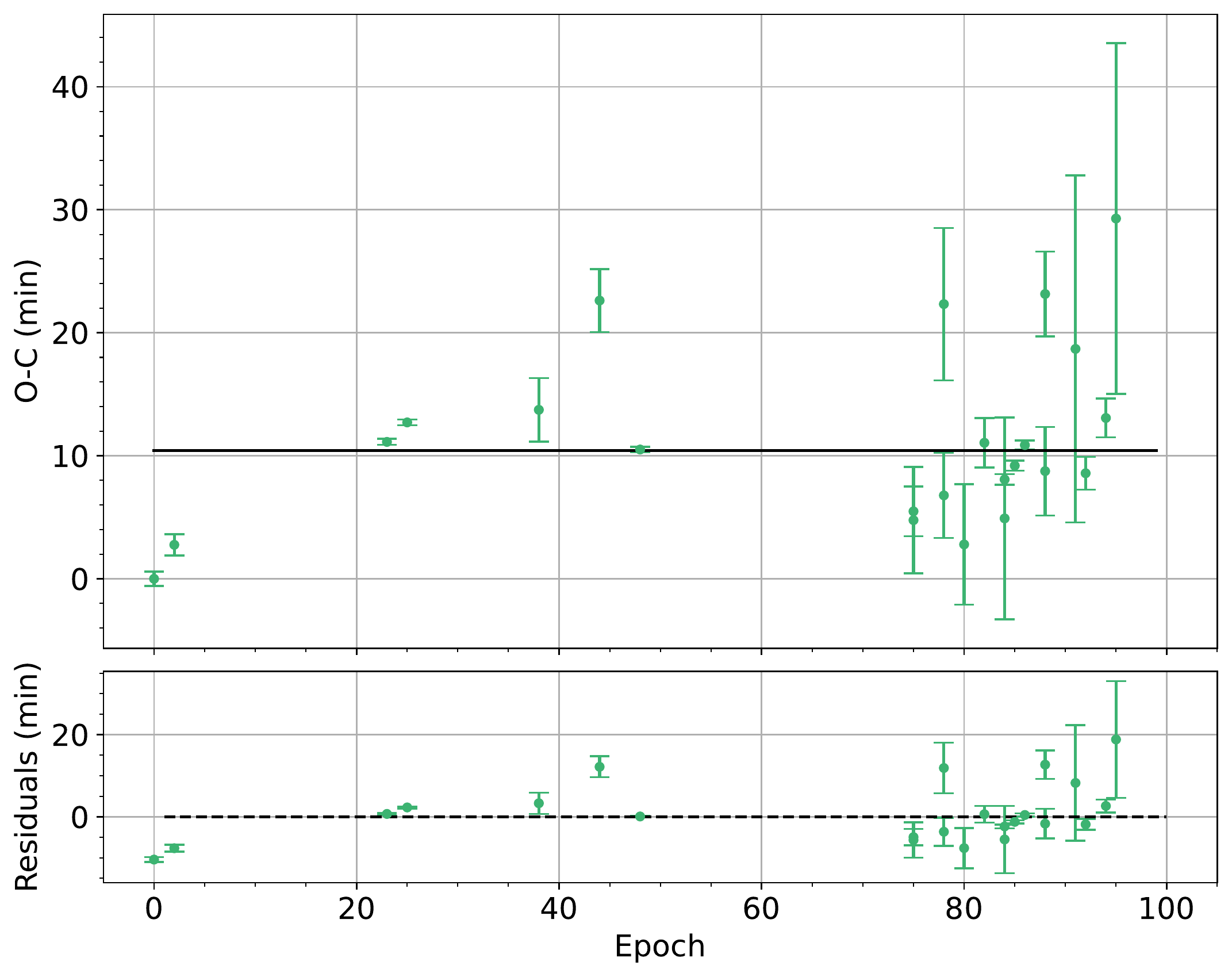} &
    \includegraphics[width=86mm]{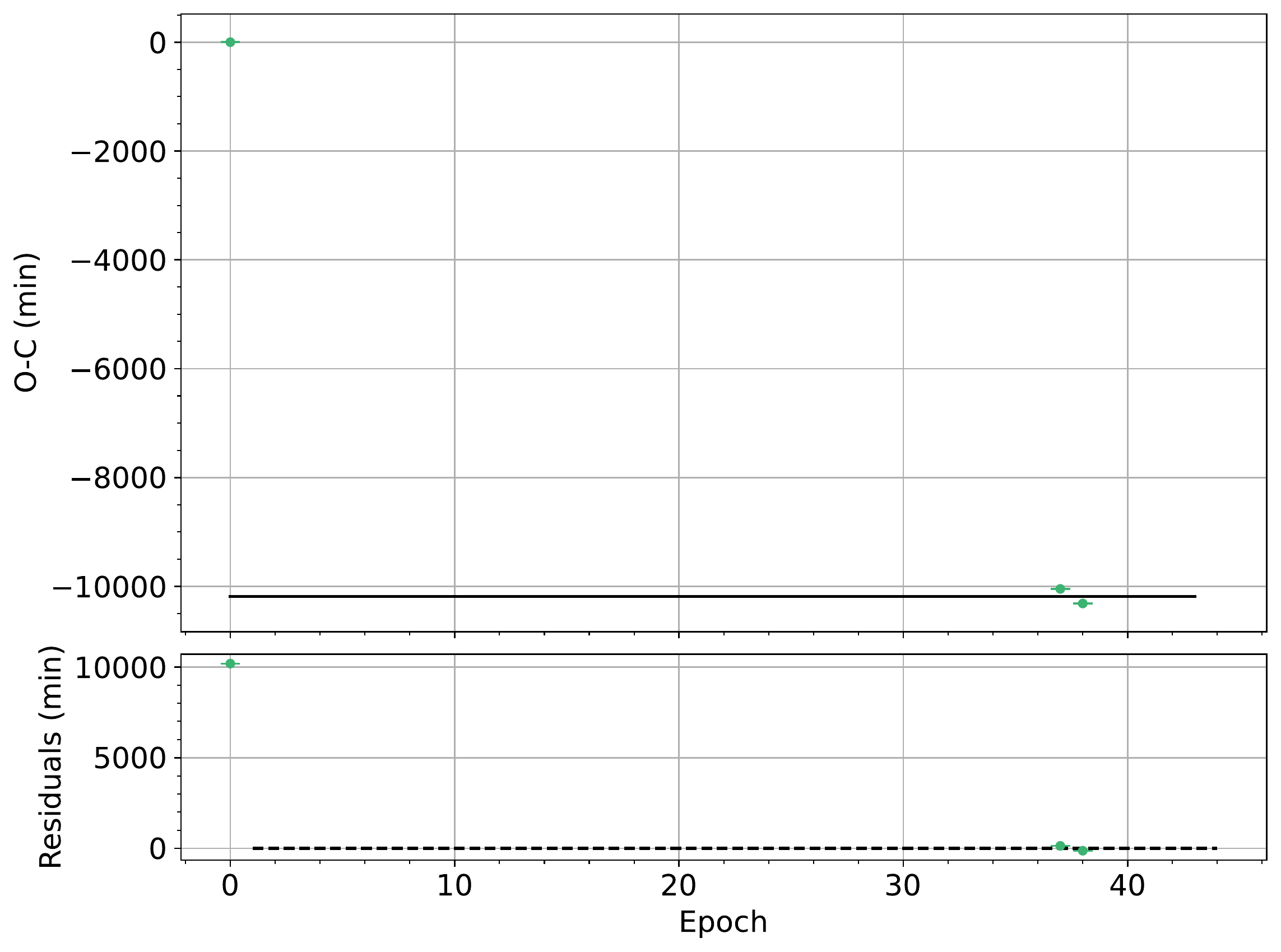}
    \end{tabular}
    \caption{Mass-less planets O--C diagram of \aumic b (left) and \aumic c (right), with comparison between TTVs (green) and \exostriker-generated best-fit models (black).}
    \label{fig:0p_all_model}
\end{figure*}

The best-fit O--C diagram (Figure \ref{fig:2p_all_model}), posteriors (Table \ref{table:2p_all_parameters}), and MCMC corner plot (Figure \ref{fig:2p_all_corner}) showcase the best possible 2-planet model. The \exostriker's Angular Momentum Deficit \citep[AMD,][]{laskar1997, laskar2000, laskar2017} criteria indicated that the 2-planet model is stable. However, it is clear from Figure \ref{fig:2p_all_model} that the model does not converge very well with the data points, especially with \tess and \spitzer. This implies that we potentially need a third planet to account for the observed TTV variability. This interpretation is one of several hypotheses that are being explored and which are discussed in greater details in $\S$\ref{sec:result}. For our control no-TTV scenario of ``mass-less'' planets b and c, we present the best fit O-C diagram (Figure \ref{fig:0p_all_model}), posteriors (Table \ref{table:0p_all_parameters}), and corner plot (Figure \ref{fig:0p_all_corner}).

\begin{deluxetable*}{l|c|cc|cc}
    \tablecaption{\label{table:0p_all_parameters}\exostriker-generated parameters for \aumic mass-less planets best-fit and MCMC modeling.}
    \tablehead{\multirow{2}{*}{Parameter} & \multirow{2}{*}{Unit} & \multicolumn{2}{c|}{Best-fit} & \multicolumn{2}{c}{MCMC} \\
        & & \aumic b & \aumic c & \aumic b & \aumic c}
    \startdata
K                 & m/s & 0.00000   & 0.00000                  & 0.00000 $\pm$ 0.00000     & 0.00000 $\pm$ 0.00000      \\
\porb             & day & 8.46293   & 19.04753                 & 8.46293 $\pm$ 0.00000     & 18.98613 $\pm$ 0.00002     \\
e                 & ... & 0.00004   & 0.00001                  & 0.47442 $\pm$ 0.09381     & 0.37308 $\pm$ 0.00000      \\
$\omega$          & deg & 223.78788 & 0.00007                  & 230.38681 $\pm$ 18.24298  & 0.00000 $\pm$ 0.00000      \\
M$_{0}$           & deg & 225.89190 & 0.00010                  & 193.13783 $\pm$ 7.76425   & 0.00000 $\pm$ 0.00000      \\
i                 & deg & 90.73560  & 88.59738                 & 89.47176 $\pm$ 0.29016    & 69.69803 $\pm$ 0.00001     \\
$\Omega$          & deg & 0.00002   & 0.00000                  & 189.38073 $\pm$ 126.18823 & 180.59530 $\pm$  125.12807 \\
\hline
$-\ln\mathcal{L}$ & ... & \multicolumn{2}{c|}{-20683716.47978} & \multicolumn{2}{c}{-6083619013.93697}                  \\
$\chi^{2}$        & ... & \multicolumn{2}{c|}{41367738}        & \multicolumn{2}{c}{12167264028}                        \\
\reduced          & ... & \multicolumn{2}{c|}{2954838}         & \multicolumn{2}{c}{869090288}                          \\
$\Delta AICc$     & ... & \multicolumn{2}{c|}{-41367384.95956} & \multicolumn{2}{c}{-12167237979.87394}
    \enddata
\end{deluxetable*}

\subsubsection{Results on \exostriker 3-Planet Modeling}

\begin{figure*}
    \centering
    \begin{tabular}{cc}
    \includegraphics[width=86mm]{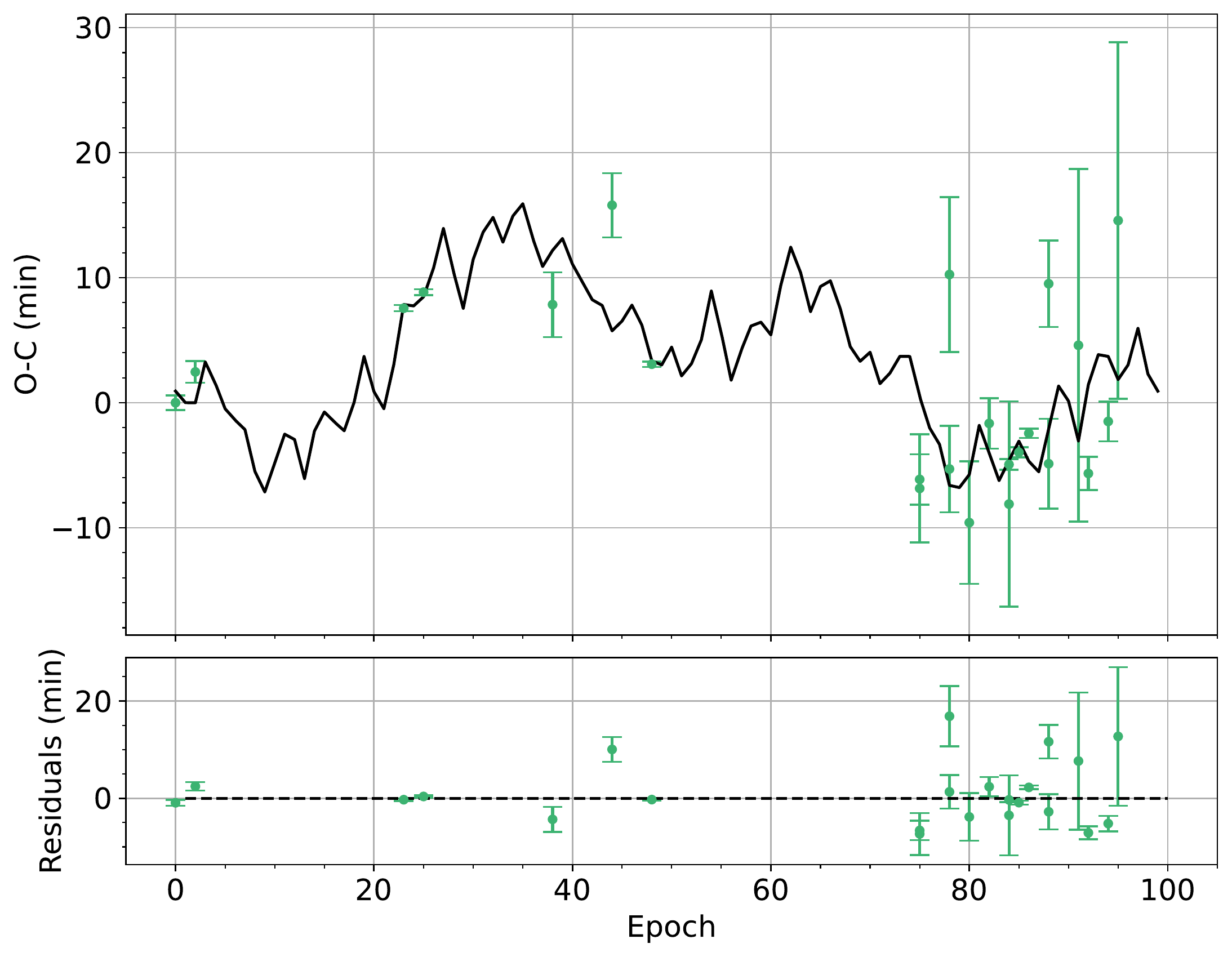} &
    \includegraphics[width=86mm]{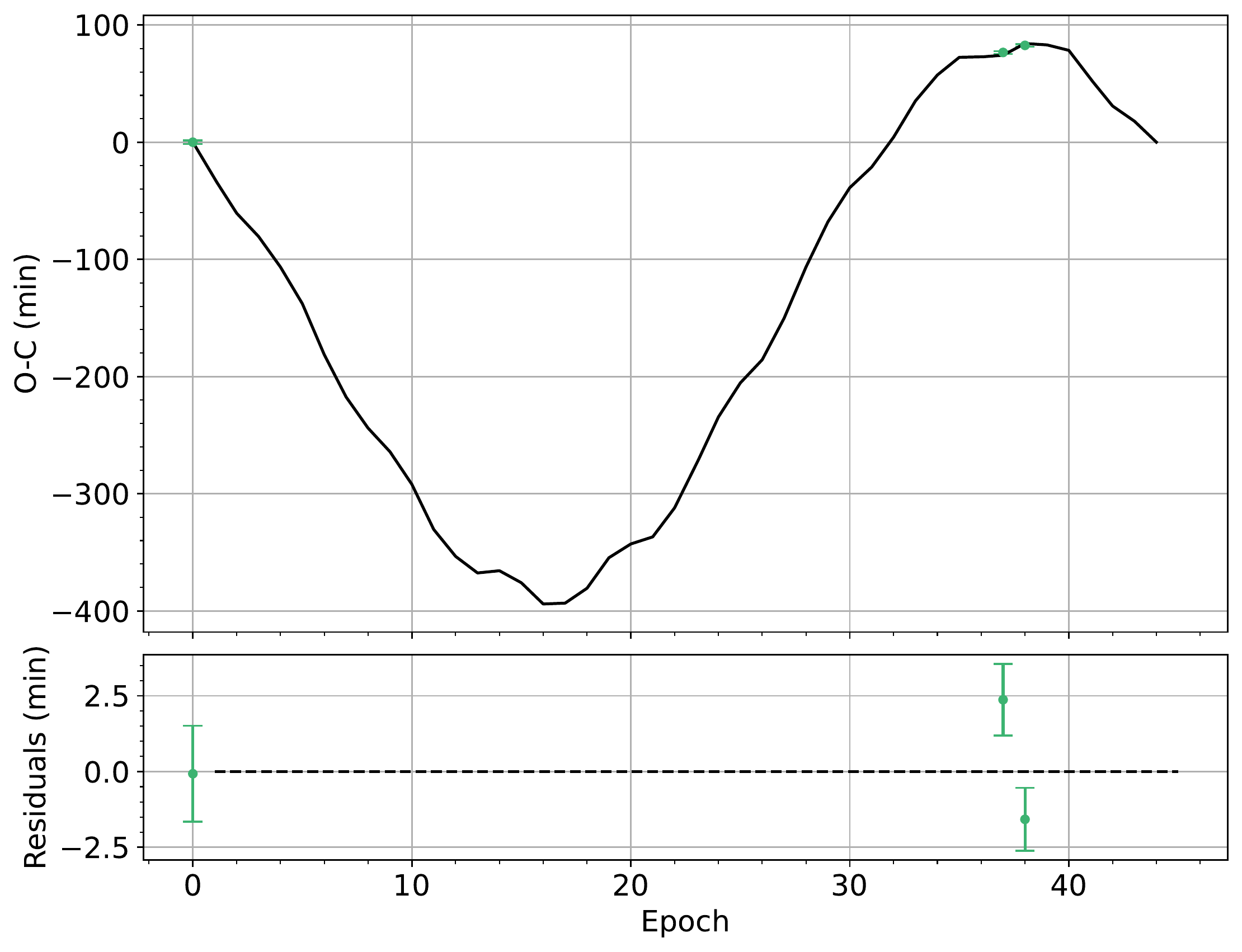}
    \end{tabular}
    \caption{3-planet O--C diagram of \aumic b (left) and \aumic c (right), with comparison between TTVs (green) and \exostriker-generated best-fit models (black).}
    \label{fig:3p_all_model}
\end{figure*}

Since the 2-planet circular orbit model does not adequately reproduce the observed TTVs as evidenced in Figure \ref{fig:2p_all_model}, we explored a {\it representative} and non-exhaustive hypothetical 3-planet scenario. We find a best-fit and MCMC model for this representative 3-planet scenario that adequately accounts for the \tess and \spitzer TTVs and is also consistent with the RV candidate signal in \citet{cale2021}, presented in Table \ref{table:3p_all_parameters} and Figures \ref{fig:3p_all_model} \& \ref{fig:3p_all_corner}. Moreover, the 3-planet case's log likelihood, $\chi^{2}$, and reduced-$\chi^{2}$ are better than those of the 2-planet case, the latter of which are better than the no-TTV ``mass-less'' scenario.  The delta log-likelihoods and corresponding $\Delta AICc$ indicate that the 3-planet scenario is strongly favored and the two planet scenario and the no-TTV scenario are statistically excluded. \exostriker indicates that this three-planet solution fails the AMD criterion. However, given the 4:6:9 orbital period commensurability for b, d, and c, respectively, we investigate the dynamical stability of this system configuration with two N-body codes, the latter of which is used for a consistency check. We generate simulations with \rebound \citep{rein2012, rein2015} and \mercury \citep{chambers1999} to test the stability of this representative 3-planet system; both indicate that this 3-planet configuration is stable (see $\S$\ref{sec:result} for more detailed discussions).

\begin{deluxetable*}{l|c|ccc|ccc}
    \tablecaption{\label{table:3p_all_parameters}\exostriker-generated parameters for \aumic 3-planet best-fit and MCMC modeling.}
    \tablehead{\multirow{2}{*}{Parameter} & \multirow{2}{*}{Unit} & \multicolumn{3}{c|}{Best-fit} & \multicolumn{3}{c}{MCMC} \\
        & & \aumic b & \aumic c & \aumic d & \aumic b & \aumic c & \aumic d}
    \startdata
K                 & m/s & 17.40198 & 7.65369   & 5.07363   & 1.91815 $\pm$ 1.27252    & 9.21910 $\pm$ 0.50712    & 1.06232 $\pm$ 0.30432    \\
\porb             & day & 8.46340  & 18.85872  & 13.48517  & 8.46329 $\pm$ 0.00027    & 18.86224 $\pm$ 0.00126   & 13.46591 $\pm$ 0.01022   \\
e                 & ... & 0.02348  & 0.00000   & 0.00000   & 0.07436 $\pm$ 0.00994    & 0.02508 $\pm$ 0.01531    & 0.00998 $\pm$ 0.00752    \\
$\omega$          & deg & 89.96574 & 223.91754 & 70.01180  & 85.48567 $\pm$ 6.11803   & 210.12425 $\pm$ 16.99204 & 43.40723 $\pm$ 31.78809  \\
M$_{\rm 0}$       & deg & 0.00000  & 0.00000   & 0.00000   & 5.18482 $\pm$ 3.84571    & 11.58305 $\pm$ 8.87570   & 42.36104 $\pm$ 29.98678  \\
i                 & deg & 89.47231 & 89.10159  & 115.68159 & 89.43632 $\pm$ 0.30223   & 88.99503 $\pm$ 0.54176   & 103.58128 $\pm$ 4.14661  \\
$\Omega$          & deg & 0.00000  & 0.17063   & 0.00000   & 123.30784 $\pm$ 87.12008 & 162.09943 $\pm$ 91.36624 & 139.42421 $\pm$ 90.81975 \\
\hline
$-\ln\mathcal{L}$ & ... & \multicolumn{3}{c|}{73.34798}    & \multicolumn{3}{c}{-848.73127}                                                 \\
$\chi^{2}$        & ... & \multicolumn{3}{c|}{158}         & \multicolumn{3}{c}{1991}                                                       \\
\reduced          & ... & \multicolumn{3}{c|}{32}          & \multicolumn{3}{c}{398}                                                        \\
$\Delta AICc$     & ... & \multicolumn{3}{c|}{419.69596}   & \multicolumn{3}{c}{-1424.46254}
    \enddata
\end{deluxetable*}

\begin{deluxetable}{l|ccc|c}
\tablecaption{\label{table:modcomp}Comparisons of best-fit model fitting parameters among the mass-less planets, 2-planet, and 3-planet cases.}
\tablehead{Best-fit Model & $\chi^{2}$ & \reduced & $-\ln\mathcal{L}$ & $\Delta AICc$}
    \startdata
No TTVs  & 41367738 & 2954838 & -20683716.480 & -41367384.960 \\
2-planet & 428      & 36      & -61.939       & -57.697       \\
3-planet & 158      & 32      & 73.348        & 419.696
    \enddata
\end{deluxetable}

\subsubsection{The Mass of AU Mic c}\label{sec:planetcmass}

We use equation (1) from \citet{cumming1999} and the best-fit parameters \& MCMC uncertainties from Tables \ref{table:2p_all_parameters} and \ref{table:3p_all_parameters} to calculate the mass of \aumic c. To simplify the equation, we made an assumption that \mstar $>>$ \mplan. For the best-fit 2-planet scenario, our calculation yields the mass of c to be 0.0781 $\pm$ 0.0039 \mjup (or 24.8 $\pm$ 1.2 \mear) at 20.7-$\sigma$ significance, which makes it roughly a Neptune-sized planet. In the case of our representative 3-planet scenario, our calculation yields the mass of c to be 0.0631 $\pm$ 0.0049 \mjup (or 20.1 $\pm$ 1.6 \mear) at 12.6-$\sigma$ significance, implying that it again has roughly the mass of Neptune.

\section{Photodynamical Analysis}\label{sec:photodyn}

\begin{deluxetable*}{l|c|cc|cc}
    \tablecaption{\label{table:photodynamical}Parameters for the photodynamical models, with model a the simultaneous fit of stellar activity variation \& flares and model b the pre-cleaned data.}
    \tablehead{\multirow{2}{*}{Parameter} & \multirow{2}{*}{Unit} & \multicolumn{2}{c|}{Model a} & \multicolumn{2}{c}{Model b} \\
        & & \aumic b & \aumic c & \aumic b & \aumic c }
    \startdata
\mplan         & \mear & 16$^{+12}_{-9}$              & 10.8$^{+2.3}_{-2.2}$          & 13$^{+8}_{-2}$               & 13$^{+5}_{-2}$             \\
\porb          & day   & 8.4626$^{+0.0001}_{-0.0001}$ & 18.8624$^{+0.0015}_{-0.0011}$ & 8.4626$^{+0.0001}_{-0.0002}$ & 18.860$^{+0.002}_{-0.002}$ \\
e              & ...   & 0.022$^{+0.005}_{-0.006}$    & 0.097$^{+0.010}_{-0.013}$     & 0.012$^{+0.003}_{-0.003}$    & 0.069$^{+0.025}_{-0.014}$  \\
$\omega$       & deg   & 84$^{+7}_{-9}$               & 99.1$^{+3.7}_{-3.5}$          & 103$^{+6}_{-7}$              & 112$^{+5}_{-9}$            \\
t$_{\rm peri}$ & day   & 4.87$^{+0.16}_{-0.21}$       & 4.89$^{+0.15}_{-0.16}$        & 5.3$^{+0.2}_{-0.2}$          & 5.5$^{+0.2}_{-0.4}$        \\
\rplan/\rstar  & ...   & 0.0529$^{+0.0002}_{-0.0002}$ & 0.032$^{+0.002}_{-0.001}$     & 0.0511$^{+0.0002}_{-0.0001}$ & 0.034$^{+0.006}_{-0.001}$  \\
i              & deg   & 89.20$^{+0.06}_{-0.06}$      & 90.8$^{+0.1}_{-0.1}$          & 89.11$^{+0.03}_{-0.03}$      & 89.4$^{+0.08}_{-0.07}$
    \enddata
    \tablecomments{Orbital elements are given for BJD 2458300.}
\end{deluxetable*}

In a second independent analysis to validate our methods, we used a photodynamical code to simultaneously fit the transit light curves and TTVs. The core of this code is based on \rebound \citep{rein2012, rein2015} with which the N-body problem is integrated using the high accuracy non-symplectic integrator with adaptive time-stepping {\tt IAS15}. At the times of the measured mid-transit times, the current orbital elements are used to calculate the corresponding transit light curve using the {\tt python} implementation of the model from \citet{mandel2002} from Ian Crossfield\footnote{\url{https://www.lpl.arizona.edu/~ianc/python/_modules/transit.html}}. Using the Markov chain Monte Carlo sample {\tt emcee} \citep{foreman-mackey2013}, we sample the parameter space maximizing the $\log$ likelihood. In addition to the transit light curves, the code also fits the radial velocity (RV) data of \aumic taken with \spirou \citep{cale2021, klein2021}, which at near-infrared wavelengths is least impacted by the stellar activity of this young system. The RV model is calculated from the same \rebound N-body integration of the planetary system. As free parameters we have the planetary masses, orbital periods, eccentricities, longitudes of periastron, inclination, and the \rplan/\rstar ratio. The stellar mass and radius, hence the stellar density, is fixed. The limb darkening for the various bands are taken from \citep{claret2012, claret2017} and are kept at the literature values. Since the \spitzer transits show depth variations, we account for this effect by a third light contribution as free parameter (to be explored further in future work). All data sets have their individual offsets, while the \spirou data are modeled by adding a ``jitter'' white noise term.

The stellar activity is modeled using \celerite \citep{foreman-mackey2017}, which has the advantage to offer a fast and scaleable implementation of Gaussian Process (GP) regression, especially important for large data sets. The \celerite package provides several built-in covariance kernels, one is representing damped oscillations driven by white noise called {\tt SHO}. It has three parameters: the undamped oscillator frequency or period $\omega_{0}$ = 2$\pi$/P, the quality factor $Q$ of the oscillator (which is reversely proportional to the damping time scale $\tau$), and the variance $S_0$. For more details, see equations (19)-(24) in \citet{foreman-mackey2017}. Star spots typically manifest in variations at the rotation period as well as on its first harmonic.  In order to represent this in a {\it rotational} kernel, we follow the idea presented in \celeritetwo \citep{foreman-mackey2017, foreman-mackey2018}\footnote{\url{https://celerite2.readthedocs.io/en/latest/api/python}} and use the sum of two {\tt SHO} kernels, where only one of the frequencies is a free hyper-parameter and the other is fixed to its first harmonic. Both oscillators are forced by a lower boundary of the quality factor to be in the underdamped regime (i.e., $Q>1/2$). For the RV data and the \tess light curves, the same rotation period and quality factors are enforced; however, the variance can differ.

We performed the photodynamical modeling in two versions. In the first (model a, using 45\,000 steps after 10\,000 burn-in steps and 400 walkers), we simultaneously fit the flares in the \tess light curves using the flare model {\tt aflare} \citep{davenport2014}. In the second version (model b, using 10\,000 steps after 10\,000 burn-in steps and 100 walkers, will be increased), we use the cleaned \tess light curve from \citet{gilbert2021}, where also the variation due to stellar activity has been corrected, i.e. no GP modeling is required here for the transit light curves. The parameters of the photodynamical fits are listed in Table \ref{table:photodynamical} and are displayed in Figures \ref{fig:photodyn_modela} \& \ref{fig:photodyn_modelb}.

\subsection{Photodynamical Modeling Results}

\begin{figure}
    \centering
    \includegraphics[width=0.5\textwidth]{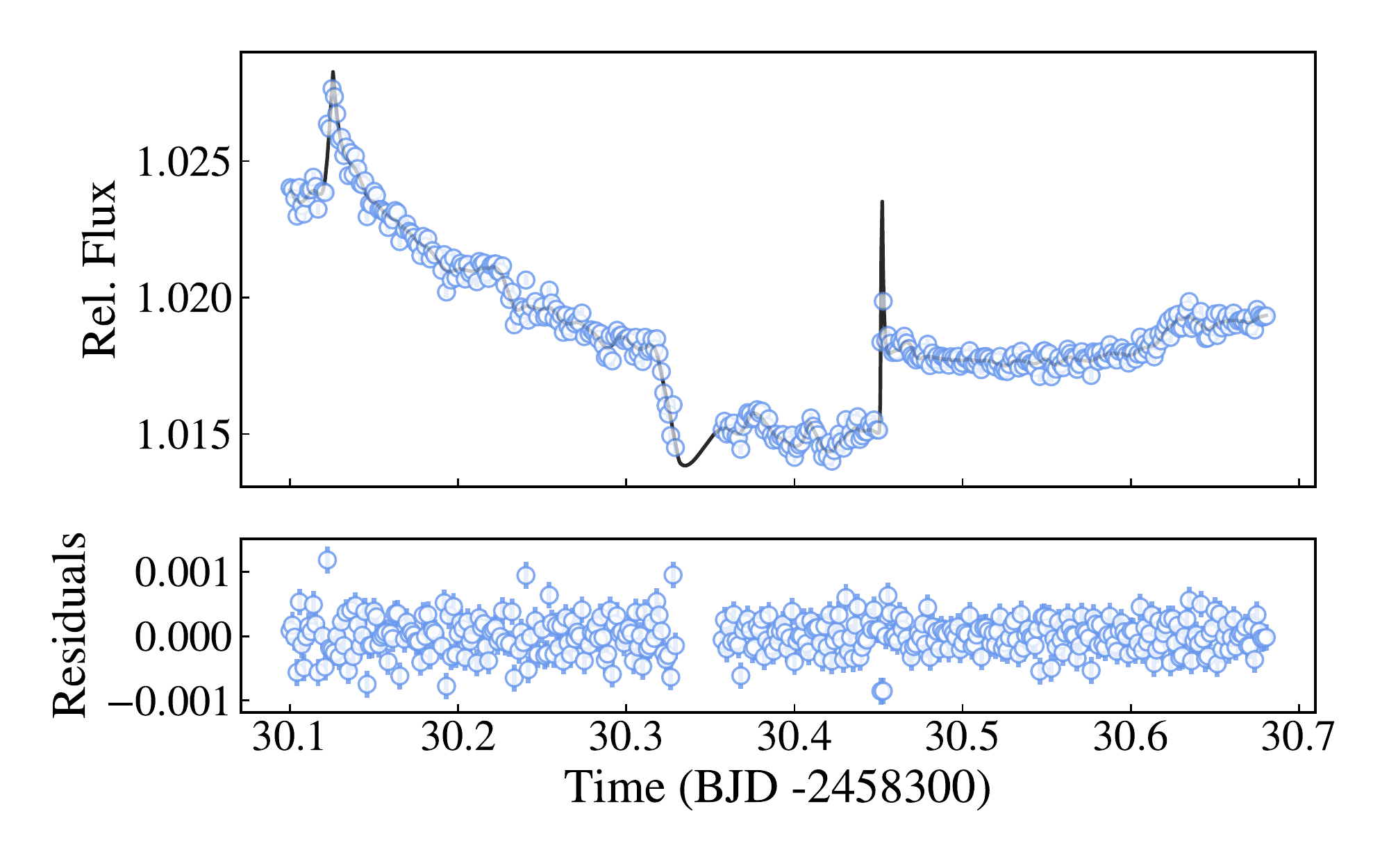}    \\
    \includegraphics[width=0.5\textwidth]{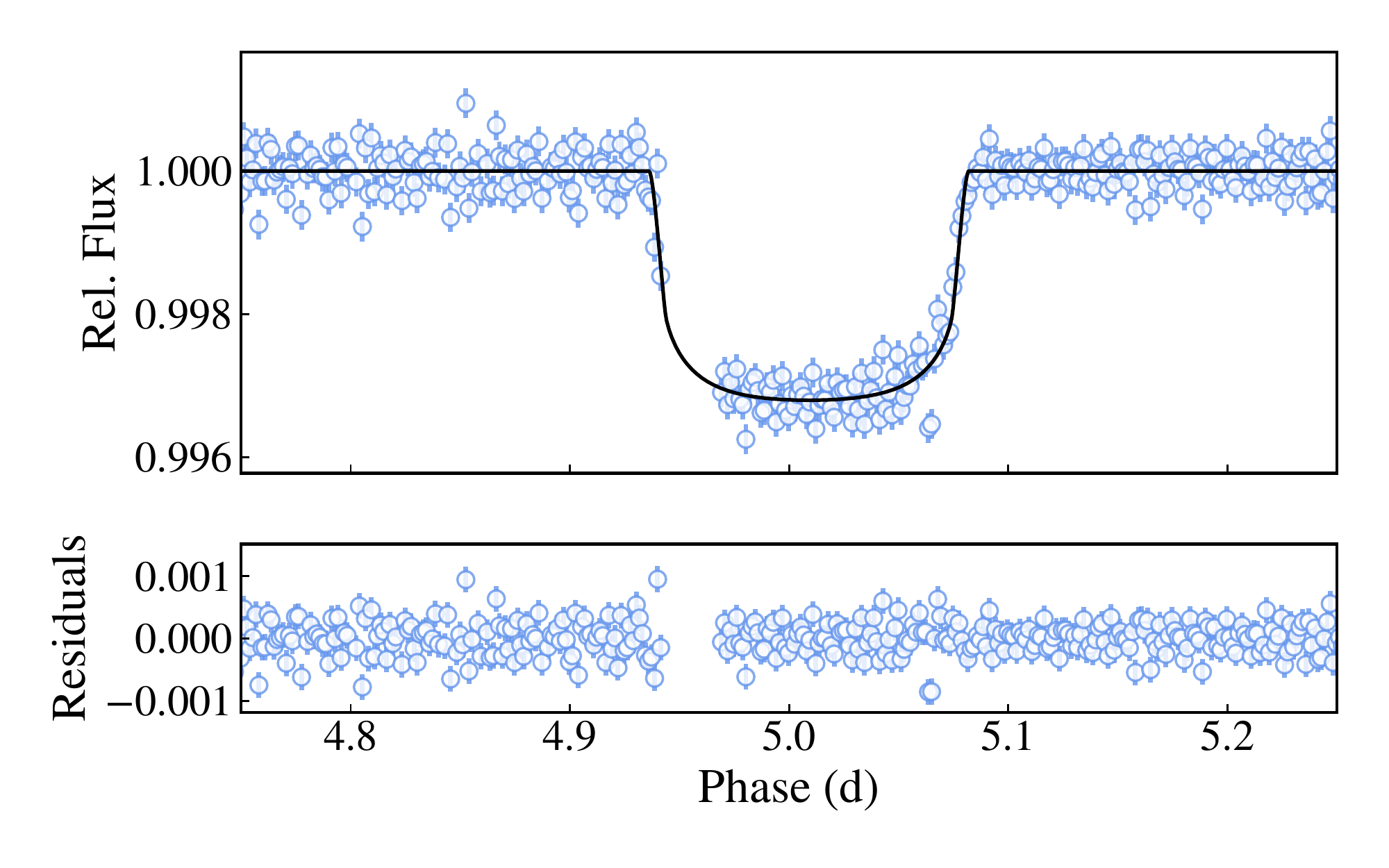} \\
    \includegraphics[width=0.5\textwidth]{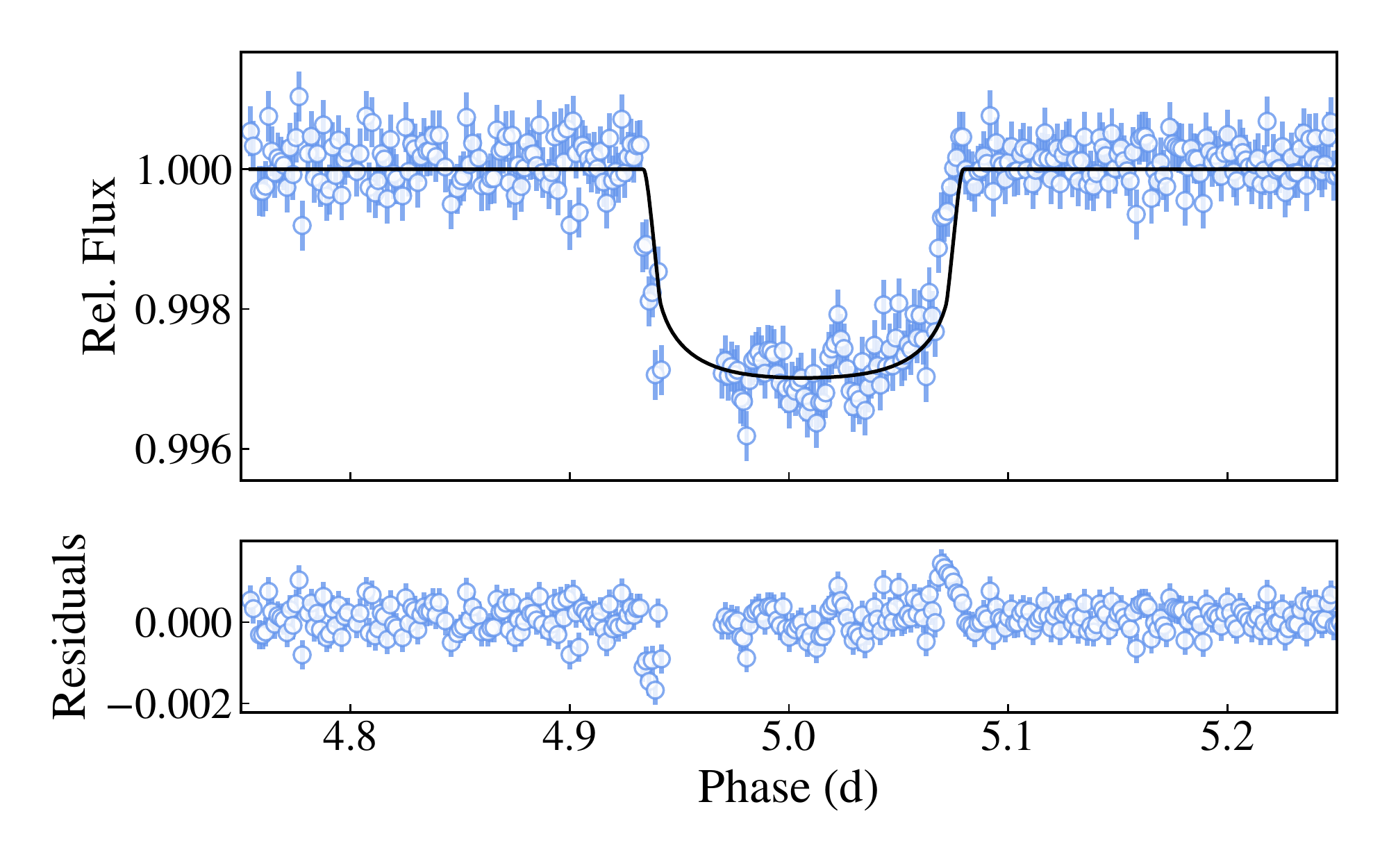}
    \caption{The first \tess transit of \aumic b in Sector 1. From top to bottom: \tess light curve with stellar activity variation and flares overlayed with the best-fit model a from the photodynamical analysis, best-fit model a with contribution of stellar activity variation and flares removed, best-fit model b.}
    \label{fig:tess_01_b}
\end{figure}

\begin{figure}
    \centering
    \includegraphics[width=0.5\textwidth]{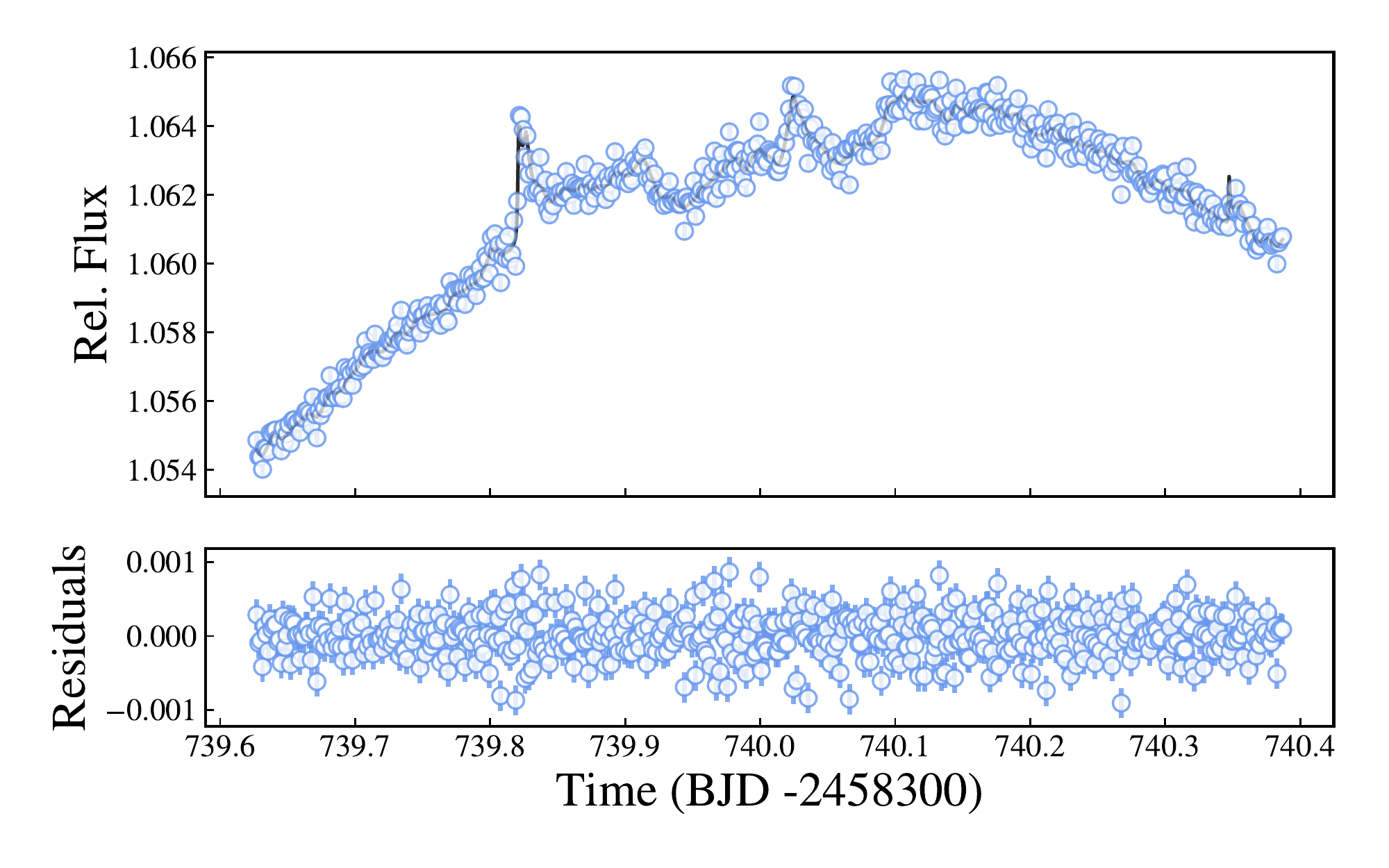}    \\
    \includegraphics[width=0.5\textwidth]{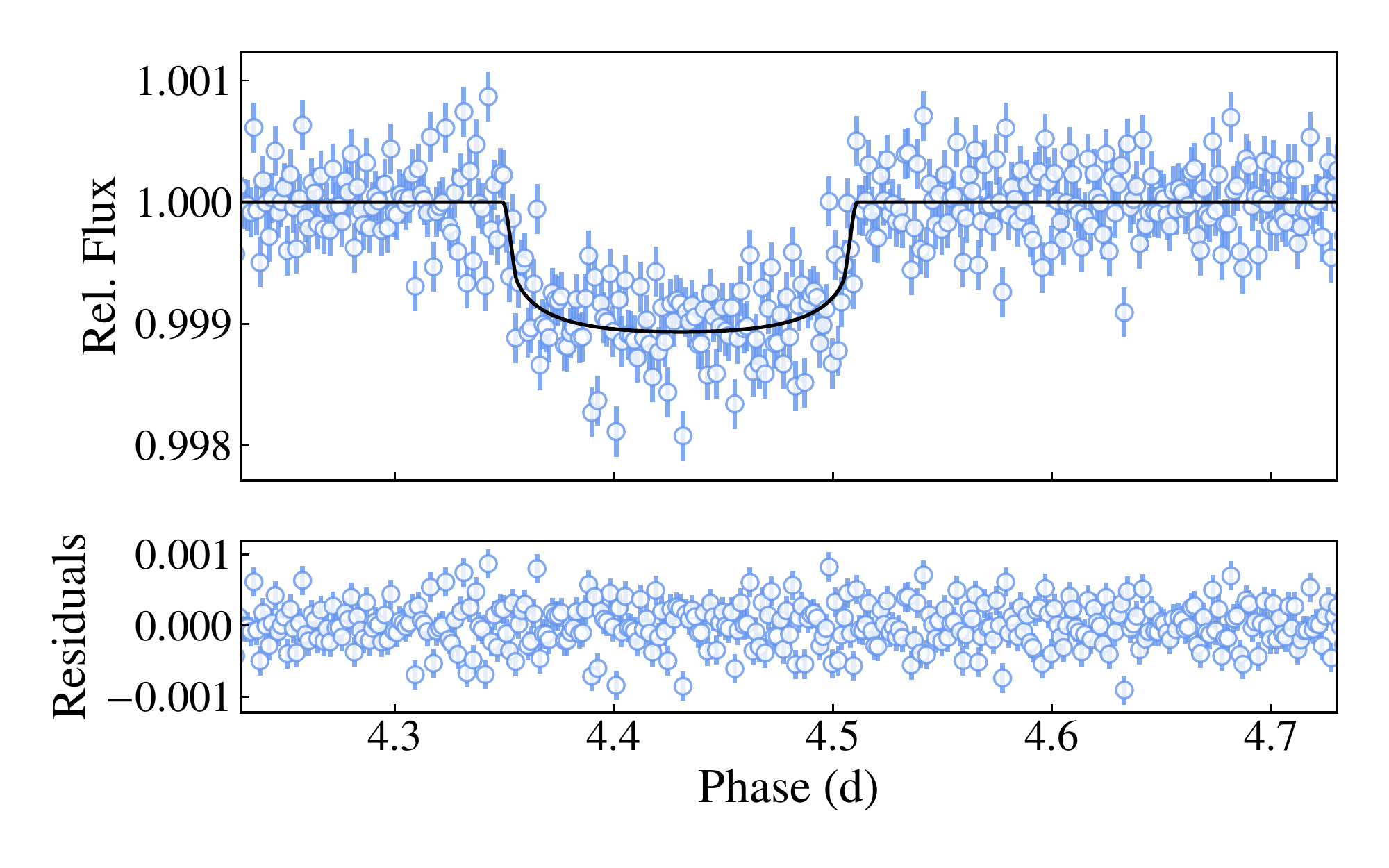} \\
    \includegraphics[width=0.5\textwidth]{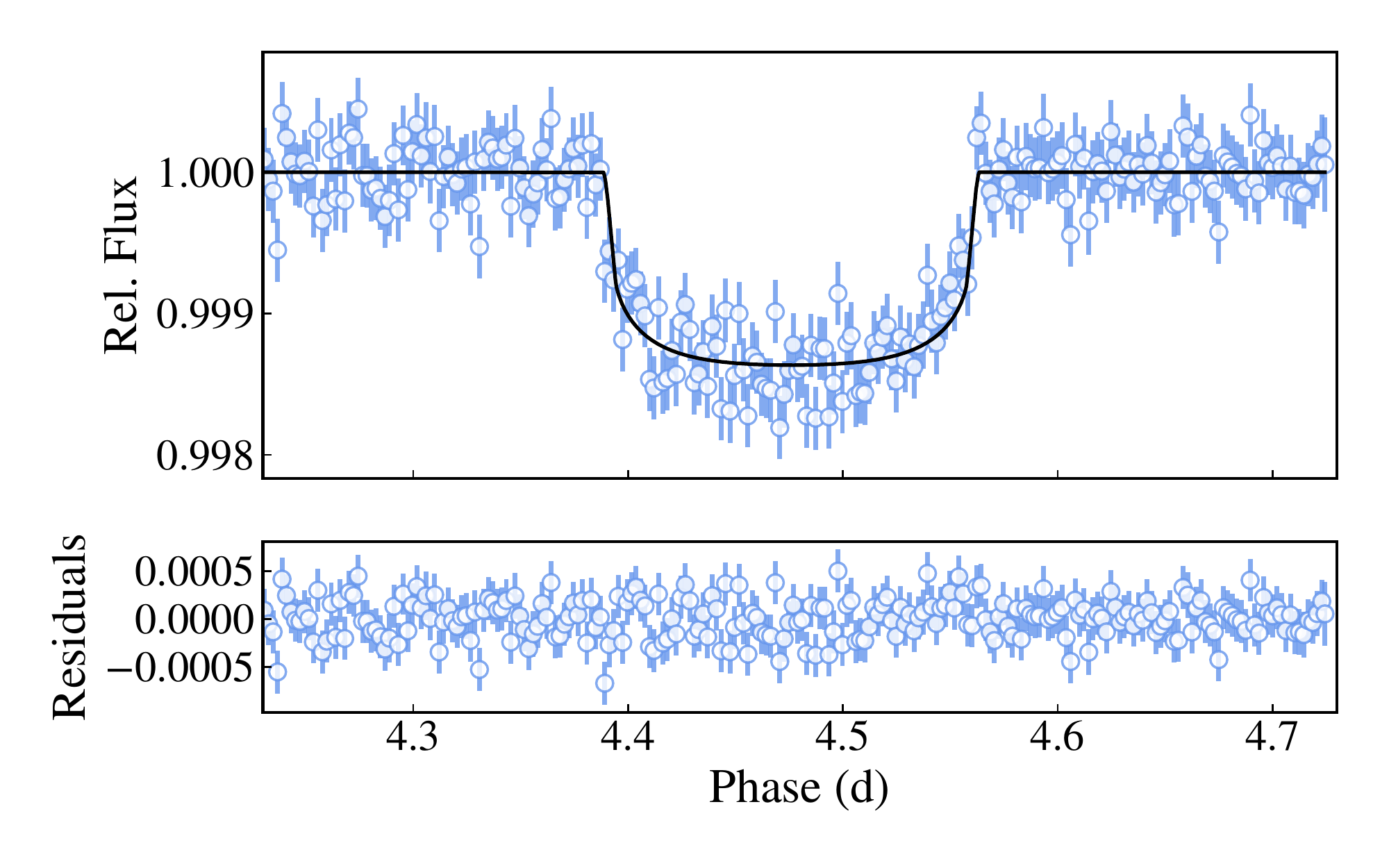}
    \caption{The first \tess transit of \aumic c in Sector 27. From top to bottom: \tess light curve with stellar activity variation and flares overlayed with the best-fit model a from the photodynamical analysis, best-fit model a with contribution of stellar activity variation and flares removed, best-fit model b, which shows by eye a clear timing offset when the activity is modeled separately from the photodynamical modeling.}
    \label{fig:TESS_27_c}
\end{figure}

\begin{figure}
    \centering
    \includegraphics[width=0.5\textwidth]{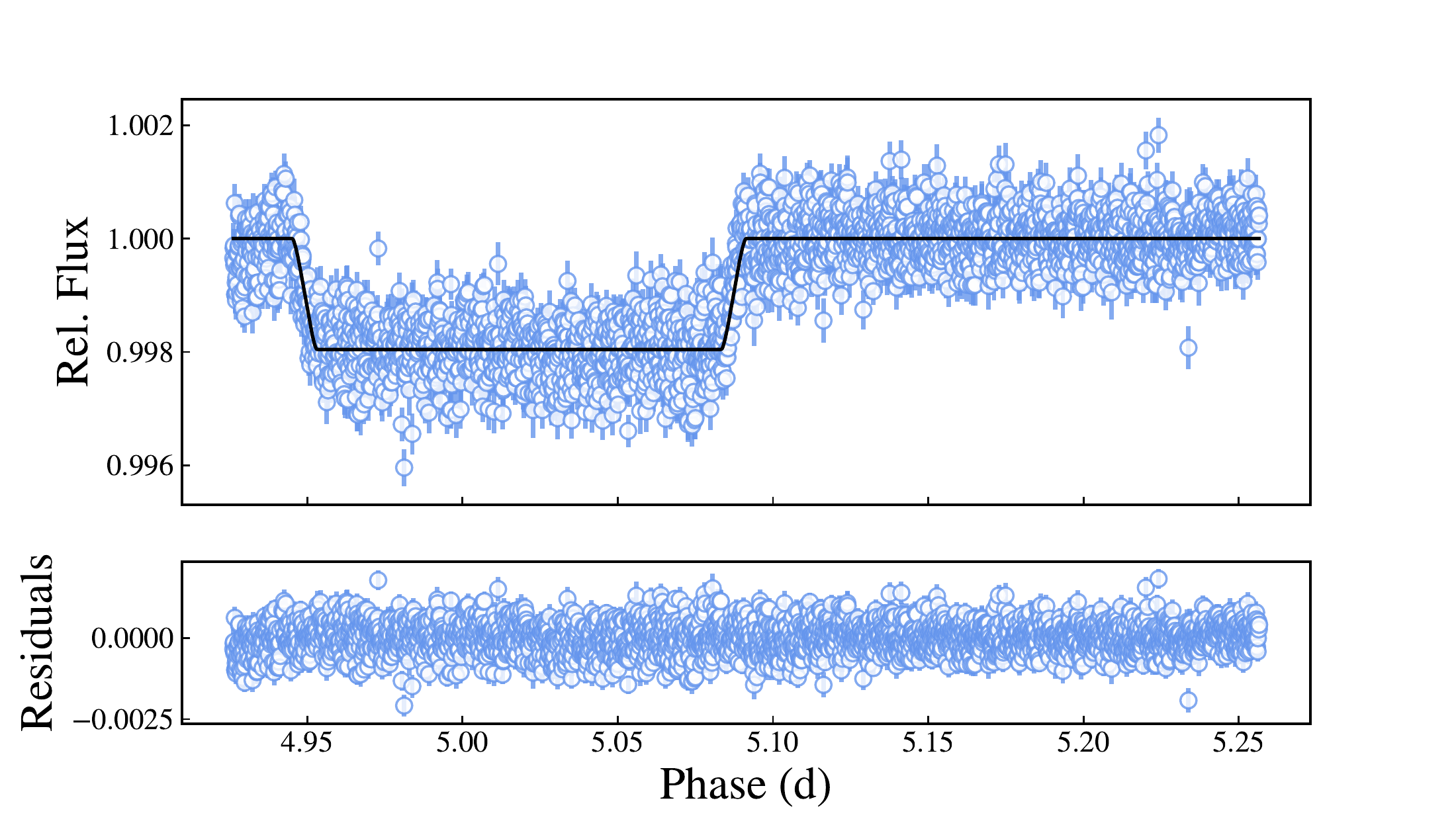} \\
    \includegraphics[width=0.5\textwidth]{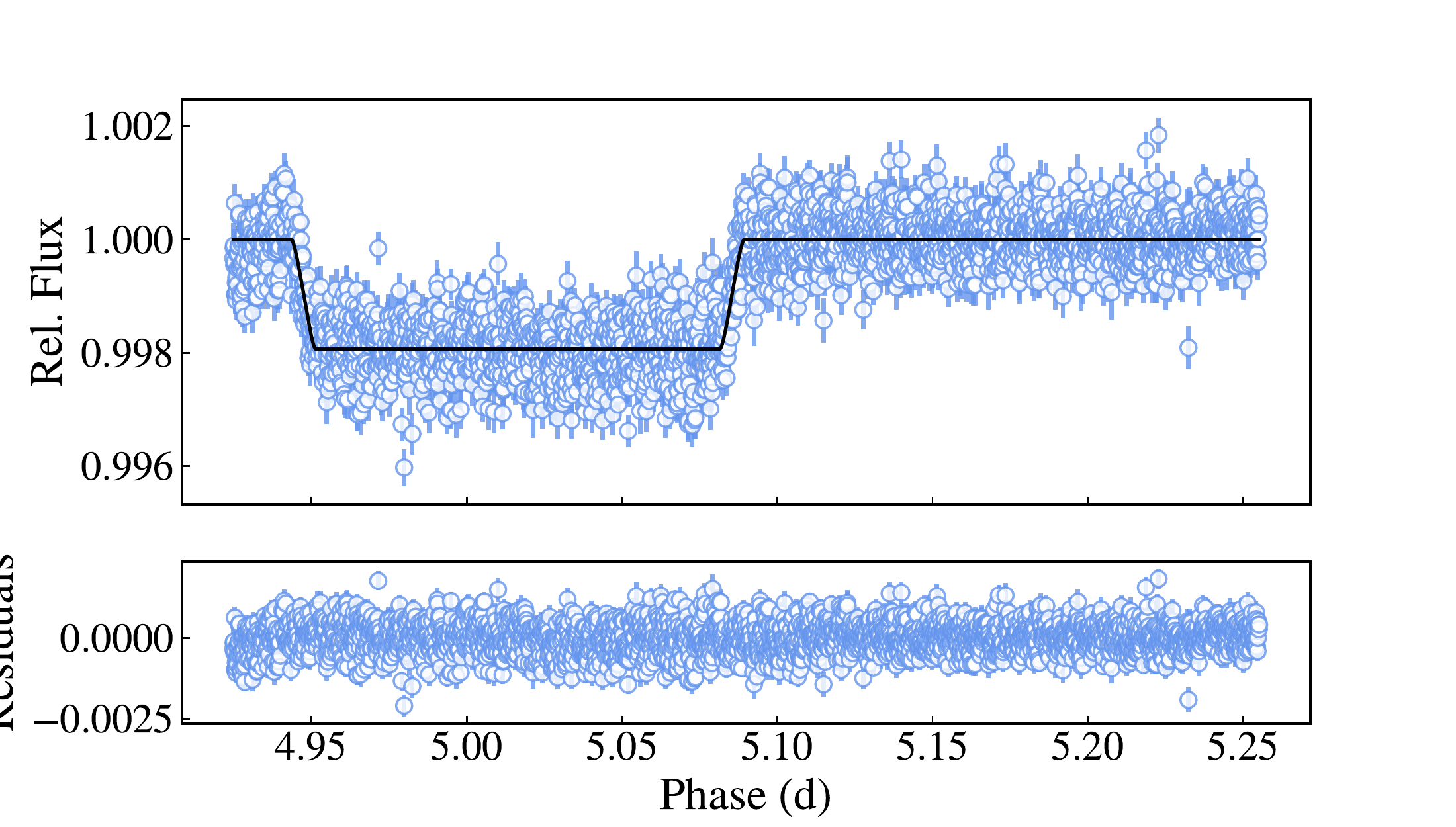}
    \caption{The first \spitzer transit of \aumic b, with the best-fit model a (top) and best-fit model b (bottom) from the photodynamical analysis.}
    \label{fig:SPITZER_01_b}
\end{figure}

In this subsection, we present a summary of the key results of our photodynamical modeling. Model a has the advantage to provide a simultaneous fit of the transits, stellar activity modulation, and flares. Since the timescales of ingress and egress are similar to those of the flares, a simultaneous fit should provide a more robust result. The flexibility of the GP kernels, however, may compensate for small deviations in time between observed and modelled transits. Comparing the calculated transit midpoints of a best fit to the {\it a priori} measured mid-transit times indeed show deviations on the order of 10 minutes. Either the transit midpoint measurements are affected by systematic errors due to the stellar activity and flaring, or the GP model is too flexible and hides these discrepancies. In model b, this is prevented due to a pre-cleaning of the light curve, however, with the caveat that the measured mid-transit times may be affected by activity modulation and flaring. A comparison between the two approaches shows that we can obtain a 2-planet model which fits all transits (\tess, \spitzer, and ground-based) if the stellar variability and flares are fitted simultaneously, while all models compared to the pre-cleaned data show noticeable discrepancies. As an example, we show the first transit of \aumic b in Sector 1 of \tess in Figure \ref{fig:tess_01_b}, on the top two panels from the simultaneous stellar variability and flare fit and on the bottom panel from the pre-cleaned data. We also show the best-fit for one of the \spitzer light curves in Figures \ref{fig:TESS_27_c} \& \ref{fig:SPITZER_01_b}.

While the best fit of the photodynamical model including stellar variability and flares is promising, we note the above mentioned compensation of deviations between models and observation due to the flexibility of the GP model. Further transits will be required to check this model. The timing discrepancies in the second version of the photodynamical fit corroborate the conclusion from our TTV fitting (i.e., a 2-planet model is not able to reproduce the small but detectable TTVs).

\section{Discussion}\label{sec:result}

In this section we present the key results of our analyses in $\S$\ref{sec:keyresults}, followed by the impact of the rotational modulation of stellar spots and plages in $\S$\ref{sec:spots} and flares in $\S$\ref{sec:flares}. In $\S$\ref{sec:dcandidate}, we assess the likelihood of an additional middle non-transiting planet candidate to explain in the observed TTVs, and in $\S$\ref{sec:implanetc} we compare our analyses' constraints on the mass of \aumic c. Lastly, in $\S$\ref{sec:limitation}, we discuss the limitation of our joint modeling.

\subsection{Key Results of Analyses}\label{sec:keyresults}

In the preceding sections, we modeled the transit observations of \aumic b and c to ascertain whether or not transit timing variations were present, and whether or not they can be accounted for by our existing knowledge of this planetary system and be used to constrain the dynamical orbits and masses of the two transiting planets. Given the inherent stellar activity in this young system, we explored two different and independent analyses: in $\S$\ref{sec:ttvs} a dynamical model of the transit midpoint times, and in $\S$\ref{sec:photodyn} a full photodynamical model. We found in both scenarios indications of transit timing variations deviating from a linear ephemeris on the order of ten minutes, particularly when comparing the \tess and \spitzer transit times, and generally consistent results, albeit with larger timing uncertainties, from the ground-based transits and R-M observations (Figure \ref{fig:ocdiagram}).

We next explored whether the two transiting planets could account for the observed TTVs. We first attempted a 2-planet dynamical model for the TTVs with only the high precision \tess transits, which we were able to find a fit akin to that derived in \citet{martioli2021}. These transit times were derived with the joint stellar activity model in \citet{gilbert2021} who find evidence for TTVs on the order of a few minutes. However, when including the \spitzer transits, which are less susceptible to stellar activity at 4.5 $\mu$m, we find that the \spitzer transits are incompatible with this model within the measured timing uncertainties, showing a significant departure from a circular two-planet model on the order of $\sim$10 minutes. We next considered whether orbital eccentricity could account for the observed TTVs, but derive high eccentricities that are incompatible with the light curves themselves which exclude high eccentricity scenarios \citep{plavchan2020, gilbert2021}.

With the full photodynamical analysis in $\S$\ref{sec:photodyn}, we were able to jointly model the activity and transits to reproduce the observed transit midpoint times. However, this left open the possibility that the flexibility of the GP for the rotational modulation of the starspots could also absorb any of the TTVs present. In fact, with a second photodynamical model using \citet{gilbert2021} activity model detrended transit light curves (e.g. the stellar activity in corrected serially rather than jointly with the photodynamical analysis), we found excess TTVs not explained by the 2-planet model alone. 

In order to explain the observed excess of TTVs than accounted for by our two-planet modeling, we next investigate in turn the following possibilities. First, that our modeling of the rotational modulation of star spots is underestimating the impact on the TTVs. Second, we consider the impact the flares have on our TTVs. Third, we consider the possibility of a third planet in the \aumic system not yet detected to transit nor through RVs. Finally, we consider the limitations of our analysis and statistical methods.

\subsection{Stellar Spots}\label{sec:spots}

Stellar spots can significantly impact the recovered stellar and planetary model parameters \citep{pont2007, czesla2009, berta2011, desert2011, ballerini2012} and can produce TTV-like signals that potentially lead to false-positive detection of non-transiting planets \citep{alonso2009, sanchis-ojeda2011-1, sanchis-ojeda2011-2, oshagh2012}. For \rplan/\rstar $\approx$ 0.05 (as is the case for \aumic b), the expected maximum amplitude of the spot-induced TTV in seconds at visible wavelengths is $AMP$ = 139 $\times$ $f$, where $f$ is the stellar spot filling factor in percent \citep{oshagh2013}. If $f =$ 0.25\%, 1\%, or 3\%, the maximum amplitude would be 34.75, 139, or 417 seconds (0.6, 2.3, or 7.0 minutes), respectively. Given the long-lived spot lifetimes for \aumic, it is not unreasonable to assume that the spot-filling fraction is $>$3\% and thus the impact of the rotational modulation of starspots could be on the order of $>$7 minutes. \citet{oshagh2013} added that if the transit duration and depth priors were fixed (and not modeled), the TTV amplitude would be smaller than if they were floating.

First, we consider the wavelength dependence of the TTVs. The \spitzer transit times have greater photometric precision due its larger aperture, a much higher cadence, and consequently more precise transit midpoint times (to within 13 seconds). Further, the \spitzer observations are less impacted by the rotational modulation of stellar activity at 4.5 $\mu$m due to the decreased flux contrast of any spots; the rotational modulation of stellar activity should be significantly decreased in amplitude. This decreased amplitude of the impacts of stellar activity with increasing wavelength is also observed for \aumic in photometry \citep{hebb2007} and in radial velocities \citep{cale2021}. However, the \spitzer transits show the most significant deviations from a linear ephemeris than those derived by the \tess data alone, running counter-intuitive to the expected impact the rotational modulation of starspots would have on the derived transit times.

Second, \citet{szabo2021} measured \aumic's rotation period with \tess data and \aumic b's orbital period with both \tess and \cheops light curves and found that the ratio between these two periods implies the 7:4 spin-orbit commensurability. Due to this resonance, for every fourth transit of \aumic b, the transit takes place over the same range of stellar longitudes of \aumic. In other words, the same stellar ``side'' is transited every fourth transit. Thus the impact of the stellar rotational modulation of spots should produce the same systematic impact on the TTVs every fourth transit. For example, if the 1$^{st}$ transit of \aumic b is 4 minutes ``late'' from the apparent effects of starspots, then this will also be true for the 5$^{th}$, 9$^{th}$, 13$^{th}$, 17$^{th}$, \ldots, N$^{th}$ transits. Given \aumic's long spot lifetime, relatively unchanged between \tess sectors 1 and 27 two years later, this pattern will have persisted over the duration of our observations.

\begin{figure*}
    \centering
    \begin{tabular}{ccc}
    \includegraphics[width=55mm]{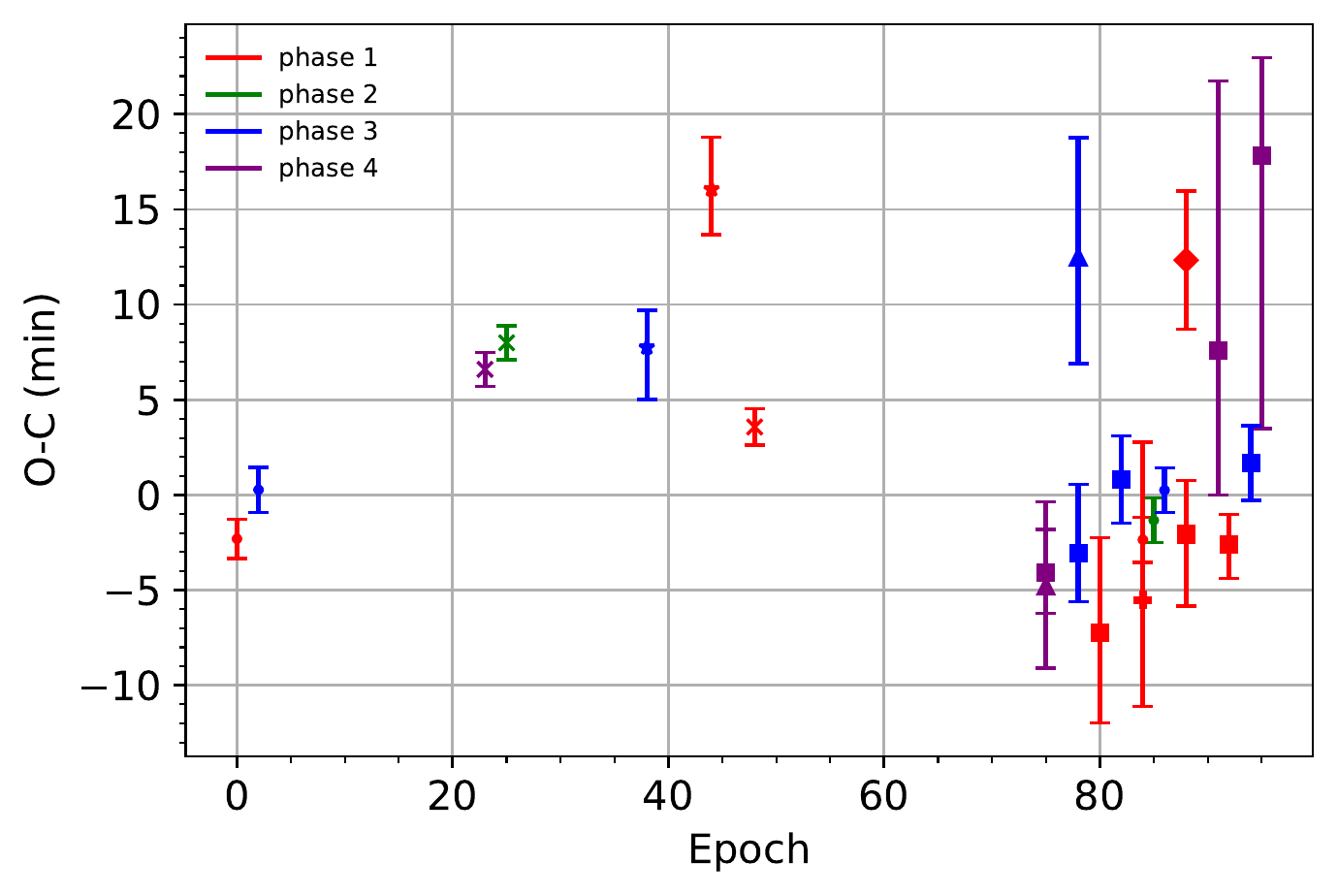} &
    \includegraphics[width=50mm]{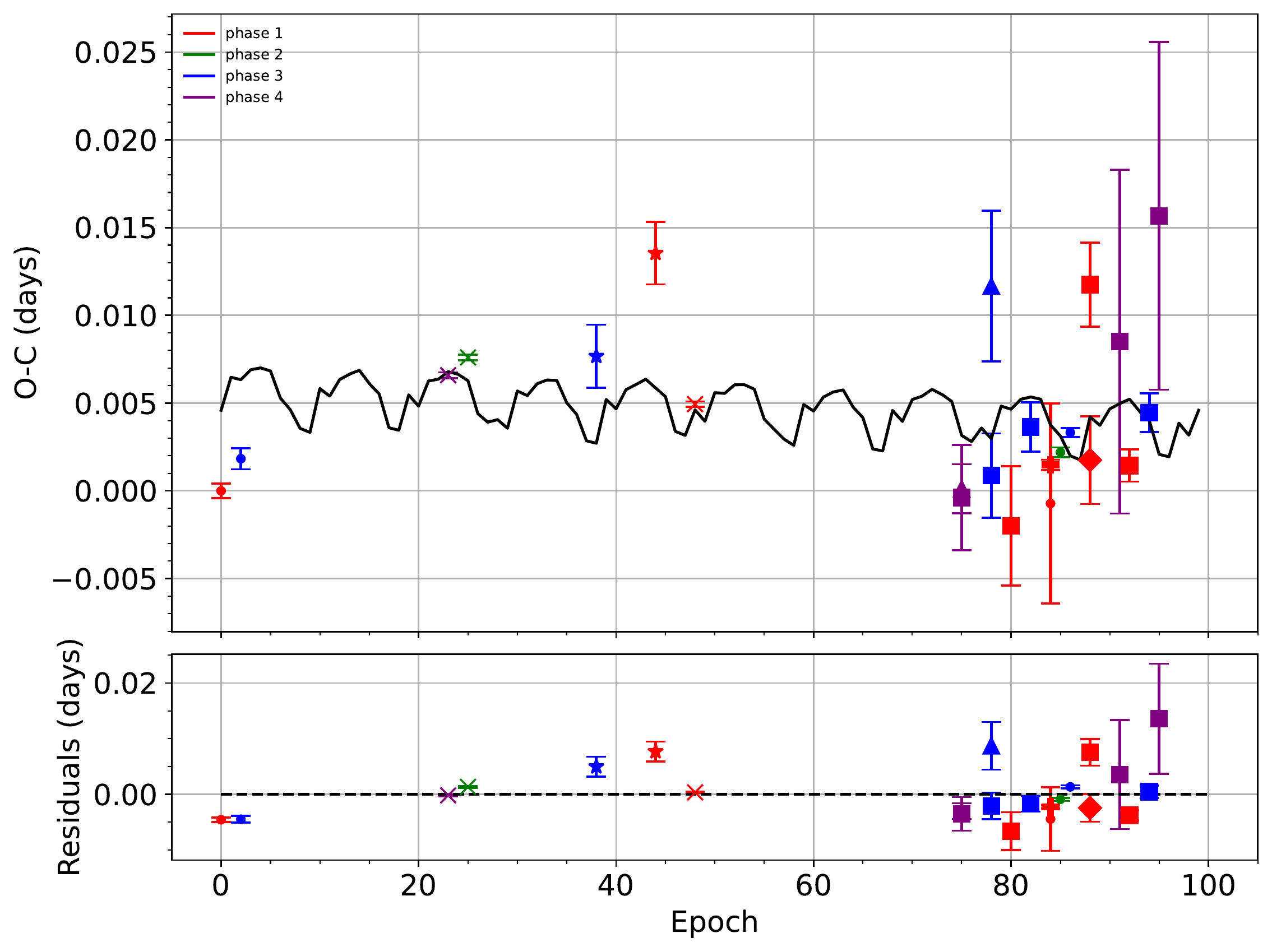} &
    \includegraphics[width=50mm]{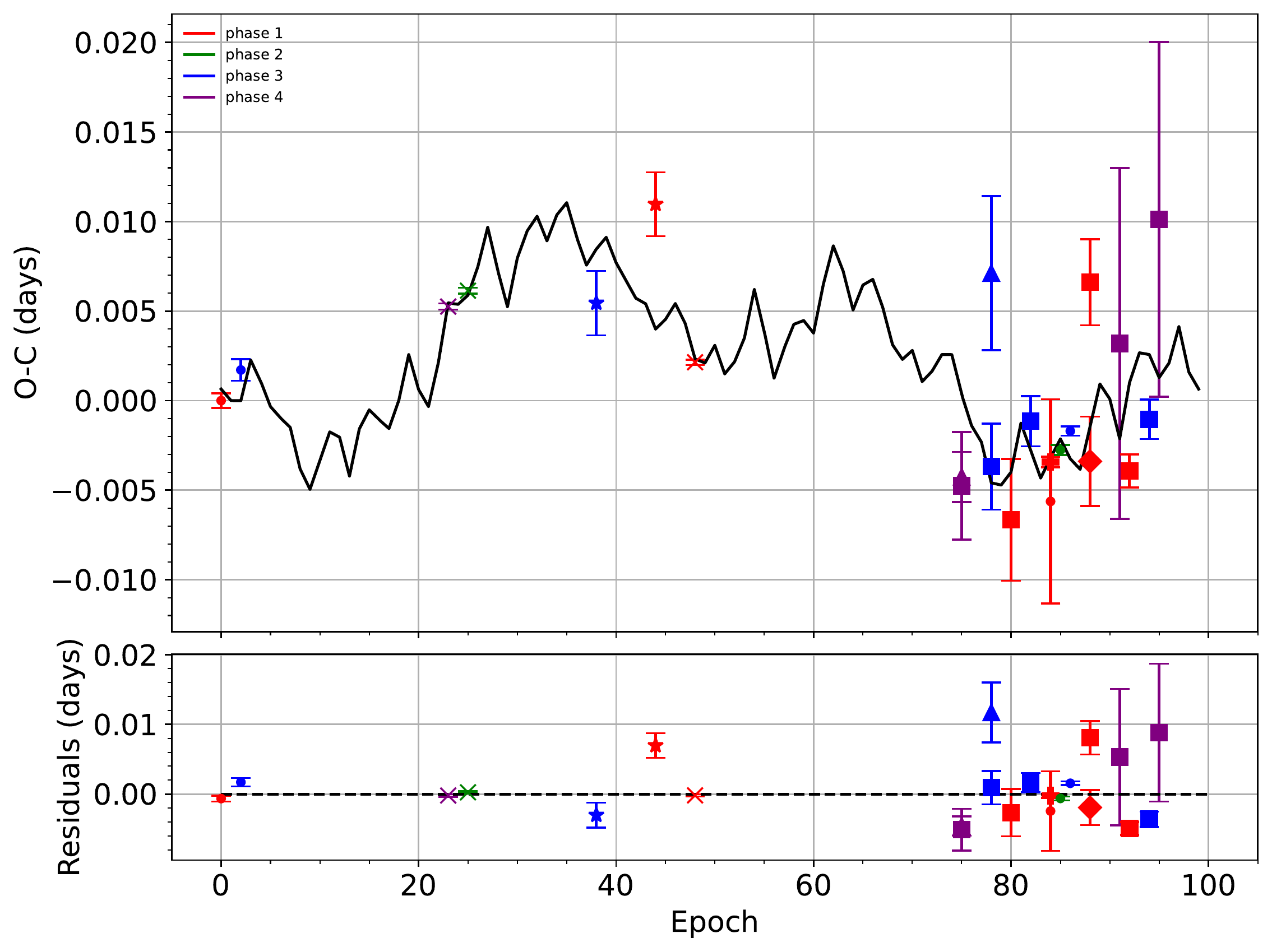}
    \end{tabular}
    \caption{O$-$C Diagram of \aumic b from initial model (left), \exostriker 2-planet model (center) and \exostriker 3-planet model (right)}, color-coded as a function of transit number modulo 4, using the \exofast-generated measured midpoint times and the calculated midpoint times for all 23 \aumic b transit data sets from Table \ref{table:datasets}. Here, we explore the impact the rotational modulation of starspots, given the 7:4 spin-orbit commensurability, may have on the observed TTVs. The weighted averages of the distribution of TTVs for each of the four stellar longitude crossing phases are in Table \ref{table:ttvmod}; they show correlations between phases 1 \& 3 and phases 2 \& 4. However, after subtracting the 2 and 3 planet models from the TTV data, the weighted averages show no clear correlation of more than $\sim$2 minutes, indicating that stellar spots of \aumic are not a significant factor in accounting for the timing of transits for \aumic b. The epochs are relative to the first \tess transit of \aumic b.
    \label{fig:ocphase}
\end{figure*}

\begin{deluxetable}{c|c|c|c}
    \tablecaption{\label{table:ttvmod}Weighted averages of the respective initial, 2-planet-model-subtracted, \& 3-planet-model-subtracted TTVs as a function of transit number modulo 4.}
    \tablehead{Phase & TTV$_{\rm i}$ (min) & TTV$_{2}$ (min) & TTV$_{3}$ (min)}
    \startdata
1 & 0.52 $\pm$ 1.17 & -0.90 $\pm$ 0.59 & -0.40 $\pm$ 0.59  \\
2 & 4.61 $\pm$ 1.00 & 1.11 $\pm$ 0.52  & 0.06 $\pm$ 0.52   \\
3 & 0.99 $\pm$ 1.28 & 0.61 $\pm$ 0.79  & 1.88 $\pm$ 0.79   \\
4 & 4.89 $\pm$ 1.13 & -0.36 $\pm$ 0.54 & -0.40 $\pm$ 0.54
    \enddata
\end{deluxetable}

To explore the impact of spot modulations on TTVs, we calculated the weighted average of the initial, 2-planet model, and 3-planet model TTVs for each of the four phases (Table \ref{table:ttvmod}); from there we can see correlations between phases 1 \& 3 and phases 2 \& 4. However, after subtracting the 2-planet and 3-planet models from the respective TTVs, these correlations disappeared. Thus, no such ``modulo 4'' pattern is observed in the \aumic b transit times (Figure \ref{fig:ocphase}), with transits being neither systematically late nor early in phase with the stellar longitude crossings. The three \spitzer transits are from three different sets of stellar longitude crossings, and one of the R-M observation midpoint times is from the fourth set of stellar longitude crossings; all four transit-midpoint times are systematically late w/r/t to the \tess transit times by $\sim$5-10 minutes. The \tess transits themselves also encompass three of the four stellar longitude crossing sets, and the variation in transit times between them is less than a couple of minutes.

Third, \citet{gilbert2021}, in deriving the \tess transit midpoint times, includes the joint modeling of the rotational modulation of the starspots in deriving the transit midpoint times. This analysis approach thus mitigates any impact the spot modulation has on the values of the derived transit midpoint times as estimated in \citet{oshagh2013}. Additionally, because of the joint modeling of the \tess transits, the impact of the rotational modulation is reflected in the posterior uncertainties and thus precision in the derived transit times. Our photodynamical analysis has shown that when modeling the rotational modulation of starspots with a GP in parallel with deriving transit midpoint times, as opposed to in series, the GP can absorb rather than introduce additional TTVs on the order of a few minutes for \aumic.

Thus, in light of the modeling presented herein and \citet{gilbert2021} the rotational modulation of starspots can likely be ruled out as an explanation for our observed TTVs between \tess and \spitzer. More detailed synthetic simulations will be needed in the future to further quantify the impact of the rotational modulation of \aumic starspots on the derived transit times, but that is beyond the scope of this work. We next turn to assess the impact of flares on the derived transit times.

\subsection{Stellar Flares}\label{sec:flares}

Like stellar spots, stellar flares can impact the midpoint timing of the transits, particularly when they occur during ingress or egress, which happened several times during the \tess transits. For the ground-based transit data, it is difficult to resolve the flares due to the relatively lower photometric precision, and additionally \aij and \exofast do not include joint modeling of flares during transits. The ground-based transit times also possess larger timing uncertainties, which may mask the impact of flares.

Several flares occurred during the \tess observations, but \citet{gilbert2021} jointly modeled them with the {\tt bayesflare} \citep{pitkin2014} and {\tt xoflares} packages, some of which do occur during egress and thus could impact the derived transit midpoint times were they not modeled jointly. Additionally, the O--C diagram from our photodynamical analysis demonstrates that the derived transit times are sensitive to the methods employed for accounting for the stellar flares. However, the analysis done by \citet{gilbert2021} shows no dependence of transit timing on activity after marginalizing over models for the flares and spot modulation, and the \tess transit times of \aumic b are fairly constant to within $\sim$4 minutes.

The second partial \spitzer transit of \aumic b contains an obvious flare during transit, which we marginalize over in our modeling, thus impacting our derived timing uncertainties. Additionally, the first and and third \spitzer transits may show some small flares pre-ingress which we do not account for in our modeling. As with spots, however, flares are smaller in amplitude at 4.5 $\mu$m than at visible wavelengths and thus should have a relatively smaller impact on the derived transit times. However, we find that the first two \spitzer transits times are consistent with one another, and the third \spitzer transit is most contemporaneous with the R-M transit observations.

For the \spirou + \ishell R-M transit observation, the analysis took into account the impact of magnetic activity and flares by constructing a model using a similar approach from \citet{donati2020}, which is then subtracted from the RV observations. Similarly, for the \espresso data, the stellar activity was modeled using the \celerite package's GP that is described by a Mat\'ern 3/2 kernel and then subtracted from the RV data.

All five R-M and \spitzer transits show significantly deviant and late transit times; it would be difficult and randomly unlucky to have randomly timed flares during the different transits all impact the derived transit midpoint times in the same way -- late, as opposed to early. While these results are based on a relatively small number of transits, it seems unlikely that flares can account for the observed TTV excesses. More detailed simulations in the future will be needed to assess the impact of flares on the transit times of \aumic, but that is beyond the scope of this work. We next turn to consider the possibility of additional planets in the \aumic system.

\subsection{Existence of a planet d candidate?}\label{sec:dcandidate}

In the previous subsections, we explored the stellar activity of \aumic and the significance of its impact on the TTVs, both through the rotational modulation of starspots and flares. It is possible that there is some unaccounted-for effect in the derived TTV uncertainties, but we deem this scenario unlikely given the above marginalization over our activity models in the derived transit time posteriors. Thus, we turn to another possibility: the presence of a third planet.

\begin{figure}
    \centering
    \includegraphics[width=0.48\textwidth]{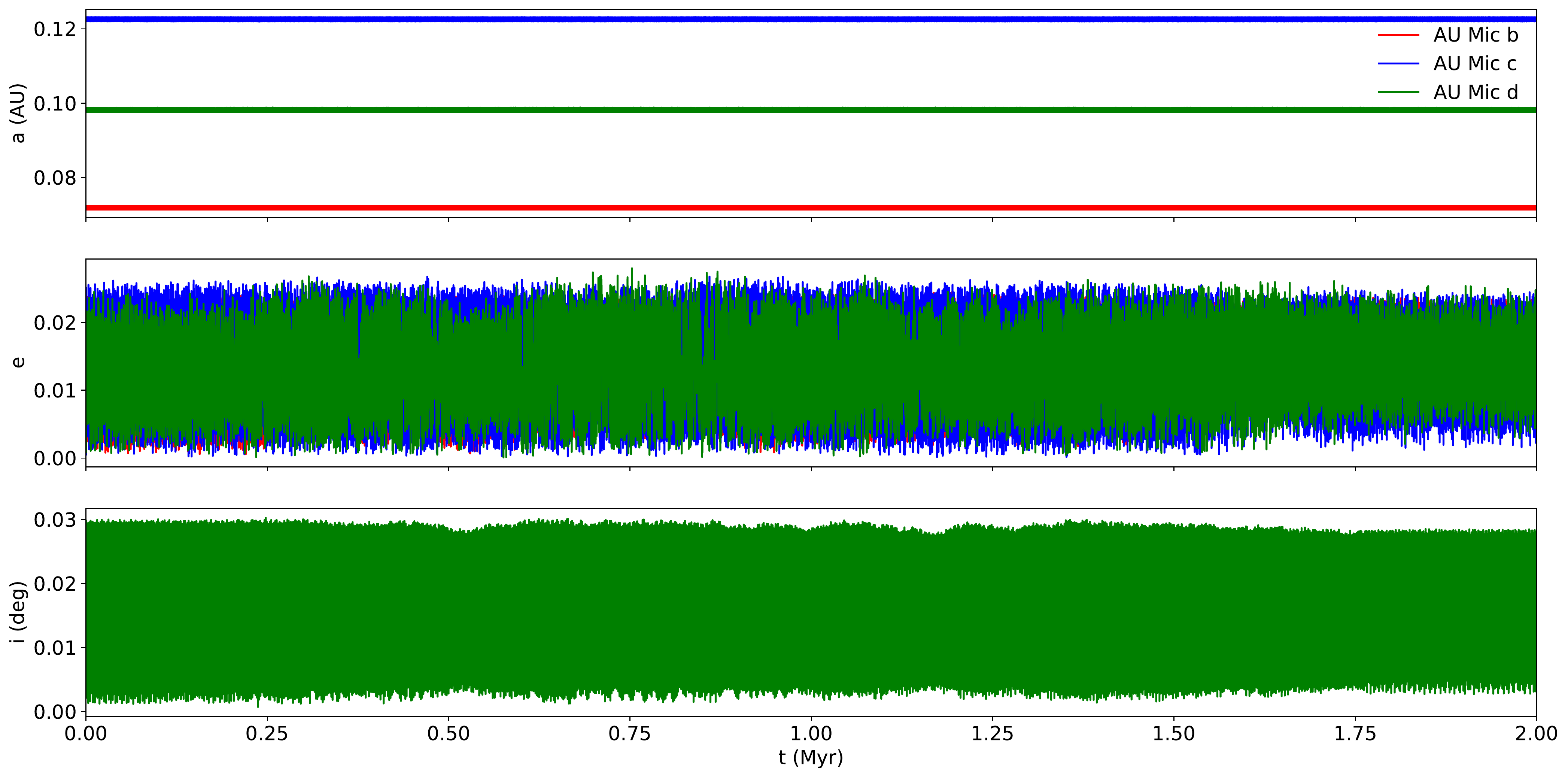} \\
    \includegraphics[width=0.48\textwidth]{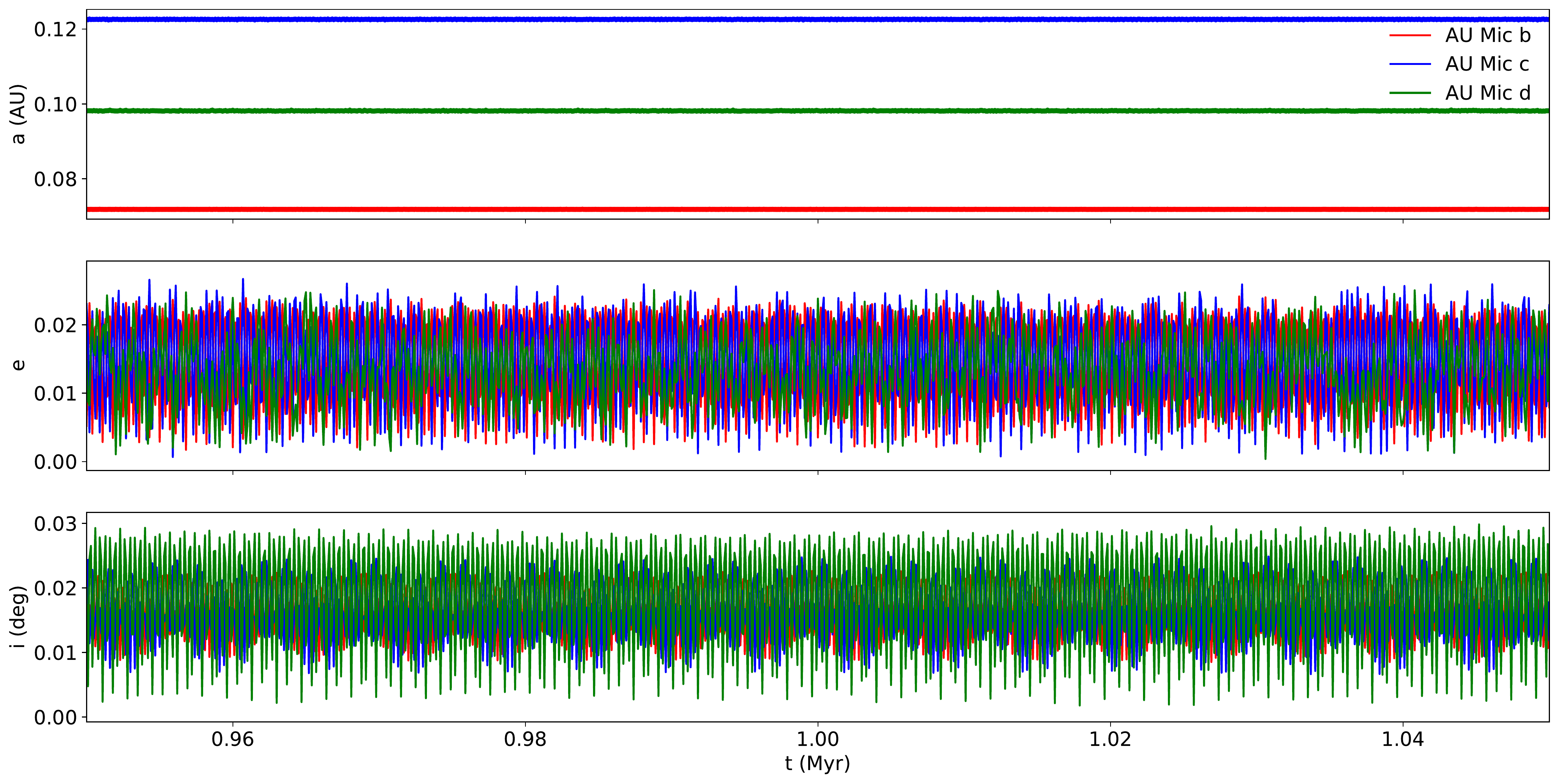}
    \caption{\rebound model of the stability of 3-planet system for \aumic. The top three plots span 2 Myrs while the bottom three plots are zoomed in between 0.95 and 1.05 Myrs.}
    \label{fig:rebound}
\end{figure}

\citet{cale2021} explored additional candidate RV planet signals when modeling the RVs of the \aumic system. One candidate period that was explored in particular was the presence of a planet in-between b and c, a ``middle-d'' non-transiting planet with a period of 12.742 days. Such a planet would establish that the \aumic system is in a 4:6:9 orbital commensurability and result in significant TTVs. In $\S$\ref{sec:3planetmod}, we modeled this 3-planet configuration with \exostriker based on this candidate RV signal from \citet{cale2021}. Since the AMD criterion indicated that this configuration is unstable, we tested its stability with N-body packages, including \rebound \citep{rein2012, rein2015} and \mercury \citep{chambers1999}, the latter of which was used as a consistency check. The \mercury simulation ran for 10 Myrs while the \rebound simulation ran for 2 Myrs; both indicated that the 3-planet configuration from $\S$\ref{sec:3planetmod} is stable (see Figure \ref{fig:rebound} for outputs from \rebound). Additional possible configurations, including having a planet beyond c, is beyond the scope of this work.  While we find that our TTVs are consistent with a possible third ``middle d'' transiting planet, we neither confirm nor rule out its existence. If this third planet does exist, our dynamical modeling implies the impact on the observed TTVs will be more readily apparent with additional TTV measurements in 2021 and the next several years, as the TTVs for \aumic b would deviate further from a linear ephemerides, and given the ``curvature'' in the TTVs from the first three years of TTVs presented herein.

\subsection{Implication on the Mass of AU Mic c}\label{sec:implanetc}

With only three transits of \aumic c, all with relatively larger transit timing uncertainties, we do not derive a TTV mass constraint for \aumic b.  However, from the perturbations of the transit times of \aumic b, we are able to place constraints on the mass of \aumic c. Table \ref{table:planetcchar} summarizes our \exostriker and photodynamical TTV masses with the RV masses from \citet{cale2021}. In both our 2-planet and 3-planet \exostriker models, the mass of \aumic c is approximately that of Neptune. Both photodynamical models determine the mass of \aumic c to be more comparable to that of Uranus instead of the more massive Neptune. The densities of \aumic c using the masses from \exostriker analyses are quite high. We instead adopt as our final mass estimate for \aumic c of $10.8^{+2.3}_{-2.2}$ \mear, the value from our ``a'' photodynamical model which jointly models the stellar activity with the transit dynamics with no third planet. This value is consistent with the mass for \aumic c as determined from RVs (as are all the masses we derive). Note that we only list the masses of \aumic c; we currently do not have enough transits nor enough precision on the timing of transits for \aumic c to place a meaningful constraint on the mass of \aumic b.

\begin{deluxetable}{l|c|l|l}
    \tablecaption{\label{table:planetcchar}Characteristics of \aumic c.}
    \tablehead{Property & Unit & Quantity & Ref}
    \startdata
\multirow{6}{*}{Mass}    & \multirow{6}{*}{\mear}      & 9.60$^{+2.07}_{-2.31}$  & K$_{J1}$ \citep{cale2021}  \\
                         &                             & 14.12$^{+2.48}_{-2.71}$ & K$_{J2}$ \citep{cale2021}  \\
                         &                             & 24.8 $\pm$ 1.2          & \exostriker 2-planet Model \\
                         &                             & 20.1 $\pm$ 1.6          & \exostriker 3-planet Model \\
                         &                             & 10.8$^{+2.3}_{-2.2}$    & Photodynamical Model a     \\
                         &                             & 13$^{+5}_{-2}$          & Photodynamical Model b     \\
\hline
Radius                   & \rear                       & 2.79$^{+0.31}_{-0.30}$  & \citet{gilbert2021}        \\
\hline
\multirow{6}{*}{Density} & \multirow{6}{*}{g/cm$^{3}$} & 2.44$^{+0.97}_{-0.98}$  & K$_{J1}$ \citep{cale2021}  \\
                         &                             & 3.58 $\pm$ 1.35         & K$_{J2}$ \citep{cale2021}  \\
                         &                             & 6.28$^{+2.12}_{-2.05}$  & \exostriker 2-planet Model \\
                         &                             & 5.08$^{+1.74}_{-1.69}$  & \exostriker 3-planet Model \\
                         &                             & 2.75$^{+1.09}_{-1.05}$  & Photodynamical Model a     \\
                         &                             & 3.3$^{+1.7}_{-1.2}$     & Photodynamical Model b
    \enddata
\end{deluxetable}

\subsection{Limitation of Our Joint Modeling \& Future Work}\label{sec:limitation}

In this subsection, we discuss some of the limitations of our analysis. First, we do not perform joint RV + TTV modeling because the RVs require a custom treatment of stellar activity that is an order of magnitude larger in effective amplitude relative to the Keplerian orbital reflex motion of the star \citep{cale2021}. We also sought to conduct an independent TTV modeling to compare with the RV analysis, given the complexity of the RV model. Additionally, to our knowledge, no prior work exists in performing a joint RV + TTV modeling in the presence of such stellar activity as exhibited by \aumic.

Second, in RVs it is common to adopt a ``jitter'' term representing the white-noise measurement error that is not captured by the formal measurement uncertainties and added in quadrature, without which Keplerian orbital fits can yield reduced-$\chi^{2}\gg1$. A similar approach may be employed for \aumic in a future study for the TTV modeling -- including a ``jitter'' term that accounts for timing uncertainties not captured by our modeling of spots, flares and orbital dynamics. We do not undertake such an analysis herein for several reasons. First, \exostriker does not include a ``jitter'' TTV error term, and thus we cannot marginalize over this parameter; e.g., we would need to assume a ``jitter'' timing term and inflate our timing measurement uncertainties. Second, our timing data is fairly in-homogeneous, spanning a range of precision and wavelength from ground and space facilities; we would need multiple independent jitter terms that are not well constrained; our \spitzer data that most deviates from a two-planet TTV model is least impacted by stellar activity. Third, as shown in \cite{szabo2021}, due to the 7:4 period commensurability of the stellar rotation to the orbital period of \aumic b, the TTV ``jitter'' from spots will not be a white noise term; given the known stellar rotation period, the timing impact is not simply random and fairly deterministic, as we have undertaken our ``mod-4'' transit analysis to identify whether or not the transit times correlation with the stellar longitude crossings. That being said, the flares are random in time and will randomly impact the measured transit times. Therefore, a ``jitter'' term would be appropriate for accounting for the impact of flares on measured the TTVs that are not captured by our flare modeling of the light curves.

Third, the uncertainties in the midpoint times from the R-M observations are likely to be underestimated. As allowing certain parameters, such as the orbital period and eccentricity, to remain free will result in degeneracy in the R-M model, setting such parameters fixed would avoid this issue but would cause the uncertainties to be underestimated.

As a consequence of these limitations in our analysis, we do not perform any statistically robust model comparisons between the 2- and 3-planet TTV models. Future observations and analyses will be necessary to reach a definitive conclusion for the hypothetical third planet candidate.

\section{Conclusion}\label{sec:conclude}

\aumic hosts a young nearby exoplanet system that serves as a useful laboratory for probing and characterizing young exoplanetary systems. We have collected 23 transits from \aumic b and 3 transits from \aumic c over the course of three years. We model the observed transits and derive new transit model posteriors. We have run two independent methods (\exostriker and photodynamical) in modeling the transits and the timing of those transits. Our observations and analyses of the transits of \aumic b and c show that \aumic b is exhibiting TTVs, consistent with \citet{szabo2021}, \citet{martioli2021}, and \citet{gilbert2021}. Our photodynamical model yields the mass for \aumic c of 10.8$^{+2.3}_{-2.2}$ \mear, consistent with the RV-determined mass in \citet{cale2021}. However, going beyond the work of \citet{szabo2021} and \citet{martioli2021}, our TTVs show timing excess of $>$5 minutes that appear to discrepant with a 2-planet model alone, particularly when comparing the \spitzer and \tess derived transit times, the former of which is presented for the first time herein. Further, we marginalize our TTV models over our models for the rotational modulation of stellar activity and flares. Consequently, stellar activity, while not excluded through statistically robust model comparison, does not appear likely to be able to account for the observed TTV excess. We mapped the O--C diagram taking into account \aumic's stellar spin 7:4 commensurability with the orbital period of \aumic b from \citet{szabo2021}, and we did not identify that the spot modulation results in a significant effect on the observed TTVs as a function of stellar longitude crossings. We explore a possible, representative, and non-exhaustive 3-planet configuration scenario that is: consistent with the identified non-transiting ``middle-d'' RV candidate signal in \citet{cale2021}, is dynamically stable, would establish the \aumic system of planets in a 4:6:9 orbital period commensurability, and can account for the observed TTV excess of our data.

Nonetheless, given the high level of stellar activity for \aumic, we cannot ignore the possibility that our modeling and marginalization over our stellar activity models does not fully account for some effects that significantly impact the observed TTVs. Thus, additional ground- and space-based TTVs in the next few years are needed to further vet the impact of stellar activity on the observed TTVs, or to confirm the possibility that the excess TTVs are due to the RV candidate highlighted in \citet{cale2021}, and to enable a more thorough search of the possible orbital periods for possible additional planets in the \aumic system. Such observations will be possible from the ground, with \cheops now, or with upcoming missions such as the {\sl Pandora} and {\sl Twinkle} missions, {\sl Ariel}, and/or {\sl JWST}.

PPP acknowledges support from NASA (Exoplanet Research Program Award \#80NSSC20K0251, TESS Cycle 3 Guest Investigator Program Award \#80NSSC21K0349, JPL Research and Technology Development, and Keck Observatory Data Analysis) and the NSF (Astronomy and Astrophysics Grants \#1716202 and 2006517), and the Mt Cuba Astronomical Foundation. DD acknowledges support from the TESS Guest Investigator Program grant \#80NSSC21K0108 and NASA Exoplanet Research Program grant \#18-2XRP18\_2-0136. EG acknowledges support from the NASA Exoplanets Research Program Award \#80NSSC20K0251. The material is based upon work supported by NASA under award \#80GSFC21M0002. LDV acknowledges funding support from the Heising-Simons Astrophysics Postdoctoral Launch Program, through a grant to Vanderbilt University.

This paper includes data collected by the TESS mission, which are publicly available from the Mikulski Archive for Space Telescopes (MAST). Funding for the TESS mission is provided by NASA’s Science Mission directorate. Resources supporting this work were provided by the NASA High-End Computing (HEC) Program through the NASA Advanced Supercomputing (NAS) Division at Ames Research Center for the production of the SPOC data products.

This work is based [in part] on observations made with the Spitzer Space Telescope, which was operated by the Jet Propulsion Laboratory, California Institute of Technology under a contract with NASA. Support for this work was provided by NASA through an award issued by JPL/Caltech. This research has made use of the NASA/IPAC Infrared Science Archive, which is funded by the National Aeronautics and Space Administration and operated by the California Institute of Technology.

This work makes use of observations from the Las Cumbres Observatory global telescope network. Part of the LCOGT telescope time was granted by NOIRLab through the Mid-Scale Innovations Program (MSIP). MSIP is funded by NSF.

This research made use of the PEST photometry pipeline\footnote{\url{http://pestobservatory.com}} by Thiam-Guan Tan. This research has made use of the NASA Exoplanet Archive and the Exoplanet Follow-up Observation Program website, both of which are operated by the California Institute of Technology, under contract with the National Aeronautics and Space Administration under the Exoplanet Exploration Program. This research has made use of the SIMBAD database, operated at CDS, Strasbourg, France. This research has made use of NASA’s Astrophysics Data System Bibliographic Services. This research has made use of an online calculator that converts a list of Barycentric Julian Dates in Barycentric Dynamical Time (BJD\_TDB) to JD in UT \citep{eastman2010}\footnote{\url{https://astroutils.astronomy.osu.edu/time/bjd2utc.html}}.

We also give thanks to Trifon Trifonov for his assistance in the use of the \exostriker package and analysis of the \aumic system.

\facilities{\brier:0.36 m ({\sl Moravian G4-16000 KAF-16803}), \cfht (\spirou), ExoFOP, Exoplanet Archive, IRSA, \lcogt ({\sl SAAO}:1 m \& {\sl SSO}:1 m; \sinistro), MAST, \pest:0.30 m (\sbig), \spitzer (\irac), \tess, \vlt:Antu (\espresso)}

\software{{\tt AstroImageJ} \citep{collins2017}, {\tt astropy} \citep{astropy2013, astropy2018}, {\tt batman} \citep{kreidberg2015}, {\tt bayesflare} \citep{pitkin2014}, \celerite \citep{foreman-mackey2017}, \celeritetwo \citep{foreman-mackey2017, foreman-mackey2018}, {\tt emcee} \citep{foreman-mackey2013}, \exofast \citep{eastman2019}, {\tt exoplanet} \citep{foreman-mackey2021}, \exostriker \citep{trifonov2019}, {\tt fleck} \citep{morris2020-1}, {\tt ipython} \citep{perez2007}, {\tt lightkurve} \citep{lightkurve}, {\tt matplotlib} \citep{hunter2007}, \mercury \citep{chambers1999}, {\tt numpy} \citep{harris2020}, \rebound \citep{rein2012, rein2015}, {\tt scipy} \citep{virtanen2020}, {\tt TAPIR} \citep{jensen2013}, {\tt xoflares} \citep{gilbert2021}}

\bibliography{bibliography} 

\appendix

\restartappendixnumbering
\section{corner plots from Main Exostriker Analysis}\label{sec:appendescor}

This section highlights the corner plots that were generated by \exostriker. All \exostriker corner plots are included here.

\begin{figure*}[b]
    \centering
    \includegraphics[width=0.98\textwidth]{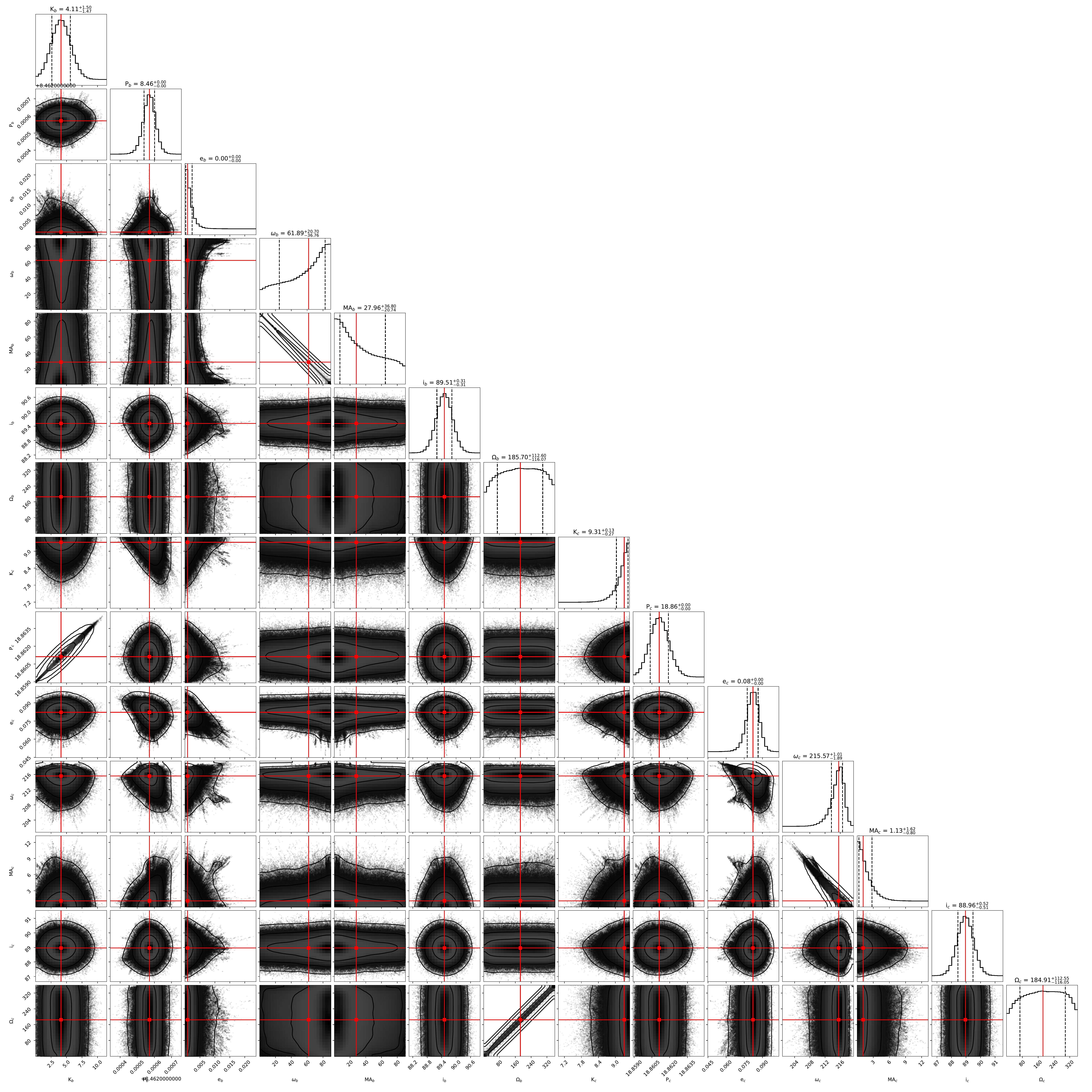}
    \caption{\exostriker-generated MCMC corner plot for \aumic 2-planet case. All TTVs are incorporated into this model.}
    \label{fig:2p_all_corner}
\end{figure*}

\begin{figure*}
    \centering
    \includegraphics[width=\textwidth]{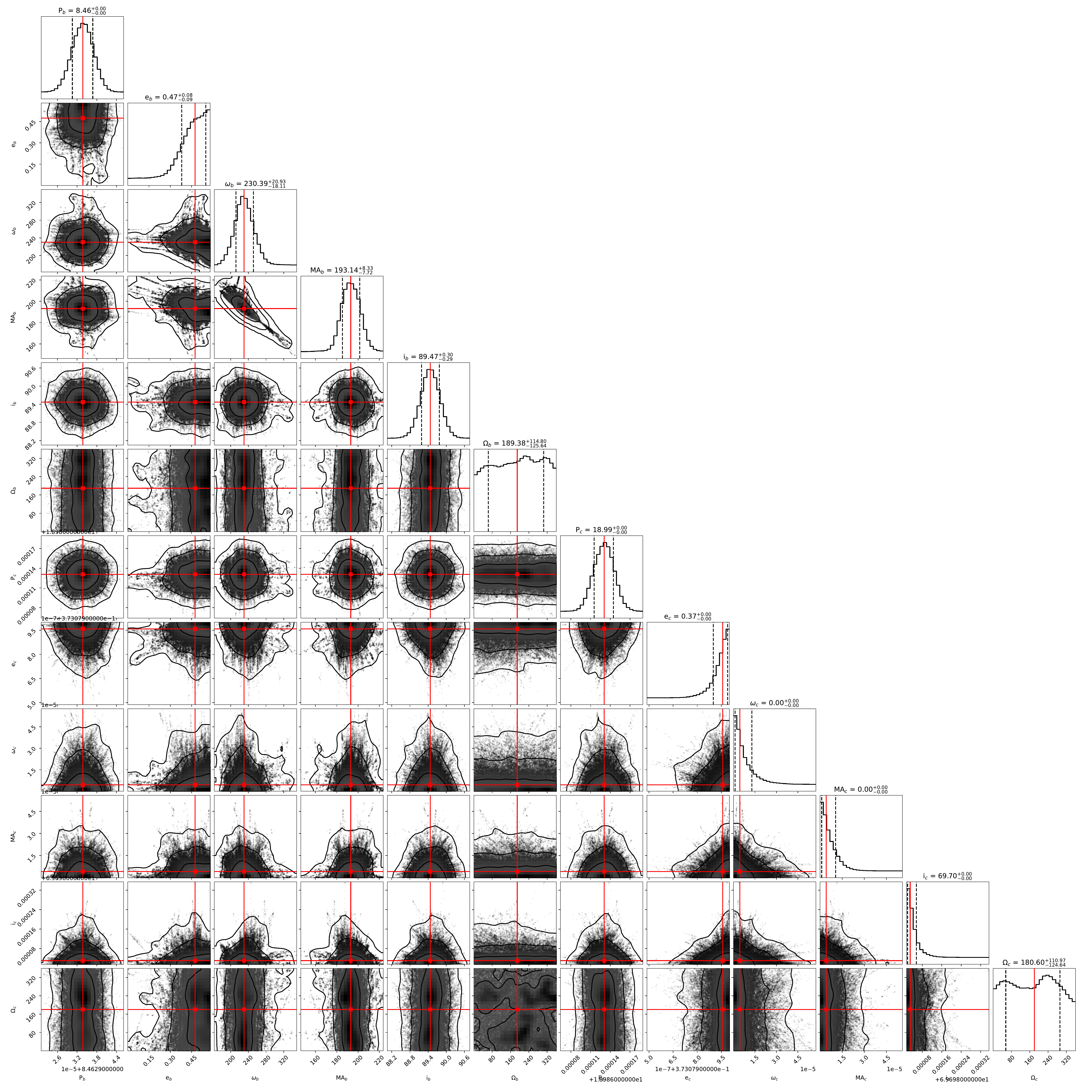}
    \caption{\exostriker-generated MCMC corner plot for \aumic mass-less planets case. All TTVs are incorporated into this model.}
    \label{fig:0p_all_corner}
\end{figure*}

\begin{figure*}
    \centering
    \includegraphics[width=\textwidth]{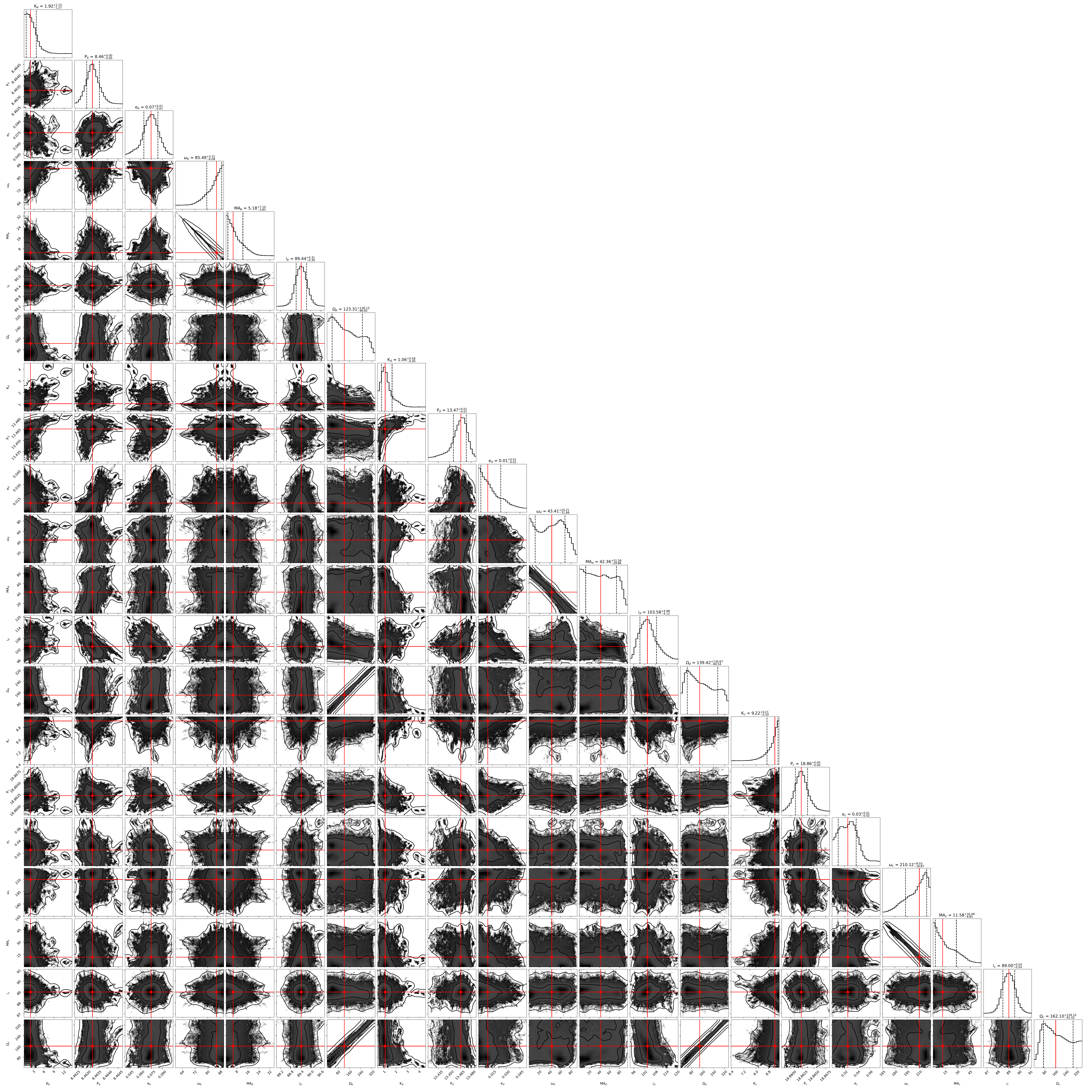}
    \caption{\exostriker-generated MCMC corner plot for \aumic 3-planet case. All TTVs are incorporated into this model.}
    \label{fig:3p_all_corner}
\end{figure*}

\clearpage
\restartappendixnumbering
\section{Corner Plots from Photodynamical Analysis}

This section highlights the corner plots that were generated by photodynamical analysis. The corner plots for both models a and b are included here.

\begin{figure*}[b]
    \centering
    \includegraphics[width=0.49\textwidth]{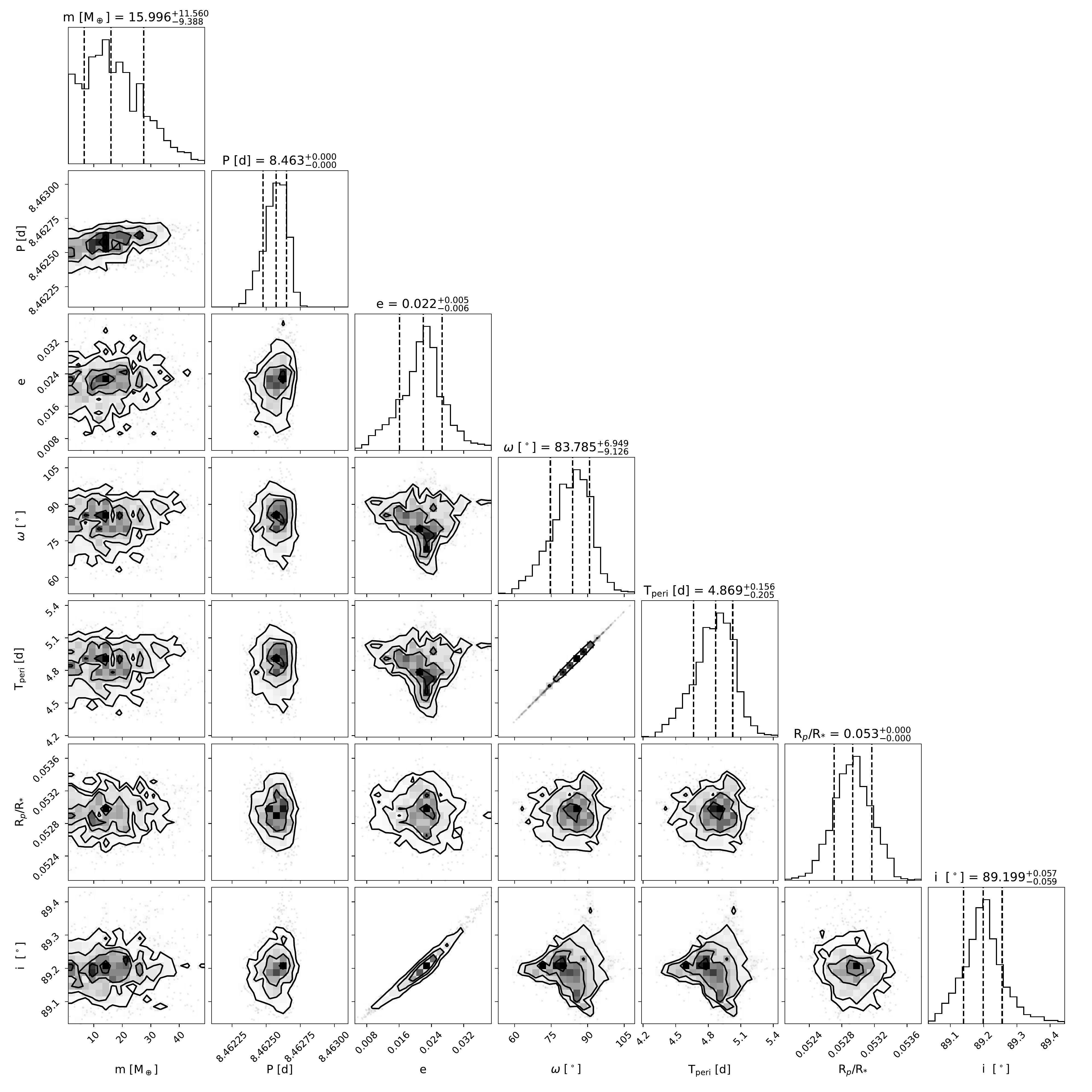}
    \includegraphics[width=0.49\textwidth]{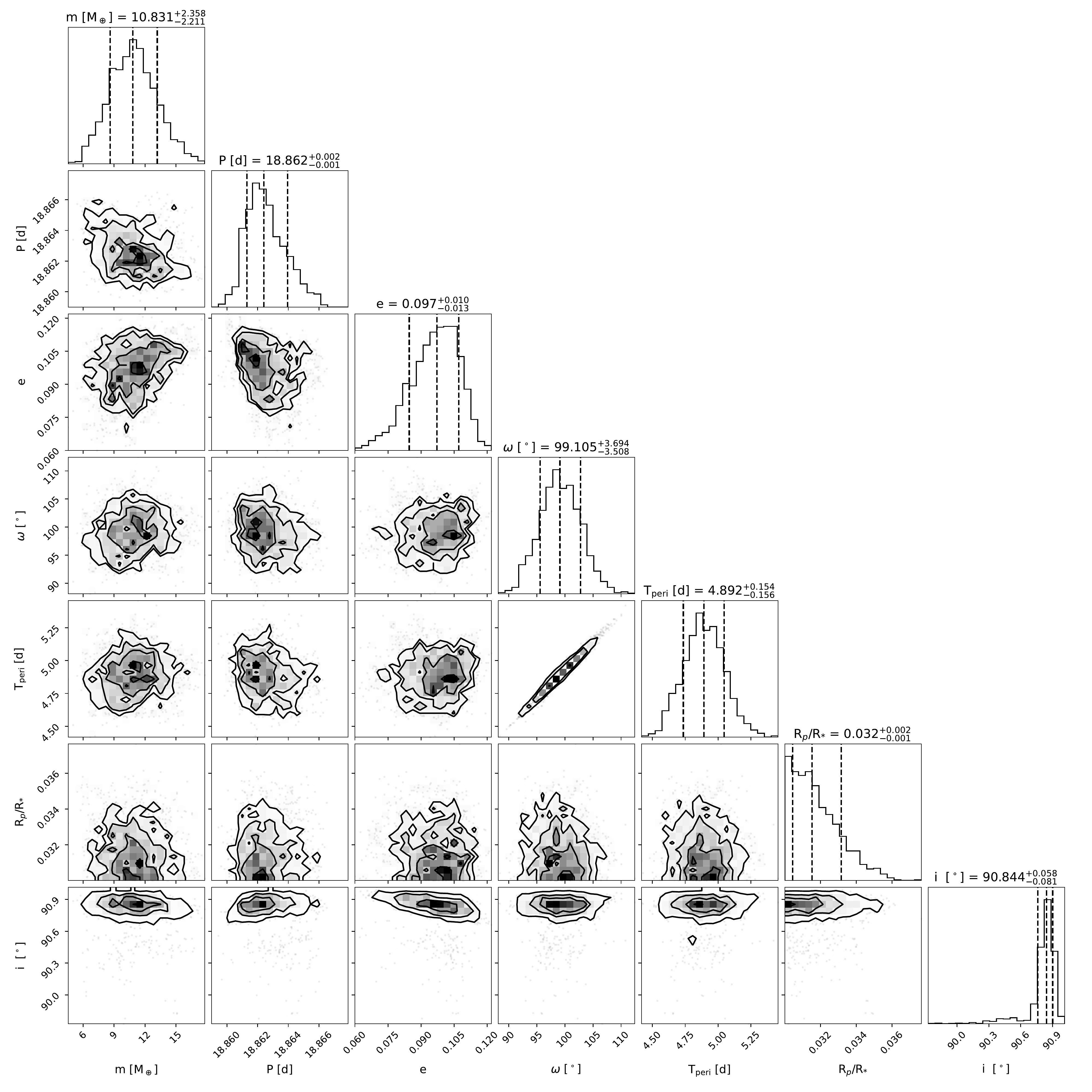}
    \caption{Corner plot for posterior distributions of orbital parameters from the photodynamical model a for \aumic b (left) and \aumic c (right).}
    \label{fig:photodyn_modela}
\end{figure*}

\begin{figure*}
    \centering
    \includegraphics[width=0.49\textwidth]{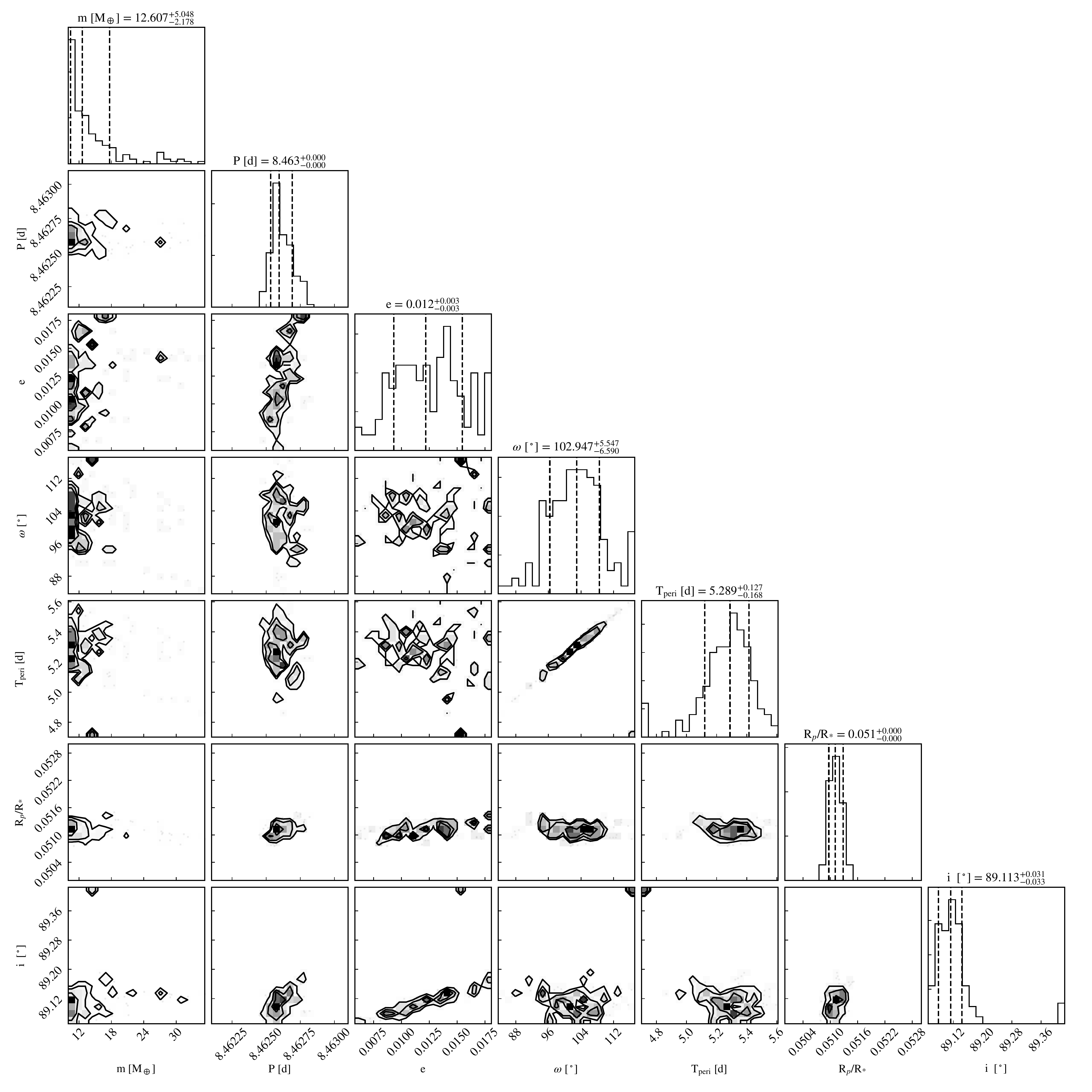}
    \includegraphics[width=0.49\textwidth]{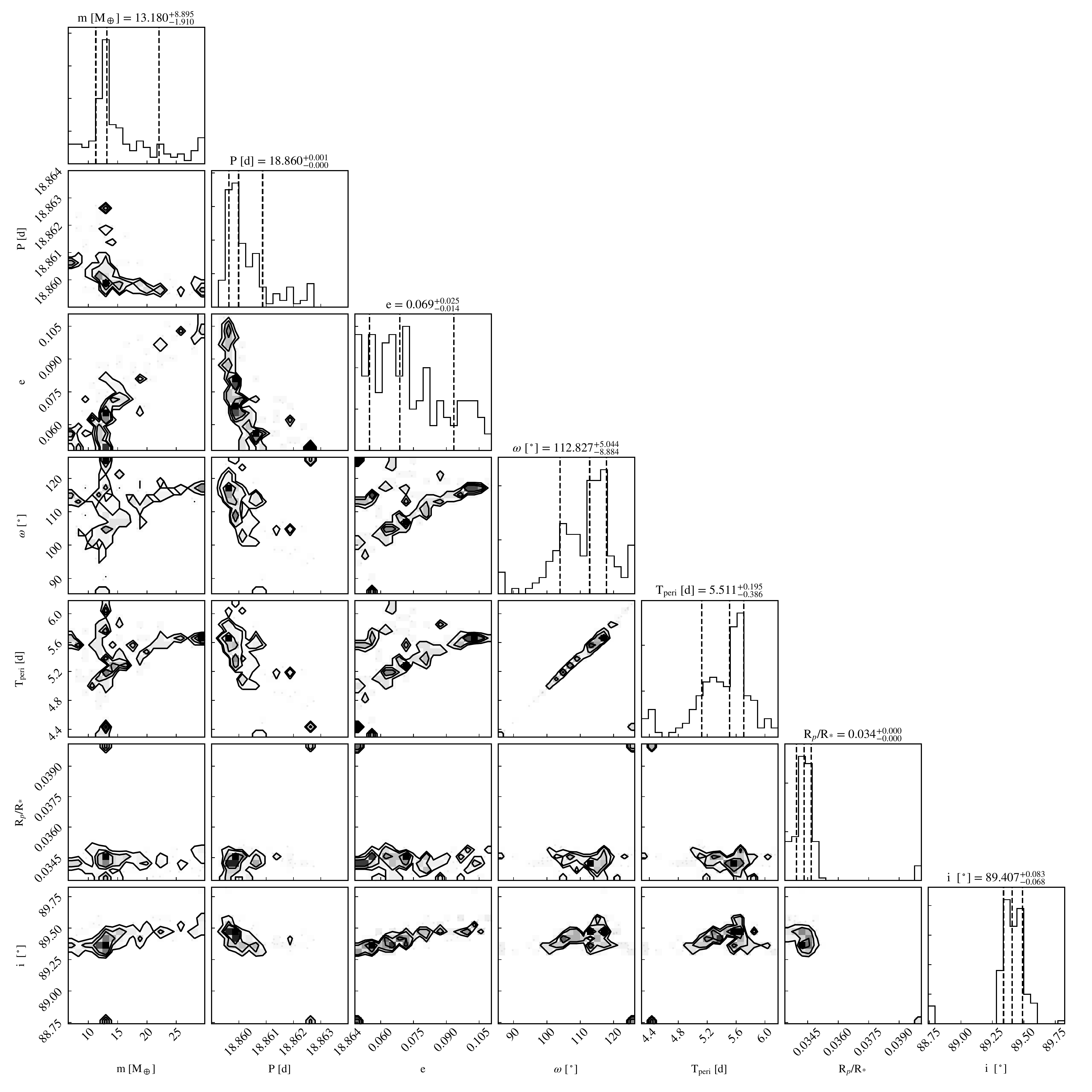}
    \caption{Corner plot for posterior distributions of orbital parameters from the photodynamical model b for \aumic b (left) and \aumic c (right).}
    \label{fig:photodyn_modelb}
\end{figure*}

\end{document}